\documentclass{article}
\usepackage{iclr2027_conference,times}
\usepackage{etoolbox}
\iclrfinalcopy
\makeatletter
\patchcmd{\@maketitle}
  {\lhead{Published as a conference paper at ICLR 2027}}
  {\lhead{Preprint}}
  {}{\PackageError{traceguard}{Unable to set the preprint header}{Check the title macro in the conference style.}}
\makeatother

\usepackage[T1]{fontenc}
\usepackage[utf8]{inputenc}
\usepackage{graphicx}
\usepackage{booktabs}
\usepackage{array}
\usepackage{multirow}
\usepackage{amsmath}
\usepackage{amssymb}
\usepackage{adjustbox}
\usepackage{xcolor}
\usepackage{hyperref}
\usepackage{url}
\usepackage{caption}
\usepackage{placeins}
\hypersetup{
  hidelinks,
  pdftitle={TraceGuard: Adaptive Multimodal Poison Filtering through Cross-Feature Rank Agreement},
  pdfauthor={Haoyang Li, Yaxin Xiao, Linyan Dai, Jiawen Fu, Zi Liang, Jason Xue, Qingqing Ye, Haibo Hu},
  pdfsubject={Multimodal data poisoning defense},
  pdfkeywords={data poisoning, multimodal learning, backdoor defense, data curation, robustness},
  pdflang={en-US}
}

\pdfmapline{+tgr TimesNewRomanPSMT <ec.enc <artifacts/followup_paper/fonts/times.ttf}
\pdfmapline{+tgb TimesNewRomanPS-BoldMT <ec.enc <artifacts/followup_paper/fonts/timesbd.ttf}
\pdfmapline{+tgi TimesNewRomanPS-ItalicMT <ec.enc <artifacts/followup_paper/fonts/timesi.ttf}
\DeclareFontFamily{T1}{tgtnr}{}
\DeclareFontShape{T1}{tgtnr}{m}{n}{<-> artifacts/followup_paper/fonts/tgr}{}
\DeclareFontShape{T1}{tgtnr}{b}{n}{<-> artifacts/followup_paper/fonts/tgb}{}
\DeclareFontShape{T1}{tgtnr}{bx}{n}{<-> artifacts/followup_paper/fonts/tgb}{}
\DeclareFontShape{T1}{tgtnr}{m}{it}{<-> artifacts/followup_paper/fonts/tgi}{}
\newcommand{\ArtifactFont}{\fontfamily{tgtnr}\selectfont}
\AtBeginEnvironment{table}{\ArtifactFont}
\AtBeginEnvironment{table*}{\ArtifactFont}
\AtBeginEnvironment{figure}{\ArtifactFont}
\AtBeginEnvironment{figure*}{\ArtifactFont}

\newenvironment{FollowupFigure*}{\begin{figure}[htbp]}{\end{figure}}

\newenvironment{FollowupTable*}{\begin{table}[htbp]}{\end{table}}

\title{TraceGuard: Adaptive Multimodal\\
Poison Filtering through\\
Cross-Feature Rank Agreement}
\author{\parbox[t]{0.96\textwidth}{\centering
Haoyang Li\textsuperscript{1}\quad
Yaxin Xiao\textsuperscript{1}\quad
Linyan Dai\textsuperscript{1}\quad
Jiawen Fu\textsuperscript{1}\\[0.3em]
Zi Liang\textsuperscript{1}\quad
Jason Xue\textsuperscript{2}\quad
Qingqing Ye\textsuperscript{1}\quad
Haibo Hu\textsuperscript{1}\\[0.6em]
{\normalfont\small
\textsuperscript{1}The Hong Kong Polytechnic University\\
\textsuperscript{2}Commonwealth Scientific and Industrial Research Organisation (CSIRO)}
}}

\begin{document}
\maketitle

\begin{abstract}
Multimodal training relies on image-text corpora collected from external sources, creating opportunities for attackers to poison the data.
Stealthy attacks can preserve plausible image-text pairs while concealing the differences used by detectors, so apparently clean data can still redirect the trained model.
We therefore ask which properties a poison set must preserve for the attack to remain effective.
A small poison set must still exert enough collective influence during training to induce the attacker's target behavior.
We analyze this influence in terms of how often an attack pattern occurs and how strongly the examples carrying it jointly affect the model.
This analysis motivates six corpus-level features that examine cross-modal neighborhoods, recurring text, and changes after text-span erasure without training the victim model.
We introduce TraceGuard, an adaptive rank-based filtering method that uses agreement among complementary feature rankings to identify suspicious examples.
It refines the selected set through shared patterns and adapts the removal threshold to each corpus without knowing the attack or poison rate.
Across 19 attack configurations spanning image-text learning, generative vision-language model fine-tuning, and encoder-transfer tests, TraceGuard removes an average of 98.4\% of poisoned examples and 5.4\% of clean examples.
After training on the filtered corpora, the residual attack metric is at most 1\% in 13 configurations.
Matched-removal controls and ablations support the contributions of sample selection and adaptive removal.
Stress tests also identify detection failures under adaptive attacks and unnecessary removal on poison-free corpora.
\end{abstract}

\section{Introduction}
Multimodal models learn from image-text corpora assembled from web crawls, public datasets, and third-party sources \citep{radford2021clip,jia2021align,gadre2023datacomp}.
This reliance on external data gives attackers an opportunity to insert examples that teach the model an incorrect association \citep{carlini2022poisoning}.
The poisoned examples need not look suspicious, as stealthy attacks can preserve natural text and plausible image-text pairings \citep{liang2024badclip,xu2024shadowcast}.
For the defender, the challenge is to remove these clean-looking poisons before training while preserving useful clean data, without knowing the attack's target or how much of the corpus is poisoned.

A natural response is to assess each pair's quality and consistency, using learned judgments \citep{fang2024datafiltering,zhu2024vdc} or agreement with nearby examples \citep{huang2025detecting,zhang2025lemon}.
These increasingly capable filters can identify suspicious data, but the differences they detect need not be essential to attack success.
Indeed, adaptive attacks can make poisoned representations resemble clean ones while retaining the target behavior \citep{tan2020bypassing,qi2023revisiting}.
As attackers optimize poisons against successive detection criteria, filtering becomes an arms race.
This motivates a different question, namely which properties sustain attack effectiveness and would weaken the attack if suppressed.

To induce the target behavior within a much larger clean corpus, a small poison set must exert a sufficiently strong collective influence on training.
Unlike an anomaly that can be concealed without changing what the model learns, this influence is necessary for the attack to succeed.
We call the properties sustaining this target-directed influence the \emph{attack pathway}.
Our first-order influence analysis connects attack effectiveness to how often an attack pattern occurs and how strongly the examples carrying it jointly affect the model.
These examples may share a trigger or a source-to-target association, and their repeated reinforcement of the same target behavior can leave observable relationships in the corpus.
For instance, repeating a target phrase across visually different examples can make their text representations similar even when their image representations differ.
We examine these relationships through six complementary \emph{pathway features} based on cross-modal neighborhoods, repeated text, and changes after text-span removal.
Fixed pretrained encoders allow us to compute these features without training the victim model or accessing its gradients.
We test their ability to distinguish poisoned from clean examples as attack strength varies.

Turning these features into removal decisions requires more than a single threshold, since the most informative feature varies across attacks and benign long-tail examples can also receive extreme values.
We introduce TraceGuard, an adaptive rank-based filtering method that compares complementary feature rankings to identify examples that consistently appear near the top.
Choosing how much data to remove must also balance leaving enough poisons to sustain an attack against discarding useful clean data.
TraceGuard uses agreement among these rankings to infer an initial removal threshold, then refines the selected set using shared text spans and differences in feature values.
The same rules adapt both the selected examples and the removal fraction to each corpus without knowing the poison rate.
We evaluate 19 attack configurations spanning image-text learning, generative VLM fine-tuning, and encoder-transfer tests.
TraceGuard achieves 98.4\% mean Poison Recall at 5.4\% mean Clean FPR, and the residual attack metric after training on the filtered corpora is at most 1\% in 13 configurations.
Equal-count removal controls and ablations distinguish the contributions of sample selection and adaptive removal from the effect of discarding more data.
Stress tests complement these results by identifying remaining failures under adaptive attacks and false positives on poison-free corpora.

\section{Related Work}

\paragraph{Multimodal Data Poisoning}
Targeted poisoning and backdoor attacks can redirect image-text associations by modifying a small portion of the training corpus \citep{carlini2022poisoning,yang2023mmpoison}, and attacks on web-hosted datasets establish practical routes for introducing such modifications \citep{carlini2024webpoisoning}.
Subsequent work extends these attacks through optimized visual triggers \citep{liang2024badclip,bai2024badclip}, text-only poisoning \citep{yao2025toxictextclip}, and concept-level manipulation \citep{hu2025c2attack,zhang2025mpnav}.
For generative VLMs, instruction-tuning backdoors induce attacker-chosen responses \citep{liang2024vltrojan,lyu2024trojvlm,yin2025badmllm}, while semantic poisoning changes how models describe concepts and relations \citep{xu2024shadowcast,zhong2025badsem,zhang2025tokenswap}.
ImgTrojan further shows that poisoning image-caption pairs can make ordinary images activate jailbreak behavior \citep{tao2025imgtrojan}.
Importantly, attack effectiveness need not depend on visibly incorrect training examples, as shown by clean-label and hidden-trigger attacks \citep{turner2019labelconsistent,saha2020hidden} and stealthy multimodal poisons \citep{xu2024shadowcast,liang2026badclippp}.
Adaptive attacks further reduce the differences exploited by detectors \citep{tan2020bypassing,qi2023revisiting}.
These developments motivate examining the collective training influence needed for attack success rather than relying on a particular visible anomaly.

\paragraph{Poisoning Defenses}
Defenses differ in both when they intervene and what information they use.
Activation clustering and spectral methods identify poisoned groups in learned representations \citep{chen2018activation,tran2018spectral,hayase2021spectre}, while poison forensics traces attacks through their effects on model parameters \citep{shan2022poisonforensics}.
Cognitive Distillation detects small input patterns that preserve a trained model's outputs \citep{huang2023cognitive}, whereas CSFF detects label consistency under cross-sample latent feature fusion \citep{zhang2026csff}.
ASSET actively separates clean and poisoned data through optimization and adaptively determines how many samples to remove \citep{pan2023asset}.
DataElixir instead repairs poisoned images and labels through diffusion-based reconstruction and relabeling \citep{zhou2024dataelixir}.
Within multimodal learning, RoCLIP and OTCCLIP revise image-text pairings during pre-training \citep{yang2023roclip,zhang2025otcclip}, whereas SafeCLIP and Semantic Shield constrain which cross-modal associations the model learns \citep{yang2024safeclip,ishmam2024semanticshield}.
Related efforts protect visual instruction tuning through regularization \citep{xun2025robustit}.
CleanCLIP, CleanerCLIP, and RVPT repair compromised models through fine-tuning or prompt tuning \citep{bansal2023cleanclip,xun2025cleanerclip,zhang2025rvpt}, while InverTune combines trigger inversion with selective fine-tuning \citep{sun2026invertune}.
CGD combines CLIP predictions with those of a potentially backdoored classifier to separate data for subsequent unlearning \citep{xu2025cgd}.
At inference time, BDetCLIP and BackdoorIDS detect triggered inputs \citep{niu2025bdetclip,huang2026backdoorids}, while PurMM suppresses visual tokens with anomalous attention \citep{jiang2026purmm}.
Subspace Detection also uses cross-modal relationships, comparing a test image with semantic variants of its predicted text in the potentially backdoored CLIP model \citep{song2026subspace}.
For dataset filtering, BYE clusters cross-modal attention entropy after fine-tuning the victim model on the supplied data \citep{rong2025bye}.
VDC and MFBD use pretrained vision-language models to identify inconsistency between visual content and textual supervision \citep{zhu2024vdc,huang2026mfbd}, and local-density methods detect poisons through their representation neighborhoods \citep{huang2025detecting}.
Our focus is to connect attack-pathway analysis to adaptive rank-based filtering before victim-model training, using agreement among complementary feature rankings to determine both which examples to remove and how much data to retain without knowing the poison rate.

\paragraph{Multimodal Data Curation}
Data curation also seeks useful training examples, with LAION, DataComp, and MetaCLIP studying large-scale selection and balancing \citep{schuhmann2021laion400m,gadre2023datacomp,xu2024metaclip}.
Beyond embedding similarity, filters use generated captions \citep{mahmoud2024sieve}, learned quality judgments \citep{fang2024datafiltering,wang2024multimodalfilters}, or fine-grained image-text alignment \citep{nam2025clip4dm}.
LEMoN detects mismatched pairs using multimodal neighbors \citep{zhang2025lemon}, while Mixture-of-Scores and FLYT combine quality estimates or learn selection criteria from downstream utility \citep{wu2025mixture,shechter2025filter}.
These approaches establish useful tools for assessing image-text data, but quality and malicious training influence are different targets.
TraceGuard builds on corpus-level comparisons to address the latter, including poisons whose individual image-text pairs remain plausible.

\section{Problem Setup}
\label{sec:threat-model}
The defender collects a training corpus \(D=\{u_i=(x_i,y_i)\}\), where each visual input \(x_i\) is paired with a caption, instruction, answer, or other training text \(y_i\).
Because the integrity of every source cannot be verified, clean and poisoned examples may enter the same corpus.

\paragraph{Attacker.}
We consider targeted poisoning and backdoor attacks.
Both aim to teach an incorrect cross-modal mapping while preserving normal behavior on clean inputs, but a backdoor activates that mapping through a trigger.
The target may be a classification label, a retrieval result, or a generated response.
The attacker can inject or modify image-text pairs, including pairs that remain semantically plausible, but cannot alter the training procedure or modify the model after training.
It knows the training task, victim architecture and objective, collection sources, and public defenses, and has resources to optimize each poisoned example.
Adaptive evaluations additionally grant complete knowledge of TraceGuard.

\paragraph{Defender and goal.}
The defender filters the corpus before training and can use public pretrained image and text encoders.
It does not know whether poisoning is present, which examples are poisoned, the poison rate, or the attack's construction and target.
Victim-model gradients and post-training behavior are unavailable at this stage.
The goal is to remove enough poisons to suppress the target behavior while preserving clean training data.
Because the poison rate is unknown, the filtering amount must be determined from the corpus rather than supplied as an attack-dependent budget.
Appendix~\ref{app:threat} gives task-specific examples and resource assumptions.

\section{Attack Pathways and TraceGuard}
\label{sec:cross-sample-evidence}
\subsection{From Training Influence to Corpus Features}
\label{sec:pathway-features}
An effective poison set must collectively change what the model learns, even when its individual image-text pairs appear benign.
To understand this requirement, we analyze how examples sharing an attack pattern, such as a trigger or a source-to-target association, contribute to the target behavior.
Influence analysis \citep{koh2017influence,pruthi2020tracin,deng2025versatile} relates their combined effect to how often the pattern occurs and whether their training gradients reinforce rather than cancel one another.
Examples with similar representations can reinforce one another when their training gradients also push model outputs toward the same target.
Image-text similarity and density assess pair plausibility and typicality rather than this combined training effect.
We call them \emph{non-pathway features} because poisons can retain values similar to clean data while still reinforcing the target behavior.
Appendix~\ref{app:pathways} formalizes the attack-success condition and derives these relationships under a first-order approximation.
The analysis concerns the training effect an attack must achieve, while detection itself requires no victim-model training.

Since neither the poisoned examples nor their gradients are known, we examine relationships in the full corpus using fixed image and text representations.
A repeated target phrase can link otherwise different images, while a source-to-target association can place the same examples in different image and text neighborhoods.
These possibilities motivate measuring text-span frequency, comparing neighborhoods across modalities, and testing whether removing a span changes their agreement.
The resulting six pathway features capture observable consequences suggested by the analysis, rather than reconstructing victim gradients.
Because natural repetitions and benign mismatches can produce similar values, their ability to separate poisons from clean data is an empirical prediction, not a guarantee of the derivation.

\subsection{Pathway Features}
\label{sec:feature-definitions}
We compute the six features below from fixed pretrained representations, with Neighbor Support contributing one feature in each direction.
A text span is a short segment of an example's paired text.

\noindent \textbf{Neighbor Support.}~
Neighbor Support asks whether examples close in one modality remain related in the other, using neighboring examples rather than only the paired image and text as the reference.
We make this comparison using fixed, \(\ell_2\)-normalized image and text representations \(e_x(x_i)\) and \(e_y(y_i)\), whose inner products give cosine similarity.
To compare \(u_i\) with semantically related samples, we let \(\mathcal{N}_x^k(i)\) and \(\mathcal{N}_y^k(i)\) denote its \(k\) nearest neighbors in the image and text representation spaces, respectively, excluding \(u_i\) itself.
In the image-to-text direction, we select neighbors of \(x_i\) and average the similarity between \(y_i\) and their paired texts.
Reversing these roles gives the text-to-image direction, a form of directional neighborhood comparison also used for image-caption label-error detection \citep{zhang2025lemon}.
We define Neighbor Support in the two directions as
\begin{equation}
\begin{aligned}
S_{x\to y}(i)
&=
\frac{1}{k}
\sum_{j\in \mathcal{N}_x^k(i)}
\left\langle e_y(y_i),e_y(y_j)\right\rangle,
\\
S_{y\to x}(i)
&=
\frac{1}{k}
\sum_{j\in \mathcal{N}_y^k(i)}
\left\langle e_x(x_i),e_x(x_j)\right\rangle.
\end{aligned}
\label{eq:neighbor-support}
\end{equation}
A small value means that neighbors selected in one modality are dissimilar to example \(i\) in the other.

\noindent \textbf{Neighbor Disagreement.}~
Neighbor Disagreement compares two ways of locating related examples, through image neighbors followed by text neighbors or in the reverse order.
For each modality, we normalize an example's weights to its \(k\) nearest neighbors to sum to one.
Collecting these weights for all examples gives the image and text transition matrices \(P_x\) and \(P_y\).
The \(i\)-th rows of \(P_xP_y\) and \(P_yP_x\) give the distributions over examples reached from \(i\) in the two orders.
The matrix commutator \(P_xP_y-P_yP_x\) compares these distributions, and its row-wise norm defines Neighbor Disagreement
\begin{equation}
D_{x\leftrightarrow y}(i)
=
\left\|
(P_xP_y)_{i,:}
-
(P_yP_x)_{i,:}
\right\|_2 .
\label{eq:neighbor-disagreement}
\end{equation}
This use of a commutator follows earlier work that compares neighborhood structures across modalities with order-dependent diffusion operators \citep{bronstein2013making,shnitzer2019recovering}.
A large \(D_{x\leftrightarrow y}(i)\) indicates that the two orders produce different distributions over nearby samples.
For example, the image may place a sample near images of one concept while its paired text is close to texts about another.

\noindent \textbf{Image-Text Support Gap.}~
Image-Text Support Gap measures how much similarity decreases across modalities for the same neighboring examples.
Within \(\mathcal{N}_x^k(i)\), write \(\rho_x(i)\) for the mean image similarity to \(x_i\), and define \(\rho_y(i)\) analogously for text neighbors.
Subtracting the corresponding Neighbor Support compares similarity to these neighbors in the two modalities.
We retain the larger directional gap so that a mismatch in either direction contributes
\begin{equation}
\begin{aligned}
\rho_x(i)
&=
\frac{1}{k}
\sum_{j\in\mathcal{N}_x^k(i)}
\left\langle e_x(x_i),e_x(x_j)\right\rangle,
\\
\rho_y(i)
&=
\frac{1}{k}
\sum_{j\in\mathcal{N}_y^k(i)}
\left\langle e_y(y_i),e_y(y_j)\right\rangle,
\\
G_{x\leftrightarrow y}(i)
&=
\max\{
\rho_x(i)-S_{x\to y}(i),
\\
&\qquad
\rho_y(i)-S_{y\to x}(i)
\}.
\end{aligned}
\label{eq:image-text-support-gap}
\end{equation}
A large \(G_{x\leftrightarrow y}(i)\) indicates that a sample is close to its neighbors in one modality but its paired content is not similarly close in the other.

\noindent \textbf{Text-Span Frequency.}~
To measure repetition across examples, Text-Span Frequency uses the fraction of the corpus \(D\) containing a given text span \(z\).
For each sample \(u_i\), \(A_y(u_i)\) contains short spans from its paired text, such as phrases, relations, or answer fragments.
Because the attack-related span is unknown, the feature keeps the largest corpus frequency among these spans
\begin{equation}
F_y(i)
=
\max_{z\in A_y(u_i)}
\frac{1}{|D|}
\sum_{u\in D}
\mathbf{1}\{z\in A_y(u)\}.
\label{eq:text-span-frequency}
\end{equation}
Related frequency statistics have been used to expose repeated textual triggers in backdoored language data \citep{wang2025datacentric,raghuram2024backdoors}.
Natural captions can also repeat common objects and templates, so a high frequency is more informative when the same samples also show unusual relationships between their image and text neighborhoods.

\noindent \textbf{Text-Span Erasure Gain.}~
Text-Span Erasure Gain examines a span's effect on cross-modal mismatch through the increase in image-to-text support after its removal.
We hold \(\mathcal{N}_x^k(i)\) and the neighbors' texts fixed, changing only the text of example \(i\).
This operation follows standard leave-one-out input erasure \citep{ribeiro2016lime,li2016erasure}, which has also been used to locate image-text misalignment \citep{nam2025clip4dm}.
Removing \(z\) from \(y_i\) gives \(y_i^{-z}\), and \(S_{x\to y}^{-z}(i)\) denotes the support in Eq.~\eqref{eq:neighbor-support} recomputed with \(y_i^{-z}\).
Text-Span Erasure Gain keeps the largest positive increase across all spans in \(A_y(u_i)\)
\begin{equation}
E_{x\to y}(i)
:=
\max_{z\in A_y(u_i)}
\left[S_{x\to y}^{-z}(i)-S_{x\to y}(i)\right]_+.
\label{eq:text-span-erasure-gain}
\end{equation}
A large \(E_{x\to y}(i)\) indicates that removing one span makes the text more similar to the texts paired with nearby images.
The maximum does not require knowing which span contributes to the attack, while the positive part excludes removals that reduce support.

The six features examine complementary consequences of repeated patterns and shared training effects.
Our empirical prediction is that successful poisons differ from clean examples under one or more of these measurements even when their non-pathway feature values overlap.
Section~\ref{sec:empirical-observations} tests this prediction before the features are used for removal.

\subsection{Adaptive Rank-Based Filtering}
TraceGuard uses these features to determine both which examples to remove and how much of the corpus to filter.
Since benign long-tail examples can also receive extreme values, it first looks for examples highlighted by several rankings, then refines that selection before deciding removal.

\paragraph{Complementary rankings.}
Raw feature values are not comparable because the six measurements have different scales and meanings.
We therefore rank examples separately under each feature and record positions as percentiles, with higher positions indicating greater suspicion.
Low Neighbor Support and high values of the other features receive high percentiles.
Averaging all six could hide a large difference in one feature when the others show little separation.
Instead, TraceGuard retains five fixed rankings, denoted by \(\mathcal V\).
The two directional Neighbor Support percentiles form one ranking through their geometric mean, emphasizing examples with weak support in both directions.
Text-Span Erasure Gain is paired separately with Text-Span Frequency and Image-Text Support Gap, using the larger percentile in each pair so either can highlight an example.
The remaining rankings use Neighbor Disagreement and the largest percentile across all six features.

An example can appear unusual under one measurement for a benign reason.
When an attack affects several corpus relationships, its examples may instead rank highly under multiple combinations.
Writing \(r_v(i)\) for example \(i\)'s percentile in ranking \(v\), TraceGuard counts these occurrences using a fixed upper-tail fraction \(\alpha\),
\begin{equation}
a_i=\sum_{v\in\mathcal V}\mathbf 1\{r_v(i)\geq1-\alpha\}.
\label{eq:main-agreement}
\end{equation}
A high \(a_i\) means that several rankings place the example in their top \(\alpha\) fraction.
The fixed \(\alpha\) sets the fraction compared across rankings, not the fraction to remove.
Since the rankings share features, this count measures agreement rather than a probability of poisoning.
To form an initial suspicious set, TraceGuard selects as its \emph{base ranking} an ordering whose top examples also have large \(a_i\).
For text-based rankings, the relevant spans must additionally recur near the top without being common throughout the corpus.
Among qualifying rankings, TraceGuard chooses the one whose top examples most often also rank highly in the others, as measured by \(a_i\).
If no ranking qualifies, the ranking based on the largest percentile across all six features becomes the base ranking.

\paragraph{Initial suspicious set.}
The base ranking provides an ordering, but the size of the initial suspicious set must still be inferred.
Without the poison rate, a fixed removal fraction could discard too many clean examples or miss poisons.
We instead search for a split separating top-ranked examples with consistently large \(a_i\) from the rest, following the statistical comparison underlying change-point methods \citep{wang2020univariate,kovacs2024optimistic}.
For each fraction \(p\) in a fixed set \(\mathcal P\), let \(S_p\) contain the top \(p\) fraction of the base ranking.
The contrast \(c(S)\) is the mean \(a_i\) inside a set \(S\) minus the mean outside it.
TraceGuard selects
\begin{equation}
p^\star=\underset{p\in\mathcal P,\ c(S_p)\geq\tau_c}{\operatorname{arg\,max}}
p\bigl(c(S_p)-\gamma\bigr).
\label{eq:main-selection}
\end{equation}
Here, \(\tau_c\) sets the minimum contrast, while \(\gamma\) penalizes increasing the set size without sufficient contrast.
Weighting by \(p\) balances separation from the remaining corpus with coverage of suspicious examples.
The selected \(S_{p^\star}\) becomes the initial suspicious set, with the smallest tested fraction used if none qualifies.

\paragraph{Refinement and final removal.}
The initial cutoff may still exclude lower-ranked poisons or include unrelated clean outliers.
TraceGuard uses spans identified by Text-Span Frequency and Text-Span Erasure Gain to find related examples across that cutoff.
A span qualifies only if it covers a prescribed fraction of the initial set while remaining uncommon in the corpus.
The first condition establishes recurrence among suspicious examples, while the second limits expansion through ordinary templates.
All examples containing a qualifying span form an alternative suspicious set, including matches below the initial cutoff.
If no single span provides enough coverage, TraceGuard seeks several uncommon spans that jointly meet the requirement.
The resulting set excludes initial examples containing none of these spans and recovers lower-ranked examples containing any of them.
Matching is restricted to the selected spans because unrestricted semantic similarity could include clean examples describing the same benign concept.

For patterns without repeated text, TraceGuard instead examines gaps between the raw feature values underlying each ranking.
A large relative drop after a small top-ranked subset defines a smaller alternative set.
Restricting the drop to the top and bounding the subset size avoids using arbitrary gaps farther down the distribution.

To decide which set to remove, TraceGuard returns to \(c(S)\), comparing how consistently examples inside and outside each set rank highly.
An alternative must retain its text-span or feature-drop conditions and satisfy \(c(S)\geq\tau_c\).
TraceGuard removes the qualifying alternative with greatest contrast, or retains the initial selection if none qualifies.
It does not merge alternatives, since combining sets linked by different patterns could include unrelated clean examples.
Since these sets can differ in size, the final selection also adapts the removal fraction without knowing the poison rate.
Appendix~\ref{app:method} gives the detailed procedure and corpus-relative conditions, which remain fixed across attacks.

\section{Experiments}
\label{sec:experiments}
\subsection{Setup}
\label{sec:setup}
We evaluate defense across 19 attack configurations spanning image-text learning, generative VLM fine-tuning, and encoder transfer.
For encoder-transfer tests, we pair image-only poisons with class-name prompts without changing the poisoned images.
The attack configurations and poisoning rates follow the original papers or official implementations.
Ten baselines cover data filtering, robust training, model repair, and test-time detection at their applicable stages.
Appendix~\ref{app:setup} details the attacks, corpora, baselines, and task metrics.

TraceGuard uses frozen CLIP ViT-B/32 encoders and \(k=10\) neighbors, with exact search for at most 30,000 examples and HNSW for larger corpora.
We fix \(\alpha=0.10\), \(\tau_c=1\), \(\gamma=2\), the fraction grid, and refinement conditions across attacks, using poison labels only for evaluation.
For filtering defenses, Poison Recall measures the fraction of poisons removed, while Clean FPR measures the fraction of clean examples removed.
End-to-end comparisons use matched undefended models, with positive attack-metric reduction and clean-utility change indicating improvement.
Follow-up controls report three-seed means and standard deviations where indicated, while trained ablations use one seed.

\subsection{Observing Attack Pathways}
\label{sec:empirical-observations}
To test whether feature separation tracks attack effectiveness, we weaken five representative attacks while fixing poison fraction and attack construction.
Appendix~\ref{app:observations} compares clean samples with weak, intermediate, and original successful poisons.

CLIP score has TPR at 5\% FPR no greater than 0.13 for any attack, despite high successful-variant ASRs.
In contrast, pathway-feature distributions move farther from clean values as the attacks strengthen (Figure~\ref{fig:eo01_feature_distributions}).
The fixed poison fraction links this progression to attack effectiveness rather than to more poisoned examples.
The clearest separation occurs under different features across attacks (Table~\ref{tab:eo01_feature_metric_summary}).
Neighbor Disagreement reaches TPR 0.90 on Shadowcast and 0.75 on MP-Nav, while Text-Span Frequency reaches 1.00 on Text-Guided Backdoor (TGB) and 0.58 on Concept-Guided Backdoor.
BadCLIP's best TPR is only 0.28, however, supporting the use of complementary measurements rather than a universal single-feature threshold.

\subsection{Large-Scale Defense Comparison}
\label{sec:large-scale-defense-comparison}
\setcounter{topnumber}{1}
We first compare poison detection and clean-data retention with all applicable filters (Table~\ref{tab:filtering_quality_main}).
TraceGuard achieves 98.44\% mean Poison Recall at 5.38\% Clean FPR, compared with 46.90\% recall at 23.04\% FPR for VDC, the strongest baseline by mean recall.
TraceGuard has the highest or tied-highest recall on 17 configurations, with recall never below 90\%.
The mean therefore reflects detection across attack constructions rather than a few favorable cases.
Clean-data removal nevertheless varies across configurations, reaching 16.17\% on BadMLLM and 10.04\% on PoisonedEncoder, where CLIPScore removes only 3.72\%.
We therefore examine whether filtering suppresses the attack while preserving the model's clean-task performance.
\begin{table}[t]
\centering
\caption{Sample-level poison detection across all attacks. Recall is Poison Recall and FPR is Clean FPR, both in percent. The final row averages each metric over the 19 attacks, assigning equal weight to each attack.}
\label{tab:filtering_quality_main}
\ArtifactFont\fontsize{9}{11}\selectfont
\setlength{\tabcolsep}{1.2pt}
\renewcommand{\arraystretch}{1.02}
\begin{tabular*}{\columnwidth}{@{\extracolsep{\fill}}>{\raggedright\arraybackslash}p{0.25\linewidth}rr@{\hspace{8pt}}rr@{\hspace{8pt}}rr@{\hspace{8pt}}rr@{}}
\toprule
& \multicolumn{2}{c}{CLIPScore} & \multicolumn{2}{c}{VDC} & \multicolumn{2}{c}{Detect-CLIP} & \multicolumn{2}{c}{TraceGuard} \\
\cmidrule(lr){2-3}\cmidrule(lr){4-5}\cmidrule(lr){6-7}\cmidrule(l){8-9}
Attack & Rec.$\uparrow$ & FPR$\downarrow$ & Rec.$\uparrow$ & FPR$\downarrow$ & Rec.$\uparrow$ & FPR$\downarrow$ & Rec.$\uparrow$ & FPR$\downarrow$ \\
\midrule
\multicolumn{9}{@{}l}{\textit{Image-text pre-training}} \\
Patch Backdoor & 97.40 & 4.55 & 98.40 & 31.01 & 4.80 & 4.95 & 100.00 & 4.53 \\
Targeted Poisoning & 100.00 & 4.82 & 32.00 & 31.01 & 0.00 & 0.00 & 100.00 & 4.81 \\
MM-Poison & 99.68 & 4.74 & 99.20 & 10.05 & 0.40 & 1.41 & 99.52 & 4.71 \\
BadCLIP & 97.73 & 3.60 & 94.87 & 31.00 & 13.20 & 4.98 & 99.20 & 3.58 \\
ToxicTextCLIP & 93.67 & 4.12 & 93.78 & 31.11 & 0.00 & 0.00 & 91.33 & 4.14 \\
C$^2$ Attack & 1.08 & 5.07 & 18.33 & 25.15 & 4.37 & 4.67 & 99.58 & 3.95 \\
BadCLIP++ & 26.33 & 4.94 & 89.67 & 31.03 & 6.33 & 4.88 & 100.00 & 4.72 \\
Text-Guided Backdoor & 9.11 & 4.82 & 74.65 & 65.52 & 4.00 & 4.67 & 100.00 & 5.26 \\
\addlinespace[3pt]
\multicolumn{9}{@{}l}{\textit{VLM fine-tuning}} \\
Shadowcast & 0.00 & 5.18 & 14.00 & 23.41 & 50.00 & 4.97 & 100.00 & 7.39 \\
MP-Nav & 0.00 & 5.09 & 0.00 & 23.41 & 10.00 & 2.68 & 100.00 & 4.45 \\
BadMLLM & 6.48 & 4.43 & 11.56 & 7.76 & 0.55 & 2.95 & 99.45 & 16.17 \\
VL-Trojan & 96.55 & 4.58 & 86.21 & 16.17 & 8.62 & 4.94 & 100.00 & 9.55 \\
TrojVLM & 6.64 & 3.52 & 18.36 & 20.06 & 4.30 & 3.91 & 100.00 & 0.00 \\
BadSem & 13.20 & 4.67 & 32.40 & 10.65 & 3.60 & 4.40 & 99.20 & 5.31 \\
TokenSwap & 6.27 & 3.87 & 18.93 & 15.80 & 5.07 & 7.13 & 92.13 & 7.87 \\
Concept-Guided Backdoor & 7.85 & 4.90 & 14.82 & 10.03 & 3.75 & 4.99 & 100.00 & 0.00 \\
CBV & 7.60 & 4.91 & 8.00 & 10.59 & 12.80 & 4.93 & 90.00 & 5.79 \\
\addlinespace[3pt]
\multicolumn{9}{@{}l}{\textit{Encoder transfer}} \\
CorruptEncoder & 6.46 & 4.88 & 50.62 & 20.02 & 8.92 & 4.95 & 100.00 & 0.00 \\
PoisonedEncoder & 18.20 & 3.72 & 35.40 & 24.04 & 0.00 & 0.00 & 100.00 & 10.04 \\
\midrule
\textbf{Mean across 19 attacks} & 36.54 & 4.55 & 46.90 & 23.04 & 7.41 & 3.76 & 98.44 & 5.38 \\
\bottomrule
\end{tabular*}
\end{table}

After filtering, the attack metric is at most 1\% in 13 of 19 configurations and at most 10\% in 17 (Appendix~\ref{app:comparison}, Figure~\ref{fig:traceguard_end_to_end}).
Tables~\ref{tab:defense_comparison_image_text} and \ref{tab:defense_comparison_vlm} compare attack reduction and clean-task change across baselines.
TraceGuard achieves the largest or tied-largest attack reduction on five of eight image-text attacks and six of nine VLM attacks.
On MM-Poison, its 19.81-point reduction is close to RoCLIP's 20.76, with clean-task changes of \(-0.49\) and \(-58.99\).
For MP-Nav, TraceGuard and RobustIT reduce the attack metric by 93.67 and 94.33 points, while clean performance falls by 14.17 and 36.83 points.
In these comparisons, TraceGuard achieves similar attack suppression with substantially less loss of clean utility.

High recall does not guarantee suppression, however, as ToxicTextCLIP's attack metric increases from 93.1\% to 94.0\% after TraceGuard and CBV increases from 4.4\% to 6.8\%.
In the separate ToxicTextCLIP failure control, removing only detected poisons leaves ASR at 78.98\%, whereas removing all poisons lowers it to 1.55\% (Appendix~\ref{app:analysis}, Table~\ref{tab:fu-failure_controls}).
The missed poisons can therefore retain enough influence to sustain the attack, even when the detected fraction is high.
The encoder-transfer tests likewise require attention to the undefended model.
PoisonedEncoder falls from 100\% to 0\%, with a 1.3-point clean-accuracy decrease, whereas CorruptEncoder starts at only 0.02\% ASR and serves as a low-ASR control rather than evidence of further suppression.

\subsection{Controls and Ablations}
To distinguish sample selection from simply discarding more data, we match the removal count of non-pathway and random filtering to TraceGuard's (Appendix~\ref{app:analysis}, Table~\ref{tab:fu-influence_removal}).
For BadCLIP++, those controls leave ASR above 87\%, compared with TraceGuard's \(4.20\pm1.91\%\).
TraceGuard also approaches the oracle that removes all poisons on TGB and BadCLIP, indicating that the benefit depends on which examples are removed, not only the removal fraction.
On TGB, TraceGuard also improves clean performance from 39.46\% to 50.32\%, while random removal leaves it at 39.82\%.
On BadCLIP++, however, clean performance falls to 94.53\%, close to the all-poison oracle's 94.07\%, indicating that removing poisons need not preserve the undefended model's clean performance.

\begin{table}[!htbp]
\centering\ArtifactFont\fontsize{8.5}{10.5}\selectfont
\setlength{\tabcolsep}{1.8pt}
\renewcommand{\arraystretch}{1.08}
\caption{TraceGuard ablations across five attacks. Entries give ASR and clean zero-shot accuracy (CA) in percent from one training seed. The oracle matches TraceGuard's removal count.}
\label{tab:fu-ablation_outcomes}
\begin{tabular*}{\linewidth}{@{\extracolsep{\fill}}>{\raggedright\arraybackslash}p{0.28\linewidth}*{10}{r}@{}}
\toprule
 & \multicolumn{2}{c}{TGB} & \multicolumn{2}{c}{Concept-Guided} & \multicolumn{2}{c}{BadCLIP++} & \multicolumn{2}{c}{MP-Nav} & \multicolumn{2}{c}{PoisonedEncoder} \\
\cmidrule(lr){2-3}\cmidrule(lr){4-5}\cmidrule(lr){6-7}\cmidrule(lr){8-9}\cmidrule(lr){10-11}
Variant & ASR & CA & ASR & CA & ASR & CA & ASR & CA & ASR & CA \\
\midrule
TraceGuard & 0.91 & 50.74 & 0.00 & 100.00 & 3.20 & 95.80 & 0.67 & 74.17 & 0.00 & 59.70 \\
Fixed 5\% removal & 58.86 & 51.69 & 68.51 & 84.00 & 84.60 & 99.60 & 74.33 & 86.00 & 100.00 & 59.42 \\
Mean of six ranks & 23.89 & 36.95 & 40.36 & 89.20 & 90.00 & 94.00 & 80.50 & 85.17 & 100.00 & 61.54 \\
Matched-size oracle & 1.67 & 51.85 & 0.00 & 100.00 & 4.20 & 95.60 & 0.33 & 71.17 & 0.00 & 60.85 \\
Without Neighbor Support & 34.27 & 39.80 & 72.55 & 73.60 & 88.80 & 96.80 & 95.00 & 87.83 & 100.00 & 58.71 \\
Without neighborhood disagreement and gap & 79.60 & 49.04 & 49.22 & 85.20 & 88.20 & 96.00 & 93.33 & 86.17 & 100.00 & 61.48 \\
Without feature-drop refinement & 1.10 & 50.96 & 66.67 & 80.20 & 88.40 & 96.00 & 96.33 & 89.00 & 0.00 & 60.02 \\
Without text-span features & 49.63 & 44.68 & 66.68 & 81.20 & 89.20 & 98.20 & 90.33 & 86.67 & 100.00 & 60.19 \\
No removal & 31.14 & 43.09 & 68.30 & 81.60 & 91.40 & 98.20 & 94.33 & 88.33 & 100.00 & 61.35 \\
\bottomrule
\end{tabular*}
\end{table}

Ablations distinguish the contributions of feature families, rank aggregation, and removal selection (Table~\ref{tab:fu-ablation_outcomes}).
A fixed 5\% removal rate leaves ASR between 58.86\% and 100\%, whereas TraceGuard leaves at most 3.20\%.
Averaging all six ranks leaves ASR between 23.89\% and 100\% at the same removal count as TraceGuard.
This gap supports preserving complementary rankings instead of averaging away differences that expose a particular attack.
Removing feature-drop refinement has little effect on TGB and PoisonedEncoder but raises BadCLIP++ ASR to 88.40\% and MP-Nav ASR to 96.33\%.
Refinement thus contributes differently across corpora, but the full method stays within one percentage point of the oracle at the same removal count on all five attacks.

Detection ablations reveal how refinement changes which examples are selected.
On MP-Nav, refinement and final selection increase recall from 5.0\% to 100\% while reducing Clean FPR from 20.1\% to 4.4\% (Figure~\ref{fig:fu-filtering_ablation}).
Extending a fixed ranking alone could not achieve both improvements, because it would retain the clean examples already selected.
Appendix~\ref{app:analysis} includes full detection ablations, gradient-influence measurements, and failure controls.

\FloatBarrier
\subsection{Stress Tests and Limits}
\setcounter{topnumber}{2}
To test deliberate evasion, an attacker with knowledge of TraceGuard modifies images, target-side wording, or both.
We evaluate image-neighborhood optimization and two, four, or eight alternative phrasings, with three training seeds per condition (Appendix~\ref{app:adaptive}, Table~\ref{tab:fu-adaptive_attacks}).
Under text adaptation, TraceGuard retains 100\% Poison Recall on BadCLIP++ and MP-Nav, with residual ASRs of 4.20--9.00\% and 0.67\%.
Joint image and text adaptation reduces BadCLIP++ recall to 1\% with four text alternatives, but also lowers undefended ASR from 87.40\% to 8.07\%.
This evasion therefore comes with a weaker attack before defense.
TGB reveals a different outcome, with two text alternatives lowering recall to 12.67\% and raising residual ASR from 1.55\% to 16.49\%.
TraceGuard thus remains vulnerable to adaptation that preserves enough attack effectiveness to exploit missed poisons.

Varying poison rates tests whether adaptive removal accommodates changes in the number of poisons (Appendix~\ref{app:rates}).
At 20\% BadCLIP++ poisoning, adaptive removal leaves 4.00\% ASR, compared with 97.60\% under fixed 5\% removal.
However, TraceGuard misses the poisons at all four tested nonzero rates below 5\%.
Adapting the removal amount can overcome an insufficient fixed budget, but cannot compensate when the features fail to identify the poisoned examples.

Poison-free corpora test whether benign relationships trigger removal.
TraceGuard removes 5--10\% of single-source corpora and 20\% of their mixed-source counterpart (Table~\ref{tab:fu-clean_only}), showing that agreement across features does not establish malicious intent or ensure abstention.
Further tests show sensitivity to encoder choice and incomplete recovery when several attacks coexist (Appendices~\ref{app:encoders} and \ref{app:simultaneous}).
Finally, computational analysis gives \(O(N\log N+Nk^2)\) for feature processing and selection after encoding and neighbor retrieval.
Measured representation extraction and initial scoring grow approximately linearly over corpora up to about 600,000 examples (Appendix~\ref{app:cost}).

\FloatBarrier
\clearpage
\section{Conclusion}
TraceGuard connects the collective training influence required by a poison set to observable relationships in an image-text corpus.
Its complementary rankings and shared-pattern refinement adapt both which examples are removed and how much data is filtered without knowing the poison rate.
The results across 19 attack configurations, matched-removal controls, and ablations show that this combination can suppress diverse attacks while retaining substantially more clean data than some competing defenses.
Adaptive attacks, encoder dependence, and false positives on clean-only corpora remain important limits.
The evidence supports adaptive corpus-level filtering as a useful defense, not a guarantee that every shared pattern is malicious.

\section*{AI Use Statement}
We used generative AI tools to assist with manuscript writing and language polishing, and to support literature search and the identification of relevant prior work.
Their use was limited to these tasks and did not extend to method development, theoretical analysis, research implementation, data generation or processing, experimental design, or interpretation of results.
We reviewed and edited all AI-assisted text and checked the accuracy and relevance of cited sources against the original publications.
We take full responsibility for the accuracy and integrity of the final manuscript, including all AI-assisted content.

\section*{Ethics Statement}
TraceGuard is intended to protect multimodal training from poisoned data, but studying the relationships that expose poisons could also help attackers design more evasive attacks.
We examine this risk through adaptive attacks and report the vulnerabilities that remain.
Filtering itself can cause harm by discarding legitimate examples, with unequal removal rates across data sources potentially changing the composition of the retained corpus.
We therefore report clean-data removal and clean-task performance alongside attack suppression, including tests on poison-free corpora.
A removal decision should not be interpreted as proof of malicious intent by a data contributor.

\section*{Reproducibility Statement}
Section~\ref{sec:cross-sample-evidence} defines the pathway features and adaptive rank-based filtering procedure, with the analytical assumptions and extended derivation in Appendix~\ref{app:pathways} and further algorithmic details in Appendix~\ref{app:method}.
Section~\ref{sec:setup} and Appendix~\ref{app:setup} describe the datasets, attack and defense configurations, pretrained encoders, evaluation metrics, and computational resources.
The construction of attack-strength variants is detailed in Appendix~\ref{app:observations}, while the controls, ablations, and stress tests specify their comparison protocols and use of training seeds in Appendices~\ref{app:analysis} and~\ref{app:stress-tests}.
Appendix~\ref{app:cost} distinguishes the complexity analysis from the measured runtime components.
We will open-source the implementation of TraceGuard upon acceptance.

\bibliographystyle{iclr2027_conference}
\bibliography{sample}
\clearpage
\appendix

\section{Threat Model Details}
\label{app:threat}

Modern multimodal models are often trained or fine-tuned on image-text datasets assembled from web crawls, public datasets, and third-party sources \citep{radford2021clip,jia2021align,schuhmann2021laion400m,gadre2023datacomp}.
Each training example pairs a visual input with text used for learning, such as a caption, instruction, or answer.
In our setting, the defender assembles such an image-text corpus for subsequent model training.
Because the provenance and integrity of every collected example cannot be fully verified, poisoned samples may enter the corpus during the collection, making data curation necessary before training.
In this section, we define the attacker, the defender's knowledge and capabilities, and the security goal for this scenario.

\paragraph{Attacker}
Data poisoning commonly includes targeted poisoning attacks and backdoor attacks, which seek to induce attacker-chosen behavior in the trained model while preserving normal behavior on clean inputs \citep{biggio2012poisoning,gu2017badnets,carlini2022poisoning}.
Our threat model considers these two types of attack.
The attacker carries out these attacks by injecting new image-text pairs or modifying existing ones, with changes made to the image, the text, or the relation between them \citep{biggio2012poisoning,gu2017badnets,carlini2022poisoning,bai2024badclip,xu2024shadowcast}.
The attacker acts only through the training data and cannot alter the defender's training procedure or modify the model after training.
In both cases, the attack causes the trained model to learn an incorrect cross-modal mapping for attacker-chosen inputs or concepts.
Targeted poisoning makes this mapping affect the intended target directly, whereas a backdoor conditions the mapping on a trigger that activates the target behavior.
Additionally, the effect of this incorrect mapping varies across model types and downstream tasks.
In zero-shot classification with CLIP, this incorrect mapping can make a targeted or trigger-bearing image align with an attacker-chosen class label \citep{carlini2022poisoning,bai2024badclip}.
In image-text retrieval, it can associate a chosen query with an attacker-chosen image or text and thereby move that target higher in the retrieval ranking \citep{carlini2022poisoning}.
In VLM fine-tuning, it can map a source visual concept to an attacker-chosen response, relation, or misleading narrative \citep{xu2024shadowcast}.

We assume a strong attacker with detailed knowledge of the training pipeline, including the training task, the victim model's exact architecture and training objective, the data sources used for collection, and publicly documented curation and defense mechanisms.
With its knowledge of the training pipeline, the attacker can employ advanced stealthy poisoning attacks to bypass any poison-detection method applied during data curation.
The resulting image-text pairs may remain semantically consistent and linguistically natural while still carrying the training signal needed to produce the target behavior \citep{turner2019labelconsistent,saha2020hidden,xu2024shadowcast}.
We also grant the attacker sufficient computational resources to optimize every poisoned example for maximum attack effectiveness.
We strengthen this threat model further by considering an adaptive attacker who uses complete knowledge of TraceGuard to design poisoning attacks specifically to evade it.
This adaptive setting provides a direct test of TraceGuard's robustness when the defense itself is known to the attacker.

\paragraph{Defender}
The defender is the model trainer who receives the corpus and will conduct data curation before training.
From the defender's perspective, the corpus contains a large number of samples from unknown sources, with no ground truth information indicating which samples are clean and which are poisoned.
With the understanding that the corpus may contain poisoned samples, the defender applies a poisoning detection method during data curation, so that any samples detected as poisoned are removed and do not enter subsequent training.
Because poison detection can also remove clean samples by mistake, the defender's objective is twofold.
It seeks to identify as many truly poisoned samples as possible while falsely flagging as few clean samples as possible.

Different from the attacker, we apply a strict knowledge constraint for the defender. 
Specifically, the defender does not know whether the corpus contains poisoned data, and if poisoning is present, it does not know how many samples are affected or what behavior the attacker seeks to induce.
Because the extent of poisoning is unknown, the defense cannot assume a removal fraction that matches the poison rate.
The defender should also has no information about how the poisoned samples were constructed and which attack is used behind.
The manipulation from the attacker may affect the image, the text, or the relation between them, but the defender will never be informed about it.

Because the defender is also responsible for the subsequent model training, we assume that it has sufficient computational resources to conduct data curation.
This assumption, however, does not leave the cost of poison detection unconstrained.
The datasets subject to detection can be extremely large, so we require the running time to grow no faster than $O(n^2)$ with the number of samples $n$.
A higher time complexity would impose a computational burden that a practical defender could not sustain \citep{schuhmann2021laion400m,gadre2023datacomp,grosse2024practical}.
We further assume that the defender has access to publicly available pretrained image and text encoders that can support the detection process.

\paragraph{Security goal}
Our security goal is to reliably detect clean-looking poisoned samples without assuming a particular poisoning algorithm.
A successful defense should remove enough poisoned samples to suppress the attacker-chosen behavior while preserving as much clean training data as possible.
Meeting these two goals without knowing the poison fraction requires the removal fraction to be determined separately for each corpus rather than follow a preset filtering budget.
This requirement applies even when the attacker deliberately optimizes each poisoned sample to resemble clean data \citep{turner2019labelconsistent,saha2020hidden,xu2024shadowcast}.

\section{Extended Attack-Pathway Analysis}
\label{app:pathways}

This appendix develops the training-influence analysis summarized in Section~\ref{sec:pathway-features}.
We first express attack success as the change induced by poisoned data, then relate the contributions of shared attack patterns to relationships among examples.

The central difficulty in poison detection is that an attacker can optimize away the features on which a detector relies.
Once a defense treats a new different feature as evidence of poisoning, a more advanced attacker can add a stealth objective or constraint that suppresses the same feature while retaining the malicious training effect.
Adaptive poisoning attacks have followed this pattern by reducing the separation between clean and poisoned representations or constraining frequency artifacts used for detection \citep{tan2020bypassing,qi2023revisiting,zeng2023narcissus}.
To move beyond the arms race between increasingly specialized detectors and increasingly evasive attacks, we begin with a more fundamental question.
That is, we ask which properties of poisoned samples are essential to attack effectiveness and cannot be suppressed without substantially weakening the attack.
This question has received far less attention than the measurable differences used by existing defenses to separate clean and poisoned samples \citep{tran2018spectral,hayase2021spectre}, even though adaptive attacks repeatedly show that these differences can be reduced without sacrificing attack success \citep{tan2020bypassing,qi2023revisiting,zeng2023narcissus}.
We argue that these harder-to-hide properties provide a more stable foundation for detecting poisoned samples even as their observable appearance changes.

\subsection{Attack Pathway}
In a successful poisoning attack, these harder-to-hide properties arise from the attack objective itself.
To induce the target behavior, poisoned samples must produce a sufficiently strong net change in what the model learns, so suppressing this change would directly weaken the attack \citep{geiping2021witches,yang2022notall}.
Poison-forensics and spectral analyses further show that successful poisons exert shared effects on model parameters or learned representations \citep{shan2022poisonforensics,tran2018spectral,hayase2021spectre}.
These findings point to a common requirement across successful poisoning attacks.
Furthermore, what makes this requirement more specific to poisoning is the attacker's additional malicious objective.
Clean image-text pairs contribute to training according to the content of each pair, so their effects are spread across many concepts and relations.
A comparatively small poison set, by contrast, must direct enough of its combined influence toward the same target behavior to alter the model despite the much larger clean corpus.
We use the term \emph{attack pathway} for the properties that sustain this collective, target-directed influence.

To characterize the attack pathway mathematically, we begin by expressing what the poison set must achieve for the attack to succeed.
Since a poisoning attack can succeed only if the model learns from the poisoned samples, the derivation begins by analyzing model training.
Notably, model training is only used to derive the pathway, whereas our defense does not require training the victim model.
Following Huber's contamination model \citep{huber1964robust}, we represent the corpus as the union \(D_c\cup D_p\) of clean and poisoned examples, which will not be observed by the defender.
Training on corpus \(D\) then produces model \(\mathcal{A}(D)\).
We quantify the attack's effect on the resulting model with \(J(\mathcal{A}(D))\), whose value increases as the model exhibits the target behavior more strongly.
With the same training randomness for the clean and poisoned corpora, we define
\begin{equation}
\Delta J
=
J\!\left(\mathcal{A}(D_c\cup D_p)\right)
-
J\!\left(\mathcal{A}(D_c)\right)
\geq \kappa .
\label{eq:reliable-attack-success}
\end{equation}
Here, the positive constant \(\kappa\) sets the minimum increase in \(J\) required for attack success.

Equation~\eqref{eq:reliable-attack-success} states how much \(J\) must increase for the attack to succeed, but it does not show how the poisoned samples collectively produce \(\Delta J\).
Influence analysis~\citep{koh2017influence,pruthi2020tracin} addresses this question by approximating how changes in individual training samples affect the trained model.
Recent work on data attribution extends this analysis to non-decomposable objectives~\citep{deng2025versatile}, including contrastive losses.
Under its first-order approximation, each poisoned sample makes an additive contribution to \(\Delta J\).
We use \(z\) to denote an attack pattern shared across multiple poisoned samples to reinforce the same target behavior, and \(\mathcal{Z}\) for the set of such patterns.
A shared attack pattern may be an explicit trigger in either modality or a semantic association between a source and the target.
Because such semantic associations can be expressed differently across samples, \(z\) need not correspond to identical tokens or pixels.
The aggregate first-order contribution from samples that rely on the same \(z\) is represented by \(I(z)\).
We write the total increase as
\begin{equation}
\Delta J
=
R
+
\epsilon
\sum_{z\in\mathcal{Z}}
I(z).
\label{eq:influence-decomposition}
\end{equation}
Here, \(\epsilon=|D_p|/(|D_c|+|D_p|)\) is the poison fraction.
For each \(z\), \(I(z)\) combines the fraction of poisoned samples in which \(z\) appears with their mean contribution to the attack objective.
The remainder \(R\) accounts for interactions, effects that cannot be assigned to any \(z\in\mathcal{Z}\), and error from the first-order approximation.

We assume that \(|R|<\kappa\), so the remainder alone cannot produce the performance increase required by Eq.~\eqref{eq:reliable-attack-success}.
When the success condition in Eq.~\eqref{eq:reliable-attack-success} holds, combining the two equations gives
\begin{equation}
\max_{z\in\mathcal{Z}}
I(z)
\geq
\frac{\kappa-|R|}
{\epsilon|\mathcal{Z}|}
>0.
\label{eq:pathway-requirement}
\end{equation}
This means that there must be at least one shared attack pattern \(z\) such that the poisoned samples carrying it collectively alter what the model learns enough to raise \(J\) and strengthen the target behavior through the frequency of that pattern, the mean contribution of those samples, or both.
The bound follows because the maximum contribution over \(\mathcal{Z}\) is at least the average contribution.

\subsection{Why Existing Data Filters Can Miss Poisons}
The requirement above exposes a basic mismatch in existing defenses.
A poisoning attack depends on the combined effect of poisoned samples, but existing methods still decide whether each image-text pair is suspicious separately.
Worse still, this limitation persists in recent work on multimodal data filtering and poisoning defense, even as these methods adopt increasingly capable filtering models \citep{yang2024safeclip,zhu2024vdc,fang2024datafiltering,huang2025detecting,zhang2025lemon,wu2025mixture,shechter2025filter}.
The most direct line filters or partitions data according to the cosine similarity between image and text representations \citep{radford2021clip,schuhmann2021laion400m,gadre2023datacomp,yang2024safeclip}.
Synthetic captions and visual answers derived from the image can instead serve as references for assessing the semantic alignment or visual-linguistic inconsistency of the original pair \citep{mahmoud2024sieve,zhu2024vdc}.
The reference can instead come from surrounding samples, where multimodal-neighbor consistency and density ratios reveal local outliers \citep{huang2025detecting,zhang2025lemon}.
Rather than fixing the filtering criterion in advance, learned filtering methods infer image-text quality from curated examples \citep{fang2024datafiltering,wang2024multimodalfilters}.
The decision can also be tied to downstream utility, with data-valuation methods combining several quality estimates or learning each example's usefulness for pretraining from downstream gradient signals \citep{wu2025mixture,shechter2025filter}.
Despite these advances, all five routes ultimately judge a property of an individual sample.
By definition, however, \(I(z)\) depends on how often an attack pattern is shared and how strongly the samples carrying it contribute to the target behavior during training.
The per-sample quantities above place no necessary upper bound on either component.
We refer to these measurements as {\it non-pathway features} because they do not constrain this shared contribution.
A poison set can therefore resemble clean data under these measurements while retaining the positive \(I(z)\) required by Eq.~\eqref{eq:pathway-requirement} \citep{turner2019labelconsistent,saha2020hidden,xu2024shadowcast}.
In other words, clean-like values under a non-pathway feature do not imply a weak attack pathway.

\subsection{Pathway Features}
\label{app:pathway-features}
We argue that pathway features should be derived directly from the two components of \(I(z)\) that non-pathway features leave unconstrained.
However, neither component can itself be observed during data curation because the defender does not know the poison set and has not trained the victim model.
The defender therefore needs to approximate these unavailable components with features computed from the corpus.
To begin with, consider a linear layer \(W\) that maps a sample representation \(h\) to the layer output.
Here, \(W\) serves only to expose the factorization of per-example gradients, which holds for any linear layer in the victim model.
For sample \(u_i\), \(h_i\) is its input to \(W\), while \(\delta_i\) is the gradient of the victim training loss at the layer output.
For any two examples \(u_i\) and \(u_j\), their gradients with respect to \(W\), denoted by \(g_i\) and \(g_j\), and their pairwise alignment factorize as
\begin{equation}
\begin{aligned}
g_i
&=
\delta_i h_i^\top,
\\
\left\langle g_i,g_j \right\rangle_F
&=
\left\langle h_i,h_j \right\rangle
\left\langle \delta_i,\delta_j \right\rangle .
\end{aligned}
\label{eq:gradient-factorization}
\end{equation}
The inner product \(\langle g_i,g_j\rangle_F\) captures whether two examples induce similar changes to the model parameters, following gradient-based influence analysis \citep{pruthi2020tracin}.
Equation~\eqref{eq:gradient-factorization} further factorizes this alignment as the product of the inner product \(\langle h_i,h_j\rangle\) between the representations and the inner product \(\langle\delta_i,\delta_j\rangle\) between the output gradients.
In practice, the term \(\langle\delta_i,\delta_j\rangle\) indicates whether two examples push the model output in the same direction.
Accordingly, poisoned examples that reinforce the same target behavior are expected to have aligned output gradients, either by moving images toward the same target text or by increasing the likelihood of the same target tokens \citep{radford2021clip,wang2020alignment}.
Suppressing this alignment for the attacker would make their output gradients less consistent and thereby weaken their combined effect on the target behavior.
When their representations are also similar, these aligned output gradients make the corresponding per-example gradients \(g_i\) point in similar directions.

To ultimately derive pathway features, we next show how the gradients of poisoned examples carrying the same attack pattern form their collective contribution \(I(z)\).
The poisoned examples containing \(z\) form \(D_z=\{u_i\in D_p:z\in u_i\}\), which occupies a fraction \(p_z=|D_z|/|D_p|\) of the poison set.
Their average per-example gradient is \(\bar g_z=|D_z|^{-1}\sum_{u_i\in D_z}g_i\).
We use a first-order Taylor expansion \citep{nocedal2006numerical} to approximate how a gradient-descent step based on \(\bar g_z\) changes the attack objective, with \(\nabla_W J\) denoting the direction in which that objective increases.
This gives \(I(z)\) and its upper bound
\begin{equation}
\begin{aligned}
I(z)
&=
-p_z
\left\langle
\nabla_W J,
\bar g_z
\right\rangle_F,
\\
\left|I(z)\right|
&\leq
p_z
\left\|\nabla_W J\right\|_F
\left\|\bar g_z\right\|_F .
\end{aligned}
\label{eq:gradient-bound}
\end{equation}
The expression for \(I(z)\) is positive when a gradient-descent step based on \(\bar g_z\) moves the model in the direction that increases the attack objective.
The upper bound follows directly from the Cauchy-Schwarz inequality \citep{hardy1952inequalities}, which bounds \(|I(z)|\) by the product of \(p_z\), \(\|\nabla_W J\|_F\), and \(\|\bar g_z\|_F\).
The norm \(\|\bar g_z\|_F\) in this bound is large only when the per-example gradients do not cancel, and its square expands as
\begin{equation}
\begin{aligned}
\left\|\bar g_z\right\|_F^2
&=
\frac{1}{|D_z|^2}
\sum_{u_i,u_j\in D_z}
\left\langle
g_i,
g_j
\right\rangle_F
\\
&=
\frac{1}{|D_z|^2}
\sum_{u_i,u_j\in D_z}
\left\langle h_i,h_j\right\rangle
\left\langle\delta_i,\delta_j\right\rangle .
\end{aligned}
\label{eq:gradient-structure}
\end{equation}
Substituting the factorization from Eq.~\eqref{eq:gradient-factorization} into each \(\langle g_i,g_j\rangle_F\) rewrites \(\|\bar g_z\|_F^2\) in terms of \(\langle h_i,h_j\rangle\) and \(\langle\delta_i,\delta_j\rangle\) among the examples carrying \(z\).
Together, Eqs.~\eqref{eq:gradient-bound} and \eqref{eq:gradient-structure} identify two parts of \(I(z)\), namely how often \(z\) occurs among poisoned examples and whether the gradients of the examples carrying it reinforce rather than cancel one another.
Neither part can be calculated directly because the defender knows neither which examples are poisoned nor their victim-model gradients.
We therefore connect each part to information available in the paired corpus.
When \(z\) is expressed in text, we represent it as a short segment of an example's paired text, which we call a text span.
The frequency of this span in the full corpus contains the poison-frequency term \(\epsilon p_z\) together with its natural occurrences, which motivates Text-Span Frequency.
For the shared training effect, Eq.~\eqref{eq:gradient-structure} shows that non-canceling gradients depend jointly on representation similarity and output-gradient alignment.
The common target behavior supplies the output-gradient alignment described above, leaving the representation relationships among the examples carrying \(z\) as the part that can be examined from the paired corpus.
An attack pattern concentrated in one modality can make these examples closely related on that side but not through their paired content, while a source-to-target relation can cause the image and text modalities to organize the same examples differently.
The two directional Neighbor Support values test whether a neighborhood formed in one modality is preserved in the other, while Neighbor Disagreement and Image-Text Support Gap measure two forms of mismatch between these neighborhoods.
Text-Span Erasure Gain further tests whether a particular text span is responsible for the observed mismatch.
These operations produce six pathway features tied to the two parts of \(I(z)\) without requiring victim-model gradients or knowledge of which examples are poisoned.

The feature definitions in Section~\ref{sec:feature-definitions} implement these corpus measurements.

Taken together, the derivation motivates the six pathway features through pattern frequency and the combined training effect of examples sharing a pattern.
The lower bound constrains \(I(z)\), but it does not guarantee separation in the features computed with fixed pretrained representations.
Whether these observable measurements distinguish successful poisons from clean data is therefore an empirical prediction, which we test in Section~\ref{sec:empirical-observations}.

\section{TraceGuard Details}
\label{app:method}

TraceGuard uses adaptive rank-based filtering to detect and remove clean-looking poisoned image-text examples from a training corpus.
We call it TraceGuard because it follows the attack pathway across examples and identifies image-text pairs that collectively reinforce the same target behavior.
The method takes the training corpus and pretrained image and text encoders as input, then identifies the examples to remove before victim-model training.
However, making this removal decision from any one pathway feature would be unreliable because benign long-tail examples can produce similarly extreme values.
Rather than collapse all pathway features into a single ranking, TraceGuard retains several pathway-feature rankings.
It selects the ranking whose highest-ranked examples most often also rank highly in the others and uses those examples to form an initial suspicious set.
The initial set is then refined using recurring text spans and gaps in pathway-feature values, and agreement among rankings guides the final selection.
Both the selected examples and the removal fraction therefore adapt to each corpus under rules fixed across attacks, without requiring the poison rate.

\subsection{Pathway Feature Rankings}

Considering that the six pathway features are computed differently, a value from one feature cannot be compared directly with a value from another.
Ranking those features thus makes the six pathway features from Section~\ref{sec:feature-definitions} comparable, and each example's position in a ranking is recorded as a percentile, with a higher position indicating that the feature provides stronger evidence of poisoning.

However, averaging all six percentiles could mask a large difference between poisoned and clean examples in one feature when the other features show smaller differences.
TraceGuard therefore retains five fixed rankings of the corpus, each capturing a different relationship measured by the pathway features.
Specifically, the Image-to-Text and Text-to-Image Support percentiles are combined by their geometric mean, so their two directions form one ranking.
Text-Span Erasure Gain contributes to two rankings by being paired separately with Text-Span Frequency and Image-Text Support Gap, using the larger percentile in each pair.
Neighbor Disagreement provides one ranking directly, and the largest percentile across all six features provides the fifth.
Together, these rankings measure how often an attack pattern recurs, whether images and texts place the affected examples in different neighborhoods, and whether a particular text span causes that mismatch.
We denote this fixed collection by \(\mathcal{V}\), and \(r_v(i)\in[0,1]\) is example \(i\)'s percentile in ranking \(v\), with larger values placing it nearer the top.

These rankings order the corpus under different pathway-feature combinations, but no individual ranking determines whether an example is clean or poisoned.
A benign long-tail example can occupy the same high-ranking region as a poisoned example under an individual feature, making the two difficult to distinguish from that ranking alone.
When an attack affects several relationships among samples, its poisoned examples may remain highly ranked under multiple pathway-feature combinations.
TraceGuard counts how many rankings place each example in their upper tails to identify examples highlighted by several combinations rather than by one feature alone.
Using a fixed upper-tail fraction \(\alpha\), we define
\begin{equation}
a_i(\alpha)
=
\sum_{v\in\mathcal{V}}
\mathbf{1}\{r_v(i)\geq 1-\alpha\},
\label{eq:traceguard-pathway-agreement}
\end{equation}
Here, \(a_i(\alpha)\) counts how many rankings place example \(i\) within their highest-ranked \(\alpha\) fraction.
Because every ranking is expressed in corpus percentiles, the same \(\alpha\) selects an equal fraction of examples from each ranking without knowing the poison rate.
The five rankings are not statistically independent, so \(a_i(\alpha)\) is a within-corpus count rather than a calibrated probability.

With \(\alpha\) fixed across corpora, we use \(a_i\) as shorthand for \(a_i(\alpha)\) during ranking selection.
Our method then identifies rankings whose top-ranked examples also have large \(a_i\) values, so that the selected examples are supported by several rankings rather than by one ranking alone.
For the two rankings built from text-span features, TraceGuard further requires the relevant spans to recur among the top-ranked examples without being common throughout the corpus.
This recurrence requirement distinguishes a text span concentrated in the upper tail from either an isolated rare phrase or a common dataset template.
When several rankings satisfy these criteria, TraceGuard selects the one whose top-ranked examples most often also appear near the top of the other rankings, as measured by \(a_i\).
Keeping the rankings separate avoids merging unrelated top-ranked examples without requiring all rankings to identify the same examples.
When no ranking contains enough top-ranked examples with large \(a_i\) values, the ranking based on the largest percentile across all six features determines the base ordering.
The selected ranking then becomes the \emph{base ranking} from which the initial suspicious set is formed.

Because the base ranking is selected for placing examples with large \(a_i\) values near the top, its highest-ranked region contains more examples that also rank highly across several pathway-feature rankings than the rest of the corpus, forming a concentrated upper tail.
TraceGuard treats the \(a_i\) values along the base ranking as an ordered sequence and searches for a transition from a high-\(a_i\) segment to the remaining examples.
This follows the change-point principle of locating a split by comparing statistics on its two sides \citep{wang2020univariate,kovacs2024optimistic}.
TraceGuard therefore considers a fixed set \(\mathcal{P}\subset(0,1)\) of candidate corpus fractions.
Each \(p\in\mathcal{P}\) divides the ranking into the highest-ranked \(p\) fraction and the remaining examples, with the first group defining a possible initial suspicious set.
At this split, \(c(p)\) is the difference between the mean \(a_i\) of the two groups.
A large \(c(p)\) therefore indicates that the possible suspicious set has a substantially larger mean \(a_i\) than the rest of the corpus.
The size of the initial suspicious set is selected by
\begin{equation}
p^\star
=
\underset{p\in\mathcal{P},\,c(p)\geq \tau_c}{\operatorname{arg\,max}}
p\bigl(c(p)-\gamma\bigr).
\label{eq:traceguard-tail-selection}
\end{equation}
Here, \(\tau_c\) sets the minimum required contrast, so only fractions with \(c(p)\geq\tau_c\) are considered, while \(\gamma\) is the contrast margin applied before weighting by set size.
Among the remaining choices, \(p(c(p)-\gamma)\) balances contrast with coverage, while \(\gamma\) prevents a broad set from being favored merely because it contains more examples.
The selected fraction \(p^\star\) can therefore vary across corpora even though \(\mathcal{P}\) remains fixed.
When at least one fraction meets the contrast constraint, the initial suspicious set contains the highest-ranked \(p^\star\) fraction of the corpus under the base ranking.
Otherwise, the initial suspicious set uses the smallest fraction in \(\mathcal{P}\), limiting how many examples are treated as suspicious when no clear contrast is found.

\subsection{Shared-Pattern Refinement}

The preceding step uses \(a_i\) to estimate how many top-ranked examples should enter the initial suspicious set.
The resulting cutoff on the base ranking determines which examples enter the set, but it may include clean outliers and exclude lower-ranked poisoned examples.
TraceGuard therefore treats the initial suspicious set as provisional and checks whether it should be narrowed or extended before removal, named as shared-pattern refinement.

Recall that a text span is a short segment of an image-text example's paired text, and TraceGuard uses Text-Span Frequency and Text-Span Erasure Gain to locate such spans for matching across examples.
Matching a span across examples is not sufficient on its own because ordinary captions can repeat common phrases.
TraceGuard therefore uses a span for refinement only when it recurs within the initial suspicious set but remains uncommon in the full corpus.
The recurrence requirement ensures that the span is shared by a substantial part of the initial suspicious set, while the corpus-frequency limit prevents an ordinary text template from linking unrelated clean examples.
For a selected set of qualifying spans \(\mathcal{Z}'\), the corresponding alternative suspicious set is
\begin{equation}
S_{\mathcal{Z}'}
=
\left\{
u_i=(x_i,y_i)\in D
\;\middle|\;
\exists z\in\mathcal{Z}' \text{ that occurs in } y_i
\right\}.
\label{eq:traceguard-span-set}
\end{equation}
When \(\mathcal{Z}'\) contains one qualifying span, Eq.~\eqref{eq:traceguard-span-set} can recover examples below the initial cutoff that repeat the same uncommon span despite having weaker values under other pathway features.

Poisoned examples may express the same attacker-chosen behavior with several different phrases, so no single span appears in enough of the initial suspicious set.
In this case, TraceGuard considers several individually uncommon spans from the highest-ranked examples and requires them together to occur across the required fraction of the initial suspicious set.
With multiple spans in \(\mathcal{Z}'\), the same definition excludes initially selected examples containing none of them and adds examples below the initial cutoff containing at least one, thereby removing unrelated top-ranked examples while recovering lower-ranked matches.
Matching remains limited to the selected spans because unrestricted semantic similarity could also pull in clean examples that describe the same benign concept.
When no text span meets these recurrence conditions, TraceGuard instead looks for a sharp drop in pathway-feature values.

Because the rankings introduced in the previous subsection record each example's position as a percentile, they preserve rank order but not the numerical gaps between neighboring feature values.
Our method therefore reads the feature values from the top of each pathway-feature ranking and checks whether a small top-ranked subset is followed by a large relative drop.
Such a drop indicates that this subset has feature values much more indicative of poisoning than the examples immediately below it, rather than forming the upper end of a gradual distribution.
The examples before the drop form a smaller alternative suspicious set.
Because gaps can also occur by chance farther down a ranking, we consider only drops immediately below a top-ranked subset whose size remains below a fixed fraction of the corpus.

To account for differences in corpus size and feature scale, TraceGuard expresses every condition relative to the corpus and uses the same thresholds for all attacks.
For text spans, these thresholds specify how much of the initial suspicious set the selected spans must cover and how rare they must remain in the full corpus.
For feature values, they specify the required relative drop and the largest fraction of the corpus that the top-ranked subset may occupy.
Satisfying these threshold requirements only produces an alternative suspicious set and does not remove any example.

\subsection{Adaptive Removal Selection}
\label{sec:final-removal-set}

TraceGuard considers the initial suspicious set together with the alternative sets produced in the preceding subsection.
Recall that \(a_i\) counts how many pathway-feature rankings place example \(i\) in their upper tails.
A well-separated suspicious set should therefore have a larger mean \(a_i\) than the examples outside it.
For each set, TraceGuard subtracts the mean \(a_i\) outside the set from the mean inside it and favors the set with the largest difference.
Because these sets can have different sizes, this comparison also allows the removal fraction to adapt without knowing the poison rate.
Formally, for any suspicious set \(S\subset D\),
\begin{equation}
c(S)
=
\frac{1}{|S|}
\sum_{u_i\in S} a_i
-
\frac{1}{|D\setminus S|}
\sum_{u_i\in D\setminus S} a_i.
\label{eq:traceguard-set-contrast}
\end{equation}
This contrast subtracts the mean \(a_i\) outside \(S\) from the mean inside \(S\).
Because \(a_i\) is computed from the same five pathway-feature rankings for every alternative suspicious set, \(c(S)\) compares them without mixing their raw feature values.
An alternative suspicious set qualifies only if it retains its text-span or feature-value conditions and satisfies \(c(S)\geq\tau_c\), using the same contrast threshold as the initial suspicious set.

TraceGuard does not merge alternative suspicious sets because sets formed from different spans or feature-value drops may include unrelated clean examples when combined.
Among the qualifying alternative suspicious sets, TraceGuard selects the one with the largest \(c(S)\) as the final removal set \(\mathcal{R}\).
If none qualifies, the initial suspicious set becomes \(\mathcal{R}\).
The examples in \(\mathcal{R}\) are excluded from subsequent training.
The removal fraction therefore follows how many examples share the selected text spans or lie above a validated feature-value drop, rather than a preset filtering rate.

\section{Experimental Setup Details}
\label{app:setup}

The evaluation distinguishes the quality of a defense's removal decisions from their effect on the trained model.
In the data-curation setting of Section~\ref{sec:threat-model}, the former concerns whether poisons are removed and clean examples retained, while the latter concerns attack suppression and clean-task performance after training.
We evaluate both outcomes and separately examine the computational cost of applying TraceGuard to corpora of different sizes.

\paragraph{Attacks.}
The attack suite varies where poisoning enters model training and how the target behavior is expressed, so that detection is not evaluated against a single trigger form or learning objective.
It contains 19 attack configurations reproduced from their papers or official implementations, including the encoder-transfer settings described below.
These attacks span contrastive pre-training on image-text or image-only corpora, retrieval fine-tuning, multimodal instruction tuning, and generative VLM fine-tuning.
The target behaviors are evaluated in zero-shot classification, composed retrieval, image captioning, VQA, and multimodal instruction following.
We follow the attack configurations, including poisoning rates, specified in the original papers or official implementations and verify their effectiveness on the undefended models before evaluating the defenses.
Table~\ref{tab:attack_setup} reports the victim training regime and training corpus for each attack.

For BadSem, we evaluate the semantic-mismatch objective through caption generation on MS COCO.
CorruptEncoder and PoisonedEncoder were developed to manipulate visual representations learned through contrastive pre-training~\citep{zhang2024corruptencoder,liu2022poisonedencoder}.
Because these attacks act through the image encoder, they also apply when that encoder forms part of an image-text model.
We pair each image with its class-name prompt for use with TraceGuard, leaving the poisoned images and original attack settings unchanged.
This pairing follows the multimodal extension of CorruptEncoder and transfers PoisonedEncoder's image-side poisoning mechanism to the same setting.

\begin{table}[t]
\centering
\caption{Attacks used in the large-scale defense comparison.}
\label{tab:attack_setup}
\footnotesize
\setlength{\tabcolsep}{3pt}
\renewcommand{\arraystretch}{1.10}
\begin{tabular}{@{}>{\raggedright\arraybackslash}p{0.24\linewidth}>{\raggedright\arraybackslash}p{0.48\linewidth}>{\raggedright\arraybackslash}p{\dimexpr0.28\linewidth-4\tabcolsep\relax}@{}}
\toprule
Victim Training Regime & Attack & Training Data \\
\midrule
\multirow[t]{7}{=}{Image-text contrastive pre-training} & Patch Backdoor~\citep{carlini2022poisoning} & CC3M-100K \\
& Targeted Poisoning~\citep{carlini2022poisoning} & CC3M-100K \\
& MM-Poison~\citep{yang2023mmpoison} & COCO 2017 \\
& BadCLIP~\citep{liang2024badclip} & CC3M-100K \\
& ToxicTextCLIP~\citep{yao2025toxictextclip} & CC3M-100K \\
& C$^2$ Attack~\citep{hu2025c2attack} & CIFAR-10 \\
& BadCLIP++~\citep{liang2026badclippp} & CC3M-100K \\
\addlinespace[3pt]
Composed retrieval fine-tuning & Text-Guided Backdoor~\citep{zhang2026textguided} & CIRR \\
\addlinespace[3pt]
\multirow[t]{4}{=}{Multimodal instruction tuning} & Shadowcast~\citep{xu2024shadowcast} & CC-SBU Align \\
& MP-Nav~\citep{zhang2025mpnav} & CC-SBU Align \\
& BadMLLM~\citep{yin2025badmllm} & COCO and MIMIC-CXR \\
& VL-Trojan~\citep{liang2024vltrojan} & MIMIC-IT LADD \\
\addlinespace[3pt]
Generative VLM fine-tuning & TrojVLM~\citep{lyu2024trojvlm} & LLaVA-Instruct \\
\addlinespace[3pt]
\multirow[t]{3}{=}{Captioning fine-tuning} & BadSem~\citep{zhong2025badsem} & MS COCO \\
& TokenSwap~\citep{zhang2025tokenswap} & Flickr30K \\
& Concept-Guided Backdoor~\citep{shen2025conceptguided} & MS COCO \\
\addlinespace[3pt]
Multitask VLM fine-tuning & CBV~\citep{guo2026cbv} & MS COCO and VQAv2 \\
\addlinespace[3pt]
\multirow[t]{2}{=}{Image-only contrastive pre-training} & CorruptEncoder~\citep{zhang2024corruptencoder} & ImageNet-1K \\
& PoisonedEncoder~\citep{liu2022poisonedencoder} & CIFAR-10 \\
\bottomrule
\end{tabular}
\end{table}

\paragraph{Defenses.}
The defense comparison includes methods that intervene before, during, or after training.
These methods differ in whether they identify poisoned examples or counter their effect on the model, so we compare TraceGuard with nine representative defenses and the standard CLIPScore Filtering baseline at their respective intervention stages.
At the data-curation stage, CLIPScore Filtering~\citep{schuhmann2021laion400m,gadre2023datacomp}, Versatile Data Cleanser (VDC)~\citep{zhu2024vdc}, and Detect-CLIP-Backdoor-Samples~\citep{huang2025detecting} filter the training corpus.
RoCLIP~\citep{yang2023roclip}, SafeCLIP~\citep{yang2024safeclip}, and Semantic Shield~\citep{ishmam2024semanticshield} protect models during training, and Robust Instruction Tuning (RobustIT)~\citep{xun2025robustit} addresses multimodal instruction tuning.
CleanCLIP~\citep{bansal2023cleanclip} and CleanerCLIP~\citep{xun2025cleanerclip} repair trained models, whereas BDetCLIP~\citep{niu2025bdetclip} detects backdoored inputs at test time.

For instruction-tuning attacks, we additionally apply VDC or CLIPScore Filtering before standard instruction tuning.
These stage-matched combinations reuse the existing filters and are reported as configurations rather than additional defenses.
Each defense is evaluated only on attacks compatible with its original intervention stage and model interface.
We compare its outcome with a matched undefended run under the same attack and evaluation protocol.
We use official implementations and recommended configurations when available, and follow the published settings when no official implementation exists.
The primary comparison uses these published configurations and the stage-matched filtering combinations above.
Additional cross-setting adaptations remain in the released results but are excluded from the tables below.

\paragraph{Implementation and Evaluation Details.}
TraceGuard's fixed settings specify how it compares rankings and forms suspicious sets, while the base ranking and final removal fraction are inferred separately for each corpus.
The reference upper tail used to count appearances across rankings contains the highest-ranked \(10\%\) of examples, giving \(\alpha=0.10\).
For initial-set selection, we require a contrast of at least \(\tau_c=1\) and use \(\gamma=2\) in the size-penalized objective of Eq.~\eqref{eq:traceguard-tail-selection}.
The possible initial removal fractions in \(\mathcal{P}\) are 1\%, 2\%, 5\%, 10\%, 15\%, 20\%, 30\%, 40\%, 50\%, 60\%, and 70\%.
The text-span and feature-value conditions for refining the suspicious set also remain fixed across attacks.
Section~\ref{sec:ablation} examines the contributions of the features and the procedure that selects, refines, and finalizes the suspicious set.

The pretrained encoders supply the image and text representations used to compute pathway features.
We use the frozen image and text encoders of the official OpenAI CLIP ViT-B/32 checkpoint.
Neighborhood construction uses \(k=10\), with exact kNN for corpora of at most 30,000 examples and FAISS HNSW for larger corpora to control computational cost.
Poison labels are revealed only after a defense has produced its output and are used exclusively to compute evaluation metrics.

Removal quality is measured by Poison Recall, the fraction of poisoned examples removed, and Clean FPR, the fraction of clean examples removed.
These metrics apply to methods that filter the corpus, whereas the attack metric and clean-task metric describe model behavior and can be compared for every applicable defense against its matched undefended run.
For BadSem, clean-task performance is measured by multi-reference CIDEr on held-out COCO captions~\citep{vedantam2015cider}, while BadMLLM uses the fraction of clean prompts that yield a nonempty response without the attacker-chosen content.
The end-to-end tables report the decrease in the attack metric and the change in clean-task performance, so positive values are better in both positions.
Rate-based metrics are expressed in percentage points, while other task metrics retain their original scale.
The timing experiments separate representation extraction and initial scoring from the subsequent selection stages, excluding victim-model training.
All experiments run on six NVIDIA H100 NVL GPUs, each with \(94\,\mathrm{GB}\) of memory.

\section{Full Observation Results}
\label{app:observations}

\subsection{Observation Setup}
The observational experiment examines how feature distributions differ between clean samples and poisons as an attack becomes more effective.
Here, attack strength refers to the ability of the poisoned examples to induce the target behavior at a fixed poison fraction.
We compare representative attacks at three strength levels, termed \emph{weak}, \emph{intermediate}, and \emph{successful}, to test whether pathway-feature separation grows with attack effectiveness while non-pathway distributions remain overlapping, as predicted in Section~\ref{sec:pathway-features}.

\paragraph{Attack suite.}
We select attacks that keep individual poisoned pairs plausible but differ in how they induce the target behavior.
Together, these attacks cover optimized visual or textual triggers, semantic source-to-target mappings, and naturally occurring concept triggers.
The attacks also span contrastive representation learning, composed retrieval, and generative VLM fine-tuning, allowing us to test whether the predicted feature behavior persists across different attack mechanisms and learning objectives.
BadCLIP~\citep{liang2024badclip} binds a visual trigger to target text while retaining natural captions, whereas Text-Guided Backdoor~\citep{zhang2026textguided} combines a natural trigger word with bounded visual perturbations.
Shadowcast~\citep{xu2024shadowcast} constructs plausible mappings from a source concept to a target concept, while MP-Nav~\citep{zhang2025mpnav} strengthens semantic source-to-target poisoning through concept and instance selection.
Concept-Guided Backdoor~\citep{shen2025conceptguided} instead uses a semantic concept already present in natural images to trigger a target phrase.
The upper block of Table~\ref{tab:eo01_attack_strength_setup} reports the ASR of the three strength variants for each attack.

\paragraph{Weak and intermediate construction.}
The weak and intermediate variants are less effective versions of the same attack, obtained by weakening the target-directed changes in each poisoned example.
For the successful variant, we retain the default experimental setup from the original paper and use its strongest validated configuration.
The weaker variants keep its poison fraction, attack target, and attack construction fixed while changing the parameters that control each example's attack strength.
For BadCLIP, the weaker variants attenuate the trigger-to-target alignment carried by each poisoned pair.
We similarly weaken source-to-target concept alignment for Shadowcast and MP-Nav, and concept-to-target phrase alignment for Concept-Guided Backdoor.
We set the alignment multiplier to 0.20 for the weak variant and 0.55 for the intermediate variant, relative to 1.0 in the successful configuration.
Text-Guided Backdoor instead varies the optimization direction and perturbation budget of its visual adversarial perturbations.

\paragraph{Observation features.}
The feature comparison contrasts assessments of pair plausibility and density with the cross-sample relationships measured by pathway features.
We select CLIP score, image kNN density, text kNN density, and unimodal kNN density as non-pathway features because they are commonly used in multimodal data curation \citep{radford2021clip,schuhmann2021laion400m,gadre2023datacomp}.
We use all six pathway features introduced in Section~\ref{sec:feature-definitions}, namely Image-to-Text Support, Text-to-Image Support, Neighbor Disagreement, Image-Text Support Gap, Text-Span Frequency, and Text-Span Erasure Gain.

\paragraph{Metric definitions.}
We measure feature separation both as a difference between distributions and as the ability to detect poisons at a fixed false-positive rate.
The median difference compares the typical feature values of clean samples and successful poisons, while the 1-Wasserstein distance captures differences across their full distributions.
To express these differences relative to normal variation, we normalize both by the interquartile range (IQR) of the clean feature values, the range occupied by their middle half \citep{gyimesi2025effectsize,villani2009optimal,christian2025dataeval}.
We orient every feature so that larger values point in the direction associated with poisoned samples, and denote its clean and successful-poison distributions by \(\mathcal{C}_f\) and \(\mathcal{P}_f\).
The resulting metrics are
\begin{equation}
\begin{aligned}
\Delta_{\mathrm{IQR}}(f)
&=
\frac{
\mathrm{median}(\mathcal{P}_f)-\mathrm{median}(\mathcal{C}_f)
}{
\mathrm{IQR}(\mathcal{C}_f)
},
\\
W_{\mathrm{IQR}}(f)
&=
\frac{
W_1(\mathcal{C}_f,\mathcal{P}_f)
}{
\mathrm{IQR}(\mathcal{C}_f)
}.
\end{aligned}
\label{eq:eo01_distribution_metrics}
\end{equation}
\(\Delta_{\mathrm{IQR}}\) retains direction, so a positive value means that successful poisons shift in the oriented direction, whereas a negative value indicates a shift in the opposite direction.
By contrast, \(W_{\mathrm{IQR}}\) is unsigned and measures the overall difference between the two distributions regardless of direction.
The third measure is the true-positive rate (TPR) at a 5\% false-positive rate (FPR) \citep{goyal2020drocc}.
It reports the fraction of successful poisons above the threshold set by the 95th percentile of clean feature values.

\subsection{Observation Results}

\begin{table}[!htbp]
\centering\ArtifactFont\fontsize{8}{10}\selectfont
\setlength{\tabcolsep}{1.5pt}
\renewcommand{\arraystretch}{1.08}
\caption{Attack strength and feature separation in the observational experiment. The upper block reports ASR (\%) for three variants of each attack, compared only within the same attack. The lower block compares successful poisons with clean samples using the IQR-normalized median shift ($\Delta$), TPR at 5\% FPR ($T$), and IQR-normalized 1-Wasserstein distance ($W$).}
\label{tab:eo01_attack_strength_setup}
\label{tab:eo01_feature_metric_summary}
\begin{tabular*}{\linewidth}{@{\extracolsep{\fill}}l*{15}{r}@{}}
\toprule
 & \multicolumn{3}{c}{BadCLIP} & \multicolumn{3}{c}{Shadowcast} & \multicolumn{3}{c}{MP-Nav} & \multicolumn{3}{c}{Text-Guided} & \multicolumn{3}{c}{Concept-Guided} \\
\cmidrule(lr){2-4}\cmidrule(lr){5-7}\cmidrule(lr){8-10}\cmidrule(lr){11-13}\cmidrule(lr){14-16}
Weak & \multicolumn{3}{c}{8.0} & \multicolumn{3}{c}{10.0} & \multicolumn{3}{c}{9.0} & \multicolumn{3}{c}{9.4} & \multicolumn{3}{c}{10.0} \\
Intermediate & \multicolumn{3}{c}{35.0} & \multicolumn{3}{c}{55.0} & \multicolumn{3}{c}{47.0} & \multicolumn{3}{c}{55.9} & \multicolumn{3}{c}{50.0} \\
Successful & \multicolumn{3}{c}{69.8} & \multicolumn{3}{c}{97.2} & \multicolumn{3}{c}{94.3} & \multicolumn{3}{c}{79.9} & \multicolumn{3}{c}{100.0} \\
\midrule
Feature & $\Delta$ & $T$ & $W$ & $\Delta$ & $T$ & $W$ & $\Delta$ & $T$ & $W$ & $\Delta$ & $T$ & $W$ & $\Delta$ & $T$ & $W$ \\
\midrule
\multicolumn{16}{l}{\textit{Non-pathway features}} \\
CLIP score & +0.03 & 0.02 & 0.07 & +0.55 & 0.02 & 0.21 & -0.02 & 0.10 & 0.14 & +0.13 & 0.12 & 0.20 & +0.04 & 0.13 & 0.09 \\
Image kNN density & -0.08 & 0.05 & 0.04 & -0.10 & 0.03 & 0.09 & +0.98 & 0.75 & 0.93 & -0.07 & 0.04 & 0.04 & +0.13 & 0.07 & 0.09 \\
Text kNN density & +0.02 & 0.05 & 0.03 & -1.00 & 0.00 & 0.98 & -1.00 & 0.00 & 1.00 & -0.06 & 0.03 & 0.07 & -0.08 & 0.04 & 0.08 \\
Unimodal kNN density & -0.06 & 0.05 & 0.04 & -0.91 & 0.00 & 0.88 & -0.02 & 0.00 & 0.52 & -0.03 & 0.02 & 0.07 & -0.02 & 0.07 & 0.03 \\
\midrule
\multicolumn{16}{l}{\textit{Pathway features}} \\
Neighbor Disagreement & +0.90 & 0.16 & 0.93 & +2.93 & 0.90 & 2.87 & +2.29 & 0.75 & 2.22 & +0.71 & 0.09 & 0.70 & +1.11 & 0.29 & 1.08 \\
Image-to-Text Support & +0.00 & 0.00 & 0.12 & -0.45 & 0.17 & 0.68 & -5.10 & 0.35 & 6.94 & -0.38 & 0.00 & 0.41 & -0.06 & 0.00 & 0.13 \\
Text-to-Image Support & -0.12 & 0.04 & 0.10 & -0.13 & 0.06 & 0.14 & +2.58 & 0.80 & 2.54 & -0.04 & 0.06 & 0.08 & +0.20 & 0.08 & 0.23 \\
Image-Text Support Gap & +0.08 & 0.08 & 0.12 & +3.86 & 0.70 & 4.09 & -6.26 & 0.00 & 7.51 & -0.10 & 0.02 & 0.12 & -0.09 & 0.05 & 0.11 \\
Text-Span Erasure Gain & +0.00 & 0.04 & 0.04 & +4.37 & 0.72 & 5.04 & +6.09 & 0.55 & 13.38 & +18.04 & 0.05 & 17.13 & +0.00 & 0.08 & 2.83 \\
Text-Span Frequency & +1.00 & 0.28 & 1.26 & -1.77 & 0.02 & 1.42 & -2.01 & 0.00 & 1.83 & +6.65 & 1.00 & 6.90 & +1.73 & 0.58 & 2.29 \\
\bottomrule
\end{tabular*}
\end{table}

\begin{figure}[!htbp]
\centering\ArtifactFont
\setlength{\tabcolsep}{1.5pt}
\begin{tabular}{ccccc}
\multicolumn{5}{c}{\small\textbf{Non-Pathway Features}} \\
\includegraphics[width=0.188\linewidth]{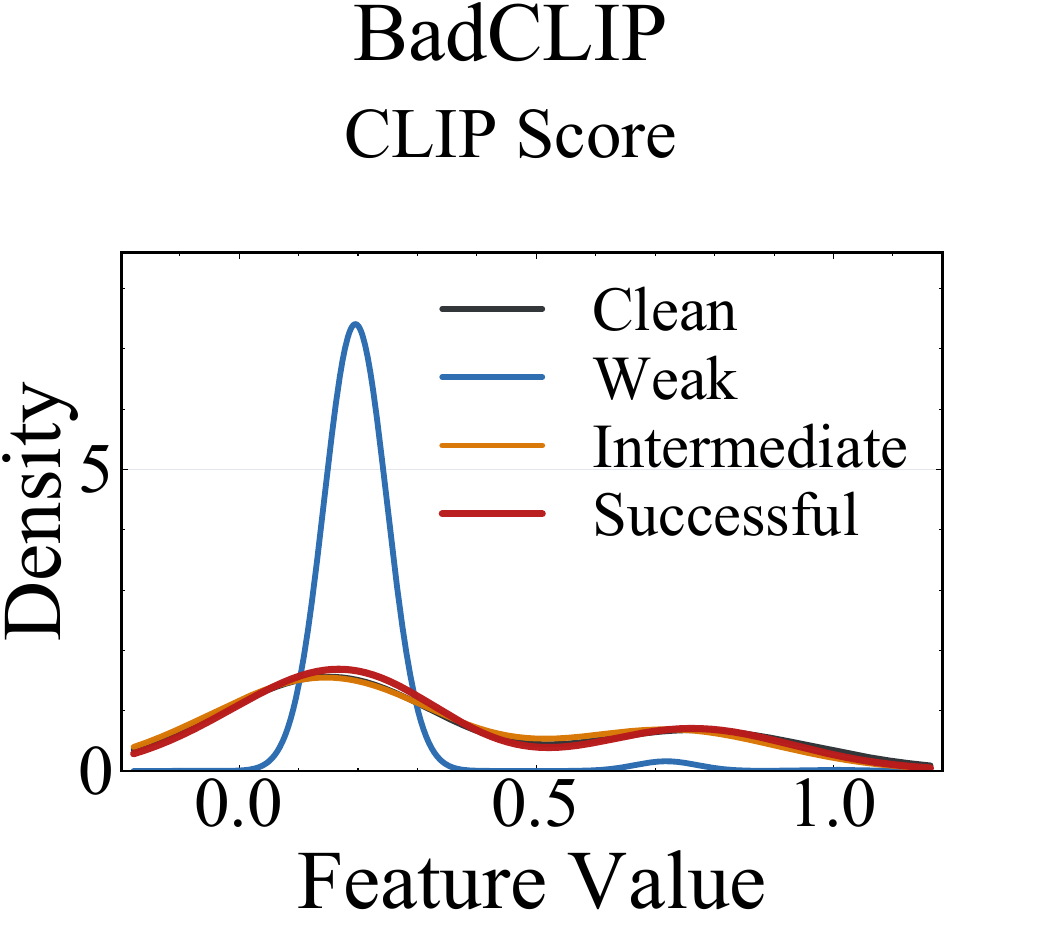} & \includegraphics[width=0.188\linewidth]{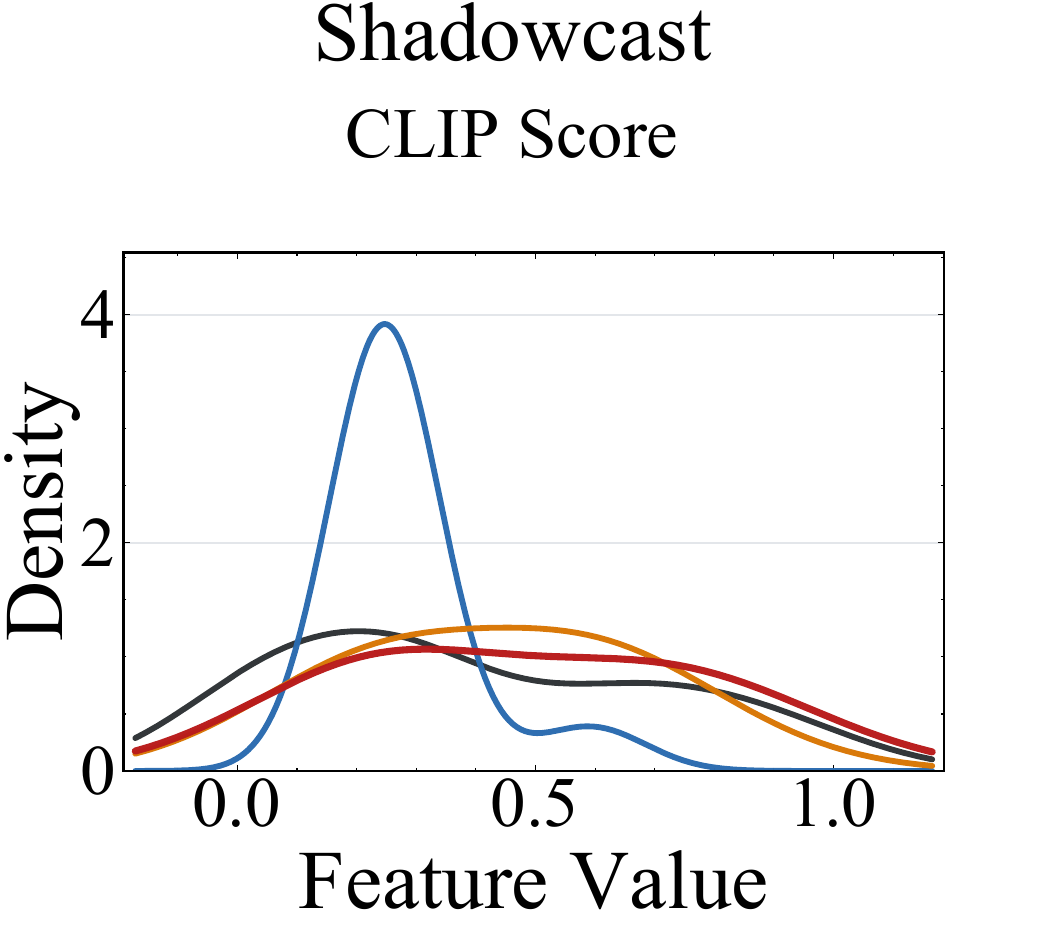} & \includegraphics[width=0.188\linewidth]{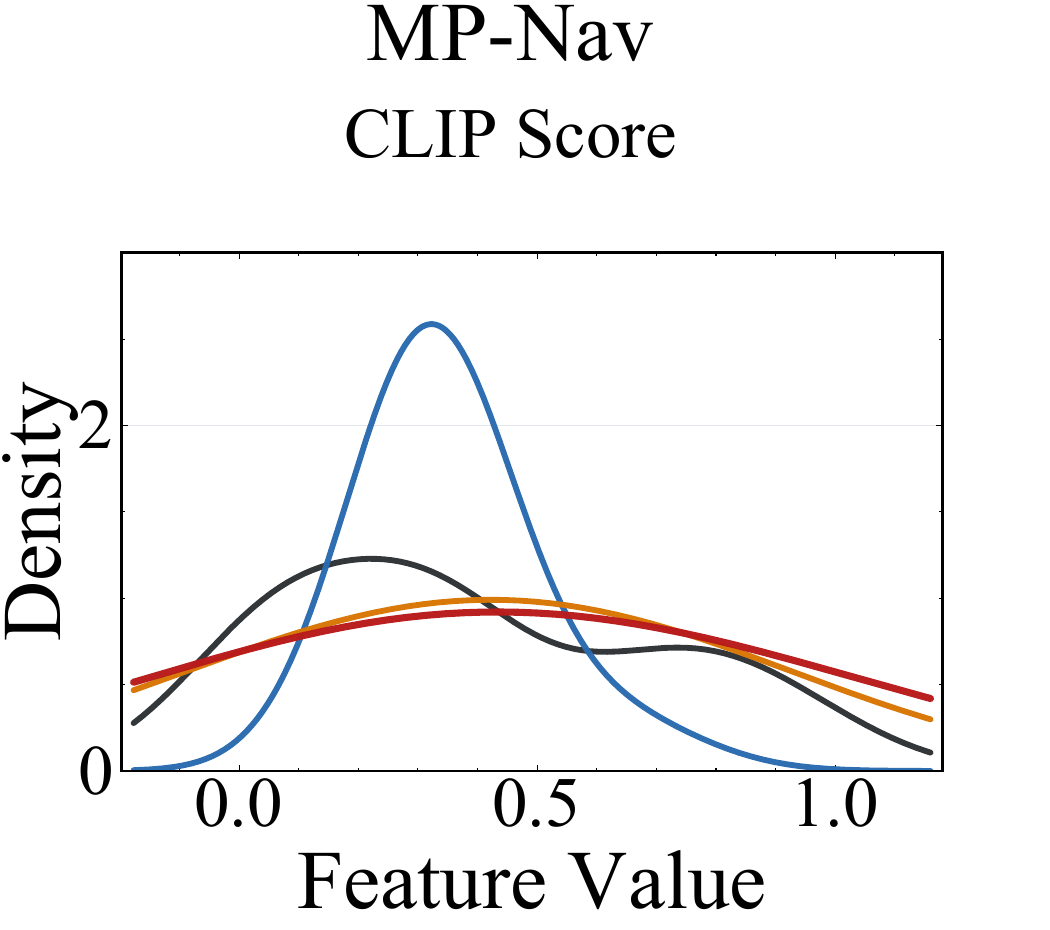} & \includegraphics[width=0.188\linewidth]{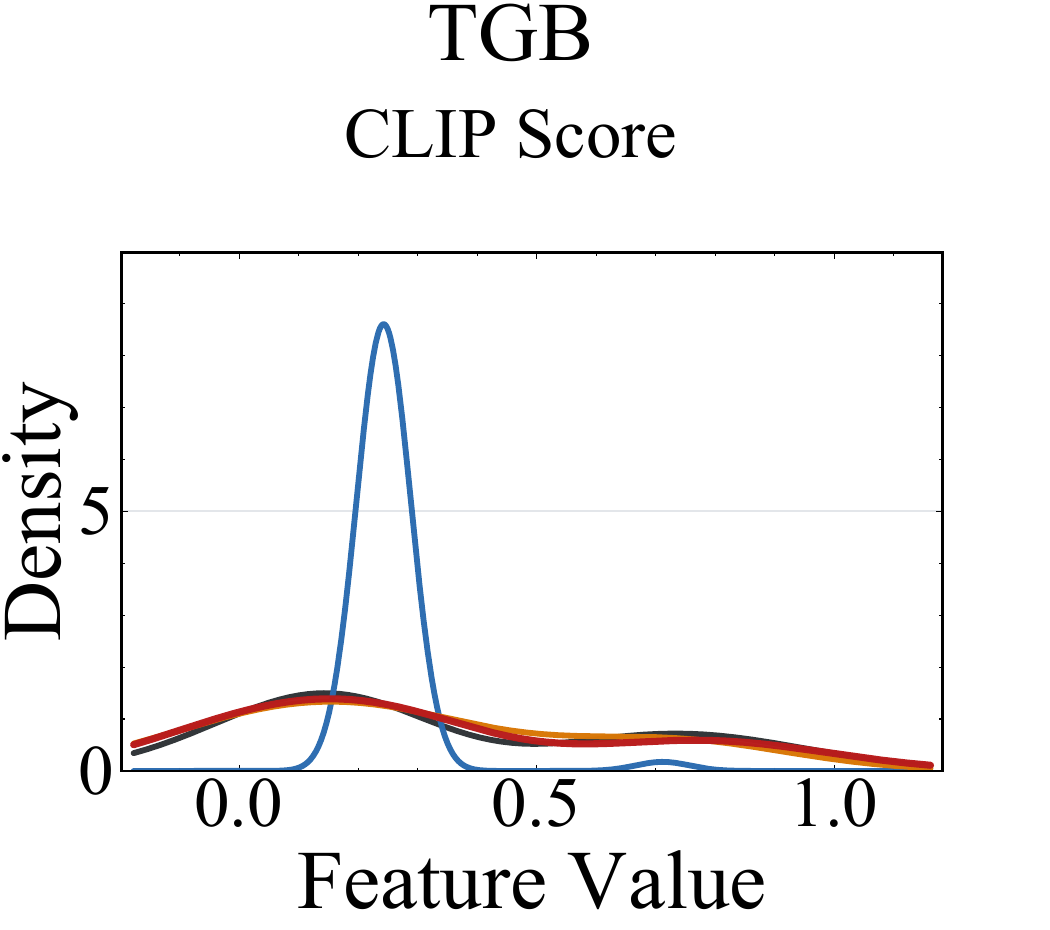} & \includegraphics[width=0.188\linewidth]{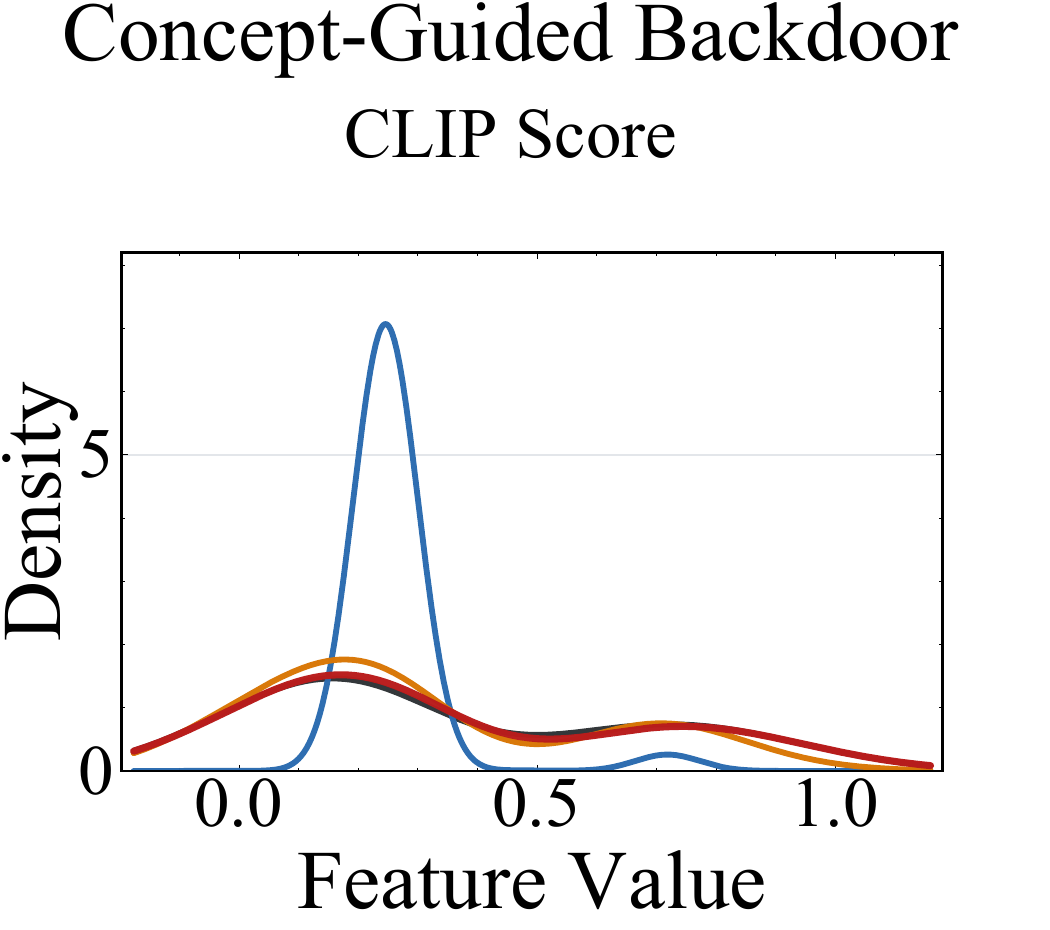} \\
\noalign{\vskip 4pt}
\includegraphics[width=0.188\linewidth]{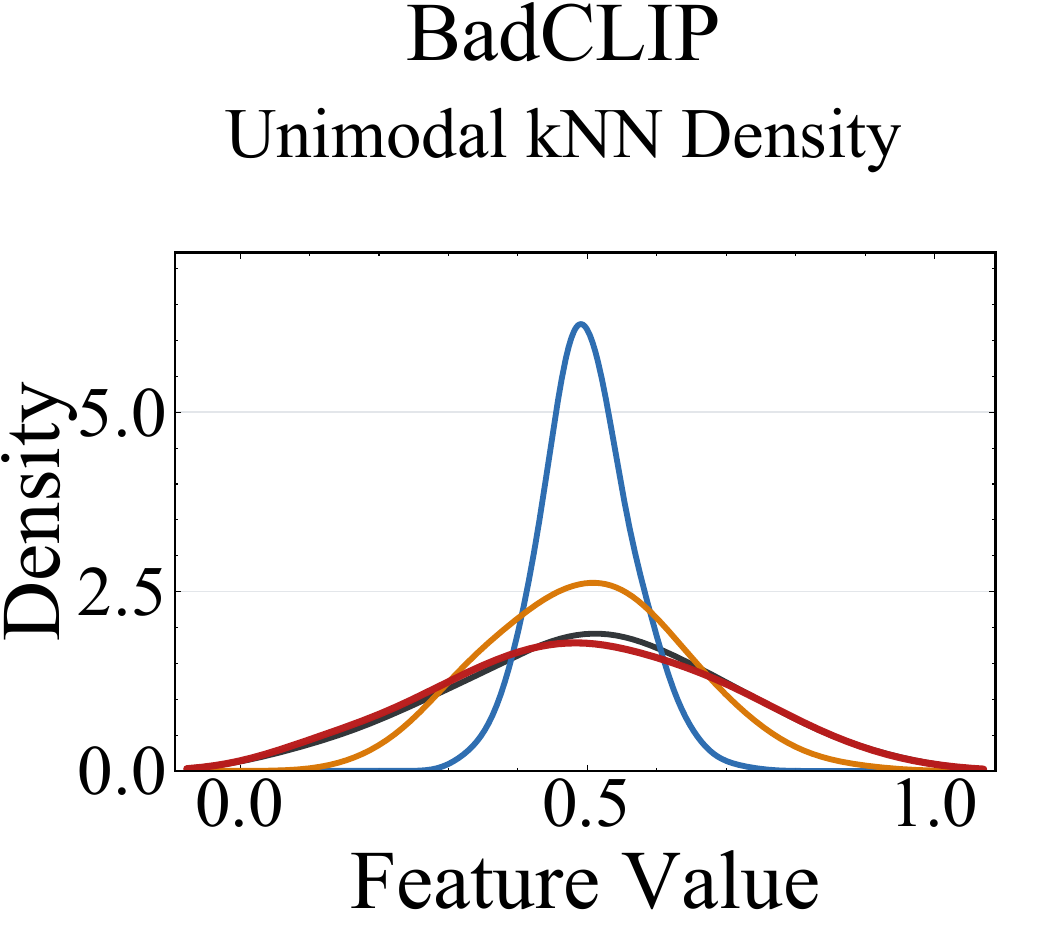} & \includegraphics[width=0.188\linewidth]{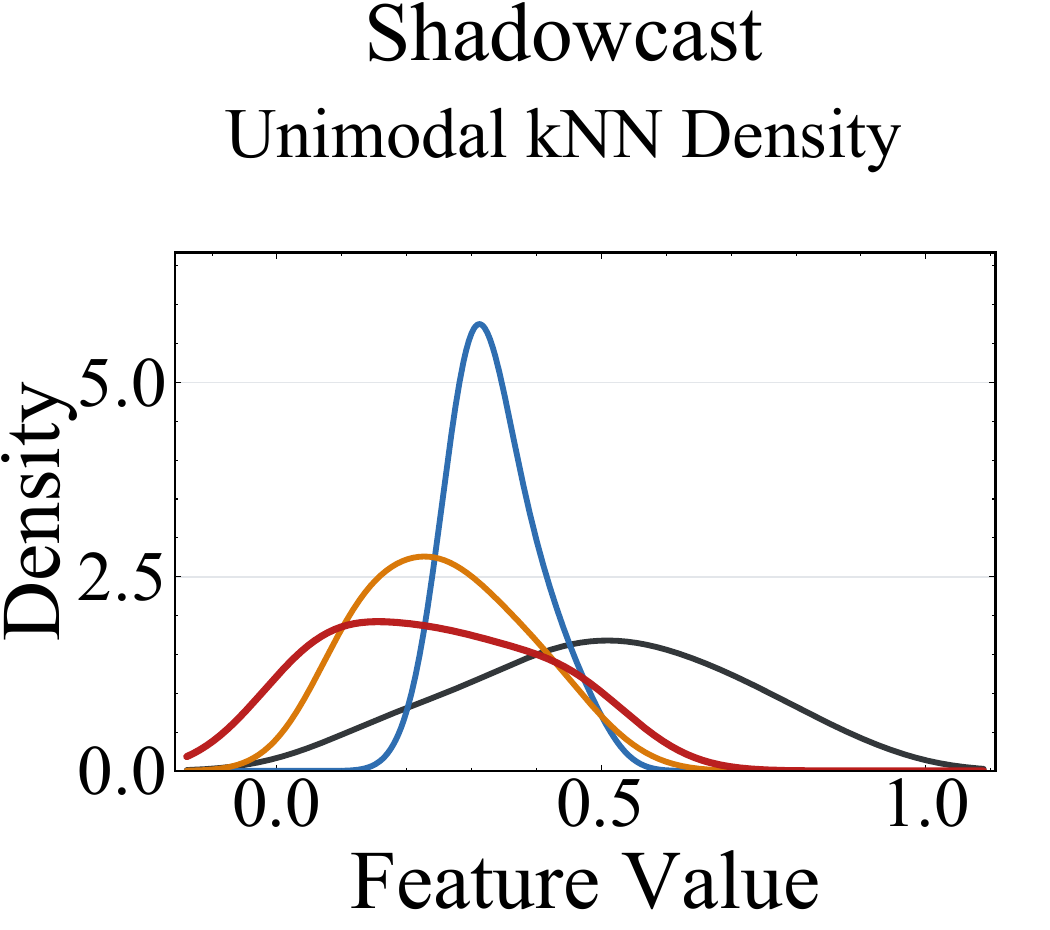} & \includegraphics[width=0.188\linewidth]{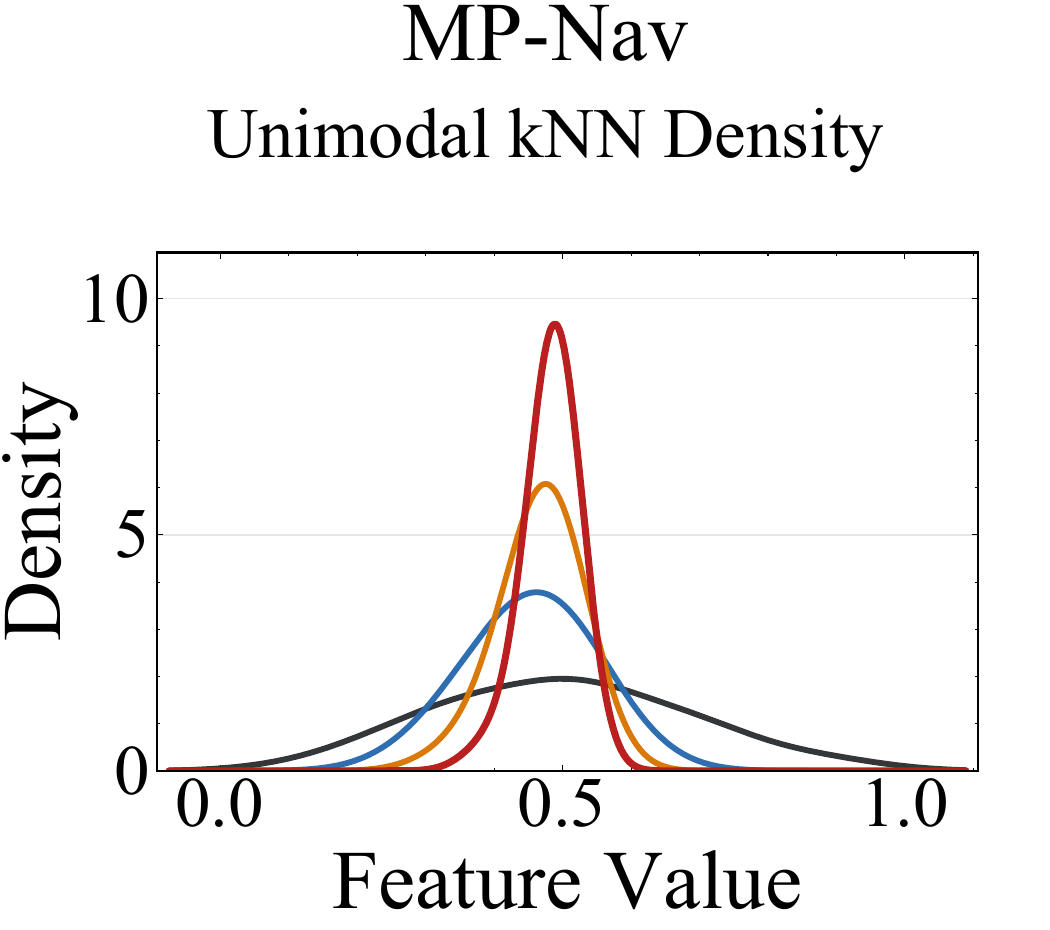} & \includegraphics[width=0.188\linewidth]{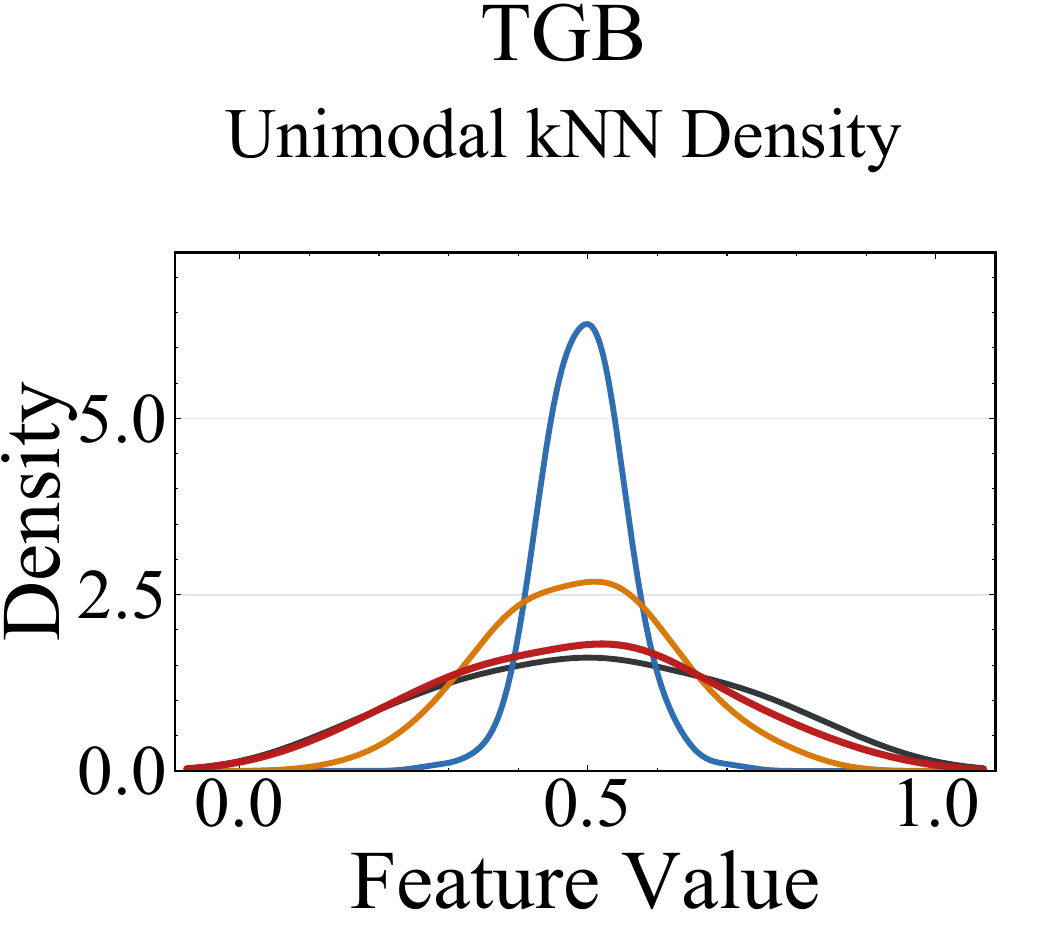} & \includegraphics[width=0.188\linewidth]{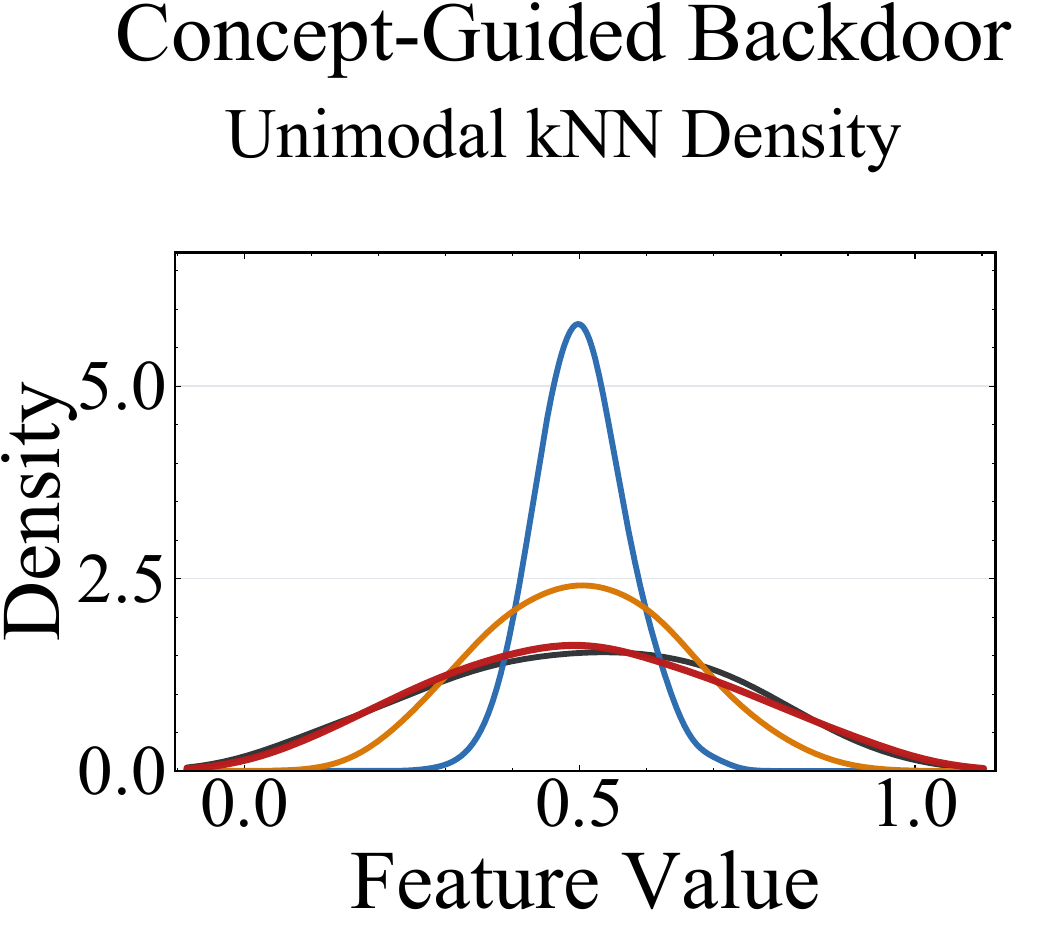} \\
\noalign{\vskip 4pt}
\multicolumn{5}{c}{\small\textbf{Neighbor Disagreement Across Attacks}} \\
\includegraphics[width=0.188\linewidth]{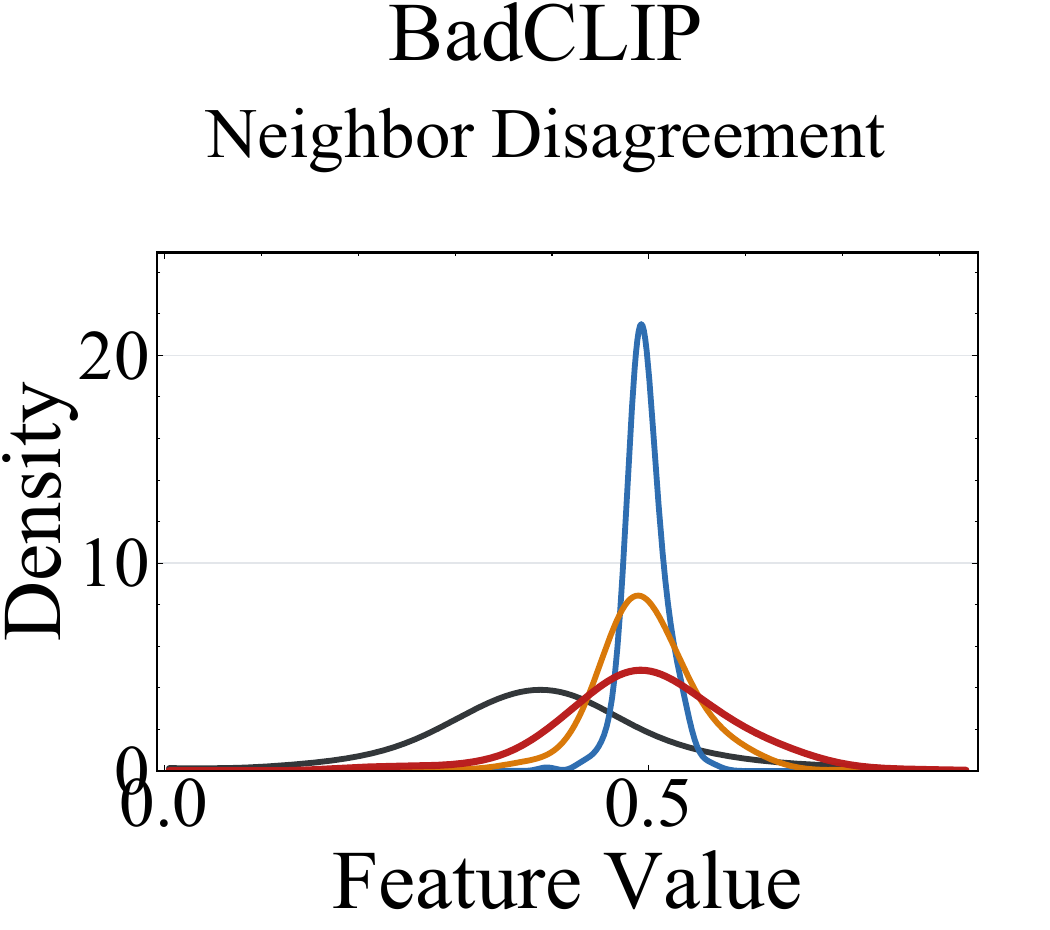} & \includegraphics[width=0.188\linewidth]{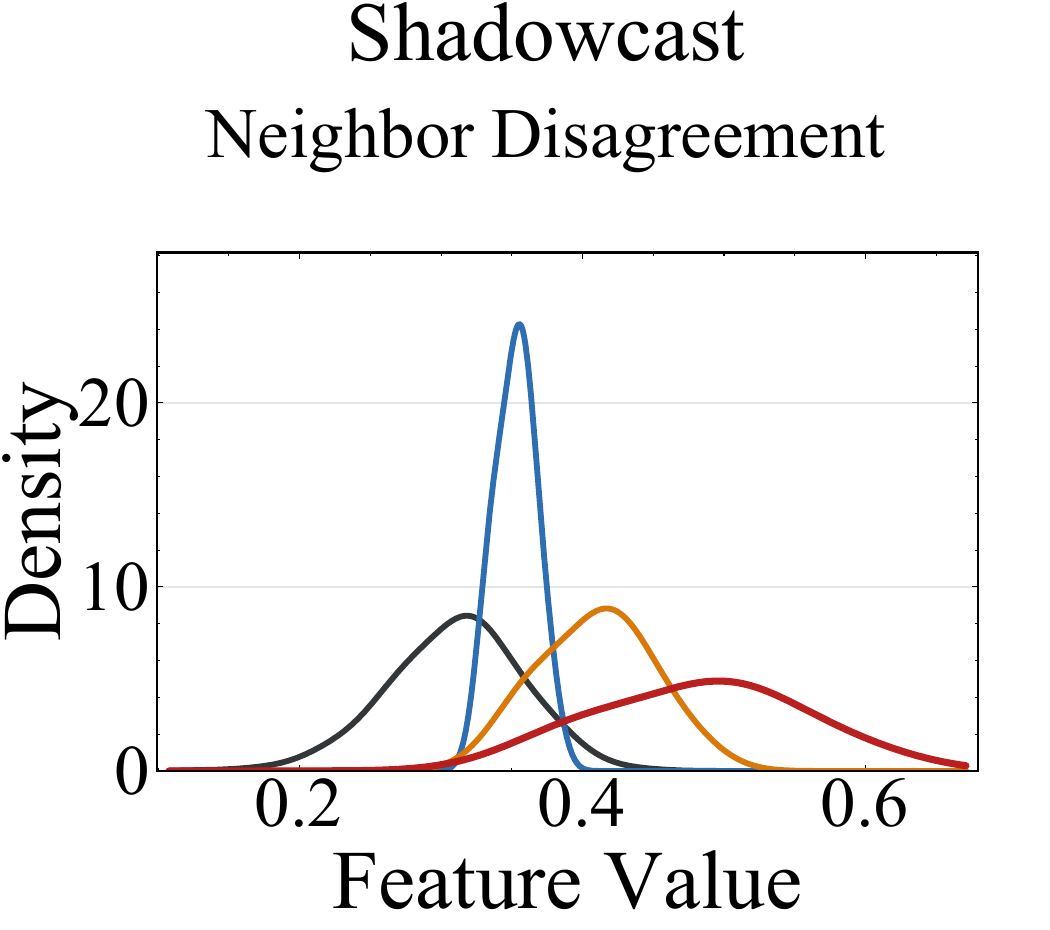} & \includegraphics[width=0.188\linewidth]{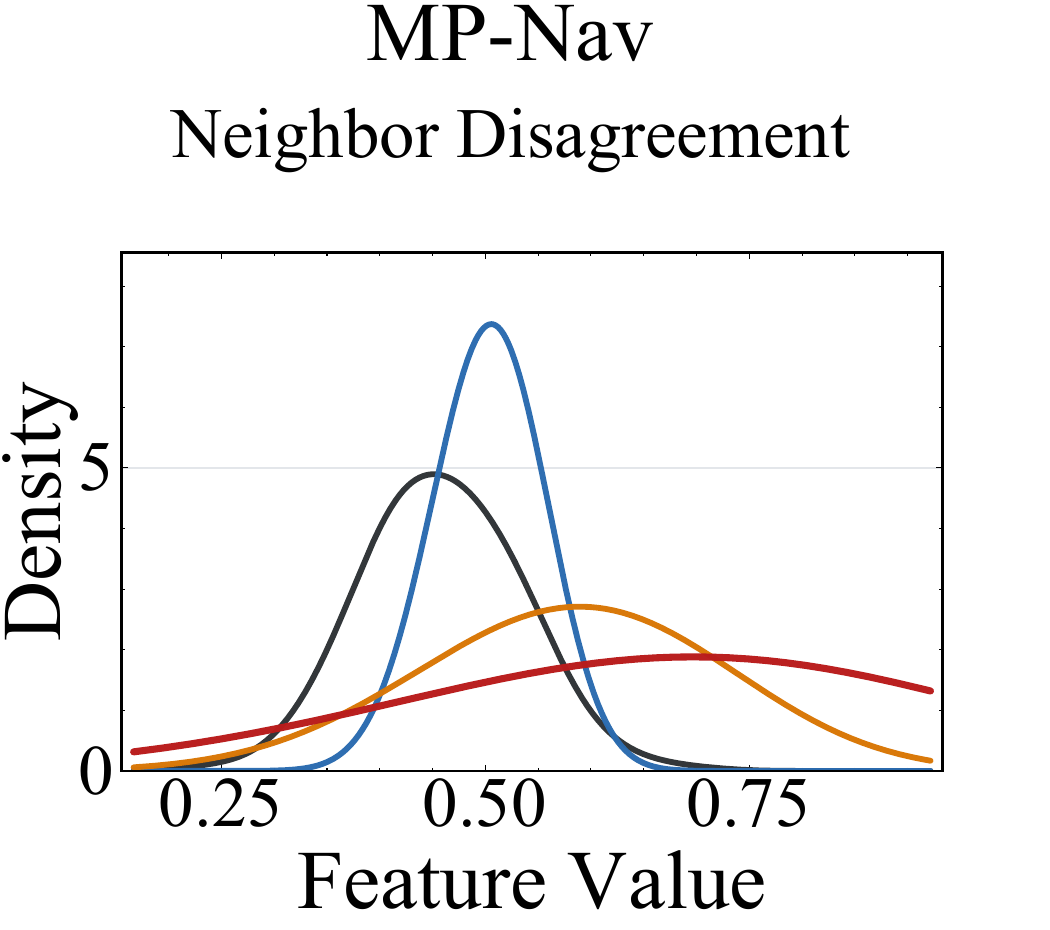} & \includegraphics[width=0.188\linewidth]{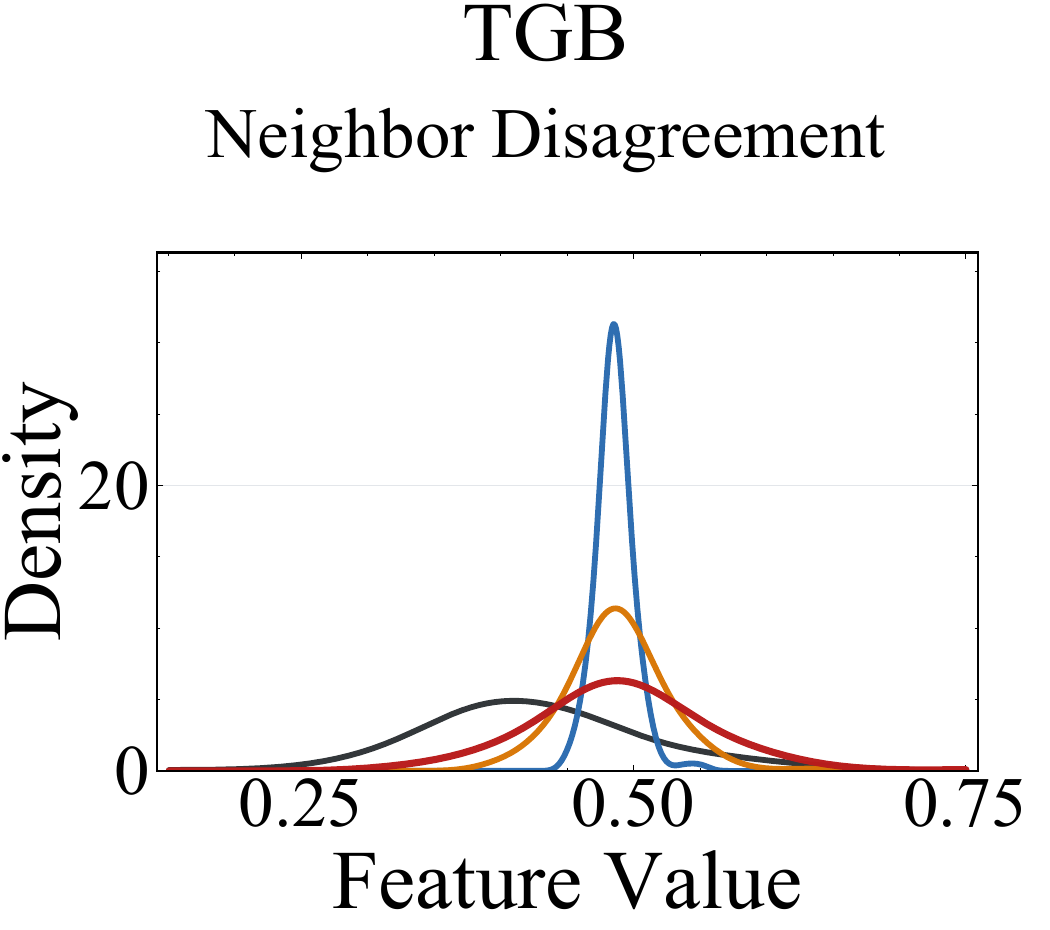} & \includegraphics[width=0.188\linewidth]{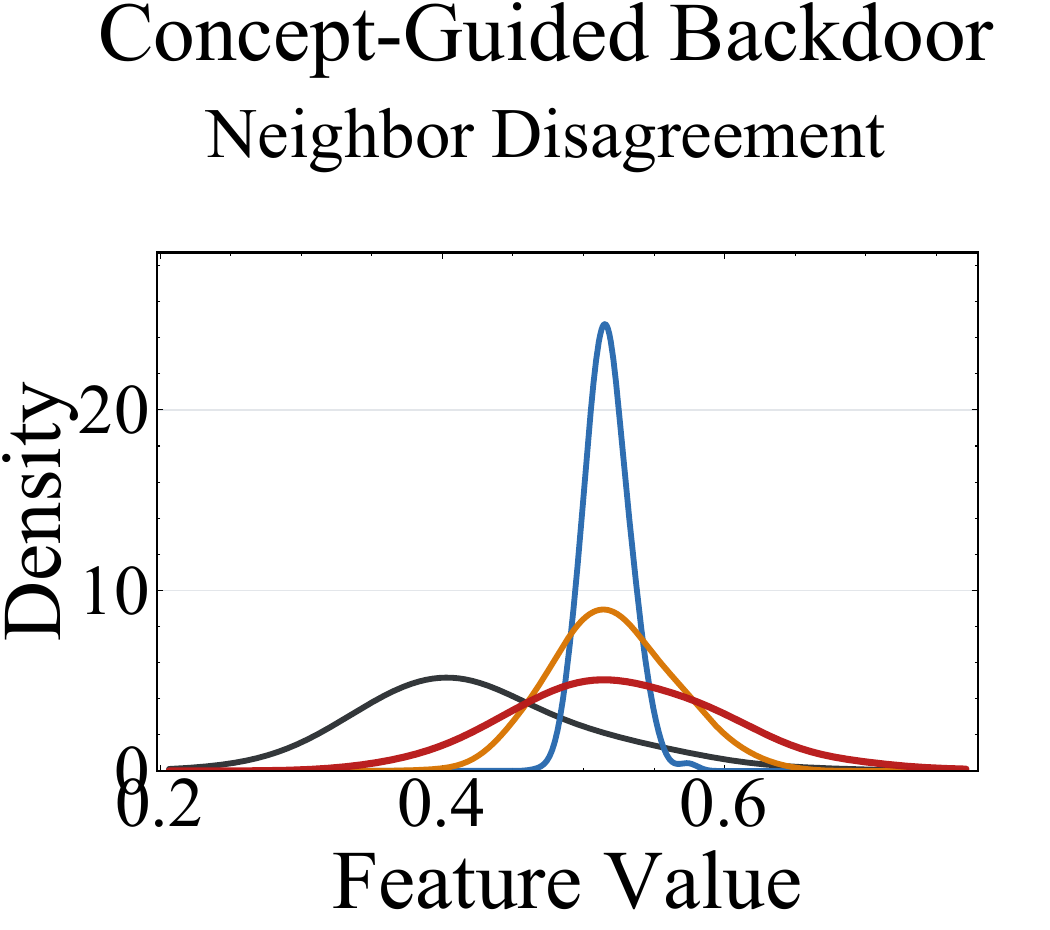} \\
\noalign{\vskip 4pt}
\multicolumn{5}{c}{\small\textbf{Strongest Remaining Pathway Feature}} \\
\includegraphics[width=0.188\linewidth]{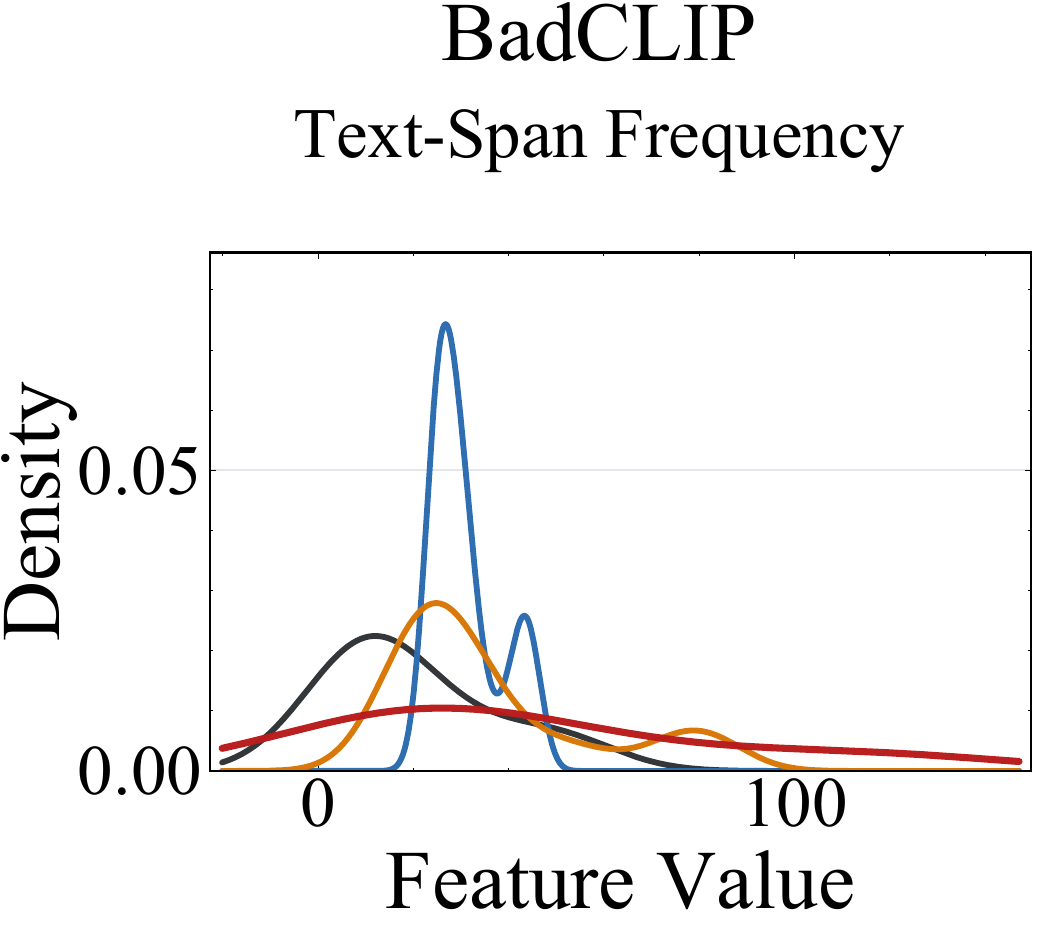} & \includegraphics[width=0.188\linewidth]{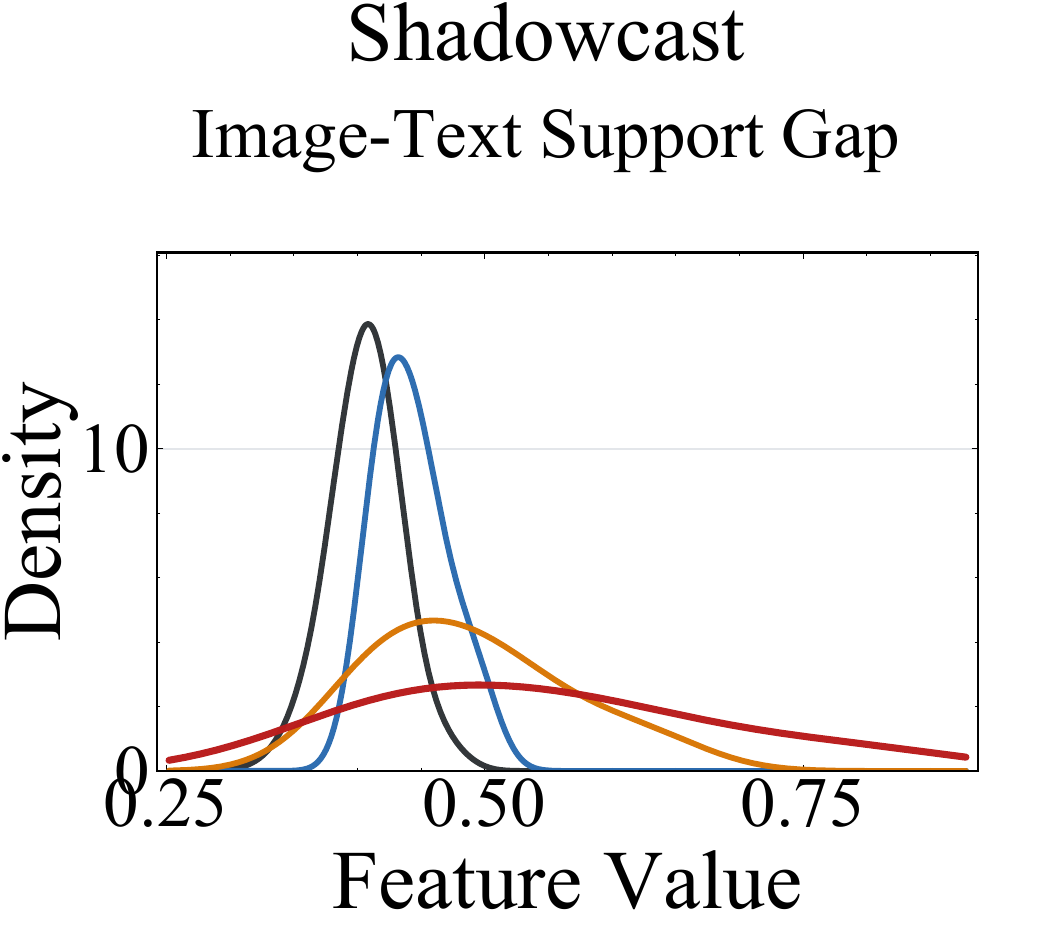} & \includegraphics[width=0.188\linewidth]{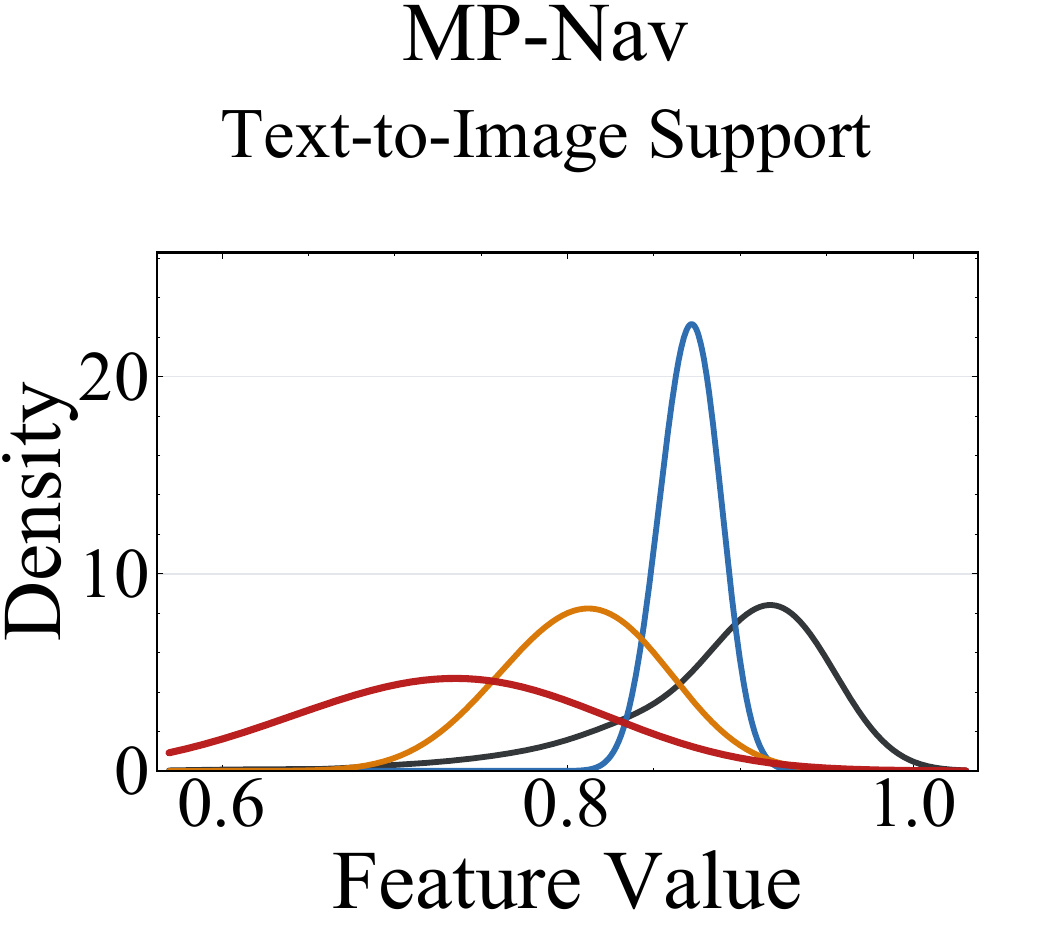} & \includegraphics[width=0.188\linewidth]{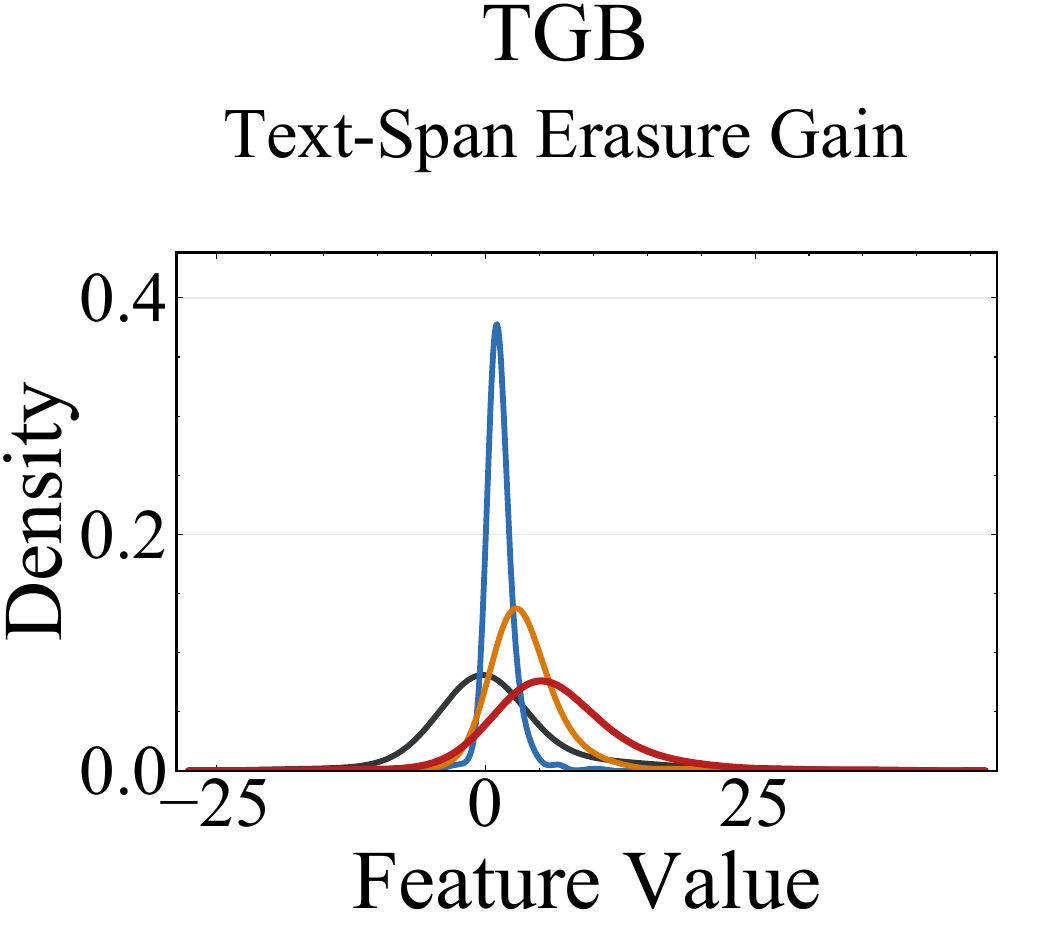} & \includegraphics[width=0.188\linewidth]{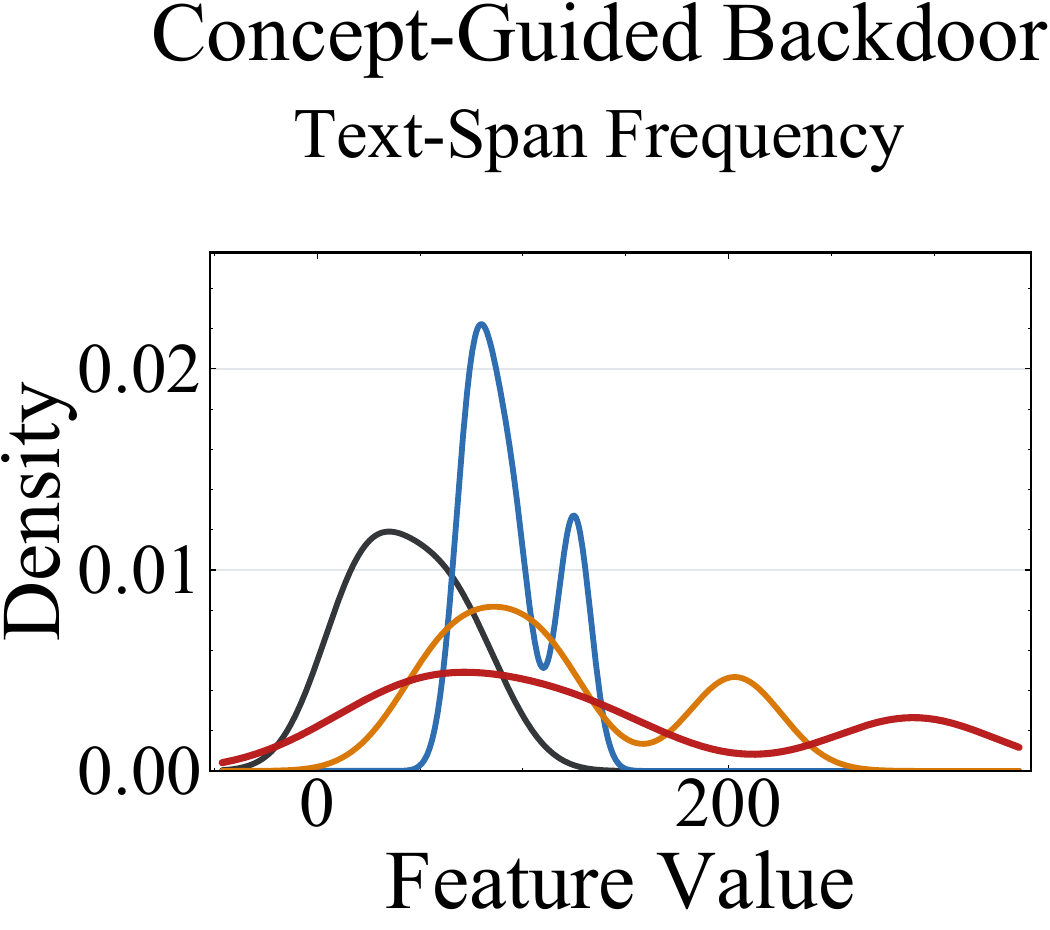} \\
\end{tabular}
\caption{Feature distributions for clean samples and weak, intermediate, and successful poisons. The first two rows show the selected non-pathway features, the third shows Neighbor Disagreement across all five attacks, and the fourth shows the strongest remaining pathway feature for each attack.}
\label{fig:eo01_feature_distributions}
\end{figure}

Figure~\ref{fig:eo01_feature_distributions} shows the full feature distributions for clean samples and weak, intermediate, and successful poisons, while the lower block of Table~\ref{tab:eo01_feature_metric_summary} quantifies the final separation between clean samples and successful poisons.
After feature orientation, a rightward ordering from weak to intermediate and then to successful poisons indicates increasing separation with attack strength, whereas overlapping curves indicate little separation.

For most attacks, image-text similarity and modality-specific density remain close to clean values because the poisoned pairs are designed to appear plausible \citep{liang2024badclip,sharma2018conceptual,xu2024shadowcast,zhang2026textguided,shen2025conceptguided}.
The CLIP score distribution remains close to the clean distribution for four attacks, with IQR-normalized median shifts between -0.02 and +0.13 and TPR at 5\% FPR no larger than 0.13.
Shadowcast has a larger median shift of +0.55, but its TPR remains 0.02, meaning that the upper-tail threshold identifies few successful poisons.
Unimodal kNN density provides similarly little separation across the attack suite, with TPR at 5\% FPR at most 0.07.
These attacks can therefore produce the target behavior without a consistent decrease in image-text similarity or shift in modality-specific density.

Some non-pathway features show a distributional difference, but that difference does not necessarily identify poisoned samples.
For Shadowcast and MP-Nav, Text kNN density has a large IQR-normalized Wasserstein distance but a negative median shift, so the poisoned samples move opposite to the oriented direction.
MP-Nav is the main exception because Image kNN density reaches an IQR-normalized median shift of +0.98 and a TPR of 0.75.
This result is consistent with its instance-level selection, which concentrates poisoned samples in a selected visual neighborhood \citep{zhang2025mpnav}.
An individual non-pathway feature can therefore respond to a particular attack mechanism without consistently tracking the collective contribution represented by \(I(z)\).

We therefore turn to the cross-sample relationships measured by pathway features.
Across the five attacks, successful poisons separate from clean samples under more than one pathway feature, so the result does not depend on a single measurement.
The poison distributions also move farther from the clean distribution as attack strength increases from weak to intermediate and successful.
Because poison count and attack construction remain fixed across these levels, this progression shows that the cross-sample separation grows with attack effectiveness rather than with the size of the poison set.

Neighbor Disagreement provides the clearest shared result, with a positive IQR-normalized median shift for every attack that ranges from +0.71 for Text-Guided Backdoor to +2.93 for Shadowcast.
Its separation is strongest for Shadowcast and MP-Nav, where TPR at 5\% FPR reaches 0.90 and 0.75.
Both attacks connect a source concept to a target concept while keeping each pair locally plausible \citep{xu2024shadowcast,zhang2025mpnav}.
The remaining features show how the cross-sample relationships associated with \(I(z)\) differ across attack mechanisms.
Text-Span Frequency captures the repeated target text in BadCLIP and Concept-Guided Backdoor, while Text-Span Erasure Gain reflects the effect of trigger text on cross-modal support in Text-Guided Backdoor \citep{liang2024badclip,shen2025conceptguided,zhang2026textguided}.
Image-Text Support Gap captures the source-to-target relation in Shadowcast, whereas Text-to-Image Support reflects the selected visual neighborhood in MP-Nav \citep{xu2024shadowcast,zhang2025mpnav}.

The comparison shows that stealth and attack success leave evidence at different levels.
An individual poisoned pair can resemble clean data, while a successful attack requires multiple pairs to concentrate their influence on the same target behavior.
The pathway feature that exposes this influence varies across attacks, and isolated clean examples can also take unusual feature values.
A defense must therefore identify which pathway feature best exposes the attack and determine where its recurring cross-sample structure gives way to isolated clean variation.
This observation motivates TraceGuard to infer both the ranking and how much of each corpus to remove rather than apply one feature threshold or a fixed removal fraction.

\FloatBarrier

\section{Detailed Defense Comparisons}
\label{app:comparison}

The evaluation first measures which examples are removed and then examines what the filtered corpus causes the victim model to learn.
These comparisons address the defender's two goals at different stages, measuring poison detection and clean-data retention before training, followed by attack suppression and clean-task performance afterward.

\subsection{Poison Detection}
For a data-curation defense, the detection outcome is the set of examples removed before training.
We compare these sets on all 19 poisoned corpora, restricting this comparison to data-curation defenses because training-time and post-training defenses do not necessarily identify examples to remove.
Poison Recall is considered together with Clean FPR to distinguish accurate detection from simply discarding more of the corpus.
Table~\ref{tab:filtering_quality_main} reports the resulting Poison Recall and Clean FPR for every attack.
Each attack receives equal weight in the reported means so that larger corpora do not dominate the comparison.

TraceGuard attains a mean Poison Recall of \(98.4\%\) with a mean Clean FPR of \(5.4\%\), and it records the highest or tied-highest recall on 17 of the 19 configurations.
Its recall never falls below \(90.0\%\), whereas the mean recall is \(46.9\%\) for VDC, \(36.5\%\) for CLIPScore Filtering, and \(7.4\%\) for Detect-CLIP-Backdoor-Samples.
VDC has the highest mean recall among these baselines, but its mean Clean FPR reaches \(23.0\%\), while the two largest Clean FPRs for TraceGuard are \(16.2\%\) on BadMLLM and \(10.0\%\) on PoisonedEncoder.
Across these 19 attack configurations, TraceGuard maintains high recall as the attack construction and victim training regime change, while the relative performance of the other filters varies substantially.
These results support using the corpus relationships motivated in Section~\ref{sec:pathway-features} to detect poisons that the other filters miss.

\subsection{End-to-End Defense}
\begin{figure}[!t]
\centering\ArtifactFont
\includegraphics[width=0.84\linewidth]{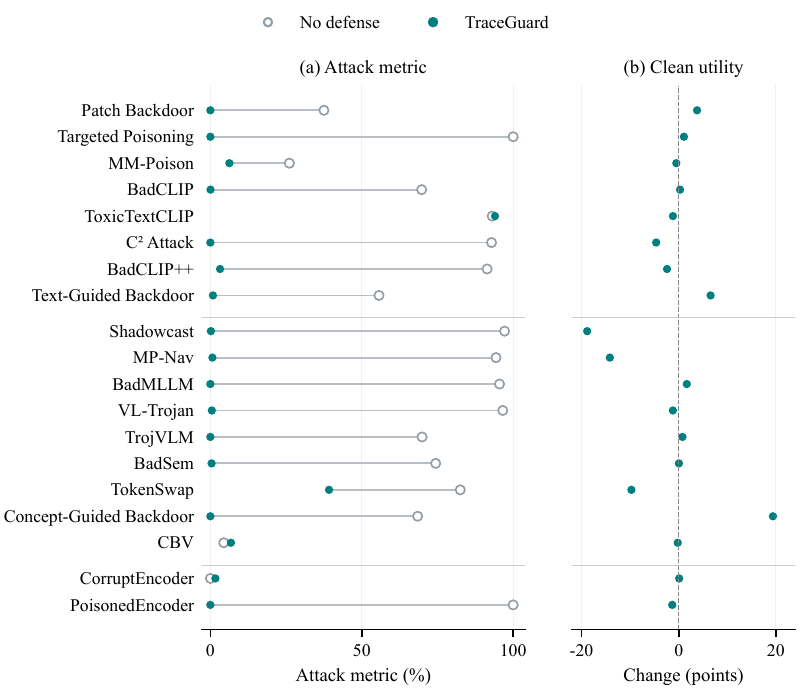}
\caption{TraceGuard outcomes across all attacks. Panel (a) compares the original attack metric before and after filtering. MM-Poison uses target-class Hit@5, while the remaining attacks use their task-specific attack success rates. Panel (b) reports the change in the corresponding clean-task metric, where zero denotes no change.}
\label{fig:traceguard_end_to_end}
\end{figure}

End-to-end defense is assessed through the behavior of the model after the defense has been applied, rather than through removal decisions alone.
The remaining poisons may still sustain an attack, while the removal of clean examples may reduce normal task performance.
We therefore train each victim model on the corpus filtered by TraceGuard and compare its attack and clean-task metrics with those of the matched undefended model.
Figure~\ref{fig:traceguard_end_to_end} summarizes these before-and-after outcomes across all attacks.

After filtering, the attack metric is at most \(1\%\) on 13 of the 19 configurations and at most \(10\%\) on 17.
The two attacks that remain above \(10\%\) are TokenSwap, with a residual attack success rate of \(39.2\%\), and ToxicTextCLIP, which changes from \(93.1\%\) to \(94.0\%\).
CBV also increases from \(4.4\%\) to \(6.8\%\), although its undefended attack success is already low.
Clean-task performance follows a separate pattern, with decreases of \(18.8\), \(14.2\), and \(9.7\) points on Shadowcast, MP-Nav, and TokenSwap.

The low residual attack metrics in most settings connect the corpus relationships used by TraceGuard to the behavior learned from the retained data.
Removing examples linked by these relationships can suppress the attacker-chosen behavior, but similar Poison Recall can still lead to different residual attack metrics after training.
Clean-task performance must likewise be considered separately, as a large attack reduction does not determine how the filtered corpus affects normal learning.

\paragraph{Image-text pre-training and retrieval.}
\begin{table}[!htbp]
\centering\ArtifactFont\fontsize{8}{10}\selectfont
\setlength{\tabcolsep}{1.5pt}
\renewcommand{\arraystretch}{1.08}
\caption{Matched end-to-end effects for image-text pre-training, retrieval, and encoder-transfer attacks. $\Delta A$ denotes attack-metric reduction and $\Delta U$ clean-utility change relative to the matched undefended run, both in percentage points. The encoder-transfer block reports $\Delta A$ / $\Delta U$ in each cell. Larger values are better; dashes denote unreported combinations.}
\label{tab:defense_comparison_image_text}
\label{tab:defense_comparison_encoder}
\begin{tabular*}{\linewidth}{@{\extracolsep{\fill}}ll*{10}{r}@{}}
\toprule
Attack &  & \shortstack{CLIP\\Score} & \shortstack{VDC} & \shortstack{Detect-\\CLIP} & \shortstack{RoCLIP} & \shortstack{Safe\\CLIP} & \shortstack{Clean\\CLIP} & \shortstack{Semantic\\Shield} & \shortstack{Cleaner\\CLIP} & \shortstack{BDet\\CLIP} & \shortstack{\textbf{Trace}\\\textbf{Guard}} \\
\midrule
\multicolumn{12}{l}{\textit{Image-text pre-training and retrieval}} \\
\multirow{2}{*}{Patch} & $\Delta A$ & +37.50 & +87.50 & +62.50 & +15.38 & -75.00 & 0.00 & +7.69 & +37.50 & 0.00 & +37.50 \\
 & $\Delta U$ & +3.78 & +3.78 & -2.16 & -14.89 & -51.35 & 0.00 & +3.55 & -0.54 & -5.41 & +3.78 \\
\addlinespace[2pt]
\multirow{2}{*}{Targeted} & $\Delta A$ & +100.00 & 0.00 & 0.00 & 0.00 & +12.50 & 0.00 & 0.00 & 0.00 & 0.00 & +100.00 \\
 & $\Delta U$ & +3.24 & +2.70 & 0.00 & -12.97 & -76.76 & +0.54 & +2.16 & -1.62 & -4.86 & +1.08 \\
\addlinespace[2pt]
\multirow{2}{*}{MM-Poison} & $\Delta A$ & +18.67 & +19.62 & -3.43 & +20.76 & +19.81 & +15.24 & -24.57 & +19.05 & -- & +19.81 \\
 & $\Delta U$ & -0.61 & -0.79 & -0.53 & -58.99 & -74.50 & -8.87 & +0.30 & -8.96 & -- & -0.49 \\
\addlinespace[2pt]
\multirow{2}{*}{BadCLIP} & $\Delta A$ & +69.54 & +62.92 & +13.04 & 0.00 & +0.06 & +10.26 & +19.00 & +1.20 & +29.62 & +69.72 \\
 & $\Delta U$ & +0.18 & +0.20 & +0.24 & +25.46 & -2.56 & -12.08 & -23.54 & -8.88 & -2.92 & +0.28 \\
\addlinespace[2pt]
\multirow{2}{*}{ToxicTextCLIP} & $\Delta A$ & +46.13 & -0.07 & 0.00 & +92.92 & +4.57 & +0.22 & -1.36 & -0.37 & -- & -0.92 \\
 & $\Delta U$ & +0.81 & -7.77 & 0.00 & -17.80 & -55.43 & +11.42 & -6.77 & +6.23 & -- & -1.19 \\
\addlinespace[2pt]
\multirow{2}{*}{C$^2$ Attack} & $\Delta A$ & +2.04 & -5.10 & +6.12 & -- & -- & +91.84 & -- & -- & +0.00004 & +92.86 \\
 & $\Delta U$ & +0.38 & -3.73 & +1.35 & -- & -- & +3.50 & -- & -- & -5.20 & -4.64 \\
\addlinespace[2pt]
\multirow{2}{*}{BadCLIP++} & $\Delta A$ & +4.60 & +24.00 & +2.00 & +85.40 & +64.40 & +4.00 & +27.00 & +2.40 & +83.26 & +88.20 \\
 & $\Delta U$ & -3.00 & -9.40 & +1.80 & -5.20 & -25.20 & -1.40 & -23.60 & -15.60 & -2.44 & -2.40 \\
\addlinespace[2pt]
\multirow{2}{*}{TGB} & $\Delta A$ & +47.67 & +2.61 & -19.97 & -- & -- & +4.14 & -- & -0.67 & -- & +54.84 \\
 & $\Delta U$ & -14.05 & -6.12 & +7.81 & -- & -- & +9.33 & -- & +9.61 & -- & +6.57 \\
\bottomrule
\end{tabular*}
\par\vspace{5pt}
\begin{tabular*}{\linewidth}{@{\extracolsep{\fill}}lrrrr@{}}
\toprule
Encoder transfer & CLIPScore & VDC & Detect-CLIP & \textbf{TraceGuard} \\
\midrule
CorruptEncoder & +0.02 / -0.70 & -1.37 / -1.80 & -0.06 / -0.64 & -1.64 / +0.10 \\
PoisonedEncoder & +100.00 / -0.81 & 0.00 / -6.61 & 0.00 / 0.00 & +100.00 / -1.32 \\
\bottomrule
\end{tabular*}
\end{table}

Model behavior provides a common basis for comparing defenses that filter data, modify training, or repair a trained model.
The image-text block of Table~\ref{tab:defense_comparison_image_text} therefore reports attack reduction and changes in clean-task performance for eight image-text pre-training and retrieval attacks.
Each defense is measured relative to its own matched undefended model to distinguish the effect of the intervention from differences in the starting attack success and clean-task performance.

Among the applicable methods, TraceGuard obtains the largest or tied-largest attack reduction on five of the eight attacks.
It completely suppresses Targeted Poisoning and reduces C$^2$ Attack and BadCLIP++ by \(92.86\) and \(88.20\) points, respectively.
On MM-Poison, its \(19.81\)-point reduction is close to RoCLIP's \(20.76\) points, while the changes in clean-task performance are \(-0.49\) and \(-58.99\), respectively.
ToxicTextCLIP is the clear exception because TraceGuard changes its attack metric by \(-0.92\) points.
The leading baseline changes from one attack to another, showing that performance does not transfer uniformly across poison constructions or intervention stages even within image-text training.
TraceGuard remains effective on most of these attacks, which is consistent with the pathway view that a defense should follow relationships required by the target behavior rather than a trigger-specific anomaly.

\paragraph{VLM fine-tuning.}
VLM fine-tuning changes the learning objective from matching images and texts to generating responses, so it tests whether the same pathway features remain useful when the target behavior is expressed through output tokens.
We evaluate nine such attacks using the same comparison with matched undefended models.
Table~\ref{tab:defense_comparison_vlm} reports the results.
\begin{table}[!t]
\centering\ArtifactFont\fontsize{9}{11}\selectfont
\setlength{\tabcolsep}{2pt}
\renewcommand{\arraystretch}{1.08}
\caption{Matched end-to-end effects for VLM fine-tuning attacks. Each cell reports attack-metric reduction / clean-utility change relative to the matched undefended run, in percentage points. Larger values are better.}
\label{tab:defense_comparison_vlm}
\begin{tabular*}{\linewidth}{@{\extracolsep{\fill}}l*{5}{r}@{}}
\toprule
Attack & Detect-CLIP & RobustIT & VDC + IT & CLIPScore + IT & \textbf{TraceGuard} \\
\midrule
Shadowcast & -0.50 / +0.83 & +89.50 / -13.67 & +0.17 / +1.67 & +0.17 / +0.50 & +97.00 / -18.83 \\
MP-Nav & +1.50 / -1.17 & +94.33 / -36.83 & +0.83 / -1.00 & +12.67 / -0.67 & +93.67 / -14.17 \\
BadMLLM & -3.73 / -5.00 & +5.60 / -0.83 & -3.67 / +1.67 & -0.67 / 0.00 & +95.49 / +1.67 \\
VL-Trojan & +20.94 / +0.05 & +96.56 / -92.40 & +95.98 / -0.51 & +96.14 / -1.52 & +96.08 / -1.19 \\
TrojVLM & +69.92 / +0.78 & +69.92 / +0.78 & +17.19 / -1.06 & +35.94 / +0.70 & +69.92 / +0.78 \\
BadSem & +2.80 / +0.03 & +62.80 / -0.67 & +19.60 / -0.09 & +8.40 / +0.05 & +74.00 / +0.05 \\
TokenSwap & +10.47 / -17.27 & +40.47 / -12.47 & +12.33 / -2.07 & +17.67 / -28.07 & +43.33 / -9.73 \\
Concept-Guided & -3.62 / -4.20 & +68.45 / -80.60 & -0.64 / -0.40 & -1.29 / +0.40 & +68.45 / +19.40 \\
CBV & -2.00 / 0.00 & +4.40 / +2.40 & -0.80 / +0.20 & 0.00 / -0.20 & -2.40 / -0.20 \\
\bottomrule
\end{tabular*}
\end{table}

TraceGuard obtains the largest or tied-largest attack reduction on six of the nine attacks.
For MP-Nav, its \(93.67\)-point reduction is within \(0.66\) points of RobustIT, while its decrease in clean-task performance is \(14.17\) rather than \(36.83\) points.
For VL-Trojan, the corresponding reductions are \(96.08\) and \(96.56\) points, but the decreases in clean-task performance are \(1.19\) and \(92.40\) points.
TraceGuard also reduces BadMLLM by \(95.49\) points while increasing its clean-task metric by \(1.67\) points.

These results extend the pathway argument beyond the embedding relationships learned by contrastive pre-training.
The same corpus-level measurements remain useful when poisoned examples reinforce attacker-chosen output tokens during generative fine-tuning.
The comparison with RobustIT further shows that data curation can approach the security benefit of a training-time defense with a smaller decrease in clean-task performance in several settings.

\subsection{Encoder Transfer}
The encoder-transfer setting applies attacks originally developed for image-only representation learning to an image-text model.
It tests whether paired text must be part of the original attack for pathway features to be useful.
Following Section~\ref{sec:setup}, we pair the image-only attack data with class-name prompts, making the cross-modal measurements available without changing the poisoned images.
We then apply the same data-curation defenses before multimodal training and report changes in attack success and clean-task performance in the encoder-transfer block of Table~\ref{tab:defense_comparison_encoder}.

For PoisonedEncoder, TraceGuard reduces the attack success rate from \(100.0\%\) to \(0.0\%\) while clean accuracy decreases by \(1.3\) points.
CLIPScore Filtering obtains the same attack reduction with a \(0.8\)-point decrease in clean accuracy.
CorruptEncoder begins with an undefended attack success rate of only \(0.02\%\), so this case serves as a transfer control rather than evidence of additional attack suppression.
After TraceGuard, its attack success rate is \(1.66\%\) and clean accuracy changes by \(+0.1\) points.

The encoder cases test whether the preceding results extend to a different attack entry point.
PoisonedEncoder shows that pathway-based curation can block an active attack that originates in image-side representation learning, while CorruptEncoder checks the same procedure when the transferred attack is inactive.
The pathway measurements therefore remain applicable even when paired text is introduced after the original poisoning process rather than serving as the attack channel itself.

Across the evaluated configurations in these three training regimes, the same pathway-based curation procedure recovers poisoned examples and usually reduces the learned target behavior.
The results support the use of several complementary corpus relationships across attack constructions, while the stress tests below examine how detection changes outside these configurations.

\FloatBarrier

\section{Additional Method Analysis}
\label{app:analysis}

The main comparison evaluates TraceGuard as a complete defense, but does not separate the effects of its removal decisions from those of reducing the training corpus.
We use controlled removal experiments and ablations to examine this distinction, investigate the remaining failures, and assess the contributions of the pathway features and selection procedure.

\subsection{Training Influence}
\begin{FollowupTable*}
\centering\ArtifactFont\fontsize{8.5}{10.5}\selectfont
\setlength{\tabcolsep}{2.5pt}
\renewcommand{\arraystretch}{1.06}
\caption{Group-removal controls for TraceGuard. ASR and clean accuracy (CA) are percentages, reported as mean $\pm$ standard deviation over three training seeds. Non-pathway and random removal match TraceGuard's removal count. The oracle removes all poisons.}
\label{tab:fu-influence_removal}
\begin{tabular*}{\linewidth}{@{\extracolsep{\fill}}>{\raggedright\arraybackslash}p{0.21\linewidth}rrrrrr@{}}
\toprule
 & \multicolumn{2}{c}{TGB} & \multicolumn{2}{c}{BadCLIP} & \multicolumn{2}{c}{BadCLIP++} \\
\midrule
Removal rule & ASR & CA & ASR & CA & ASR & CA \\
\midrule
TraceGuard & 1.61\,$\pm$\,0.32 & 50.32\,$\pm$\,0.49 & 0.06\,$\pm$\,0.00 & 58.31\,$\pm$\,0.10 & 4.20\,$\pm$\,1.91 & 94.53\,$\pm$\,2.19 \\
Non-pathway removal & 26.76\,$\pm$\,10.31 & 37.16\,$\pm$\,6.11 & 61.95\,$\pm$\,3.44 & 57.95\,$\pm$\,0.41 & 87.87\,$\pm$\,3.01 & 98.47\,$\pm$\,0.61 \\
All-poison oracle & 1.27\,$\pm$\,0.05 & 51.58\,$\pm$\,1.45 & 0.06\,$\pm$\,0.00 & 57.99\,$\pm$\,0.32 & 5.60\,$\pm$\,1.11 & 94.07\,$\pm$\,0.70 \\
Random removal & 28.36\,$\pm$\,22.41 & 39.82\,$\pm$\,3.62 & 57.51\,$\pm$\,2.60 & 58.15\,$\pm$\,0.38 & 88.80\,$\pm$\,1.80 & 95.47\,$\pm$\,5.08 \\
No removal & 34.37\,$\pm$\,12.52 & 39.46\,$\pm$\,5.77 & 67.19\,$\pm$\,3.51 & 58.00\,$\pm$\,0.12 & 88.67\,$\pm$\,3.16 & 98.13\,$\pm$\,0.31 \\
\bottomrule
\end{tabular*}
\end{FollowupTable*}

Training influence concerns how poisoned examples change the target behavior learned by the victim model.
We examine this effect by training on corpora from which different sets of examples have been removed and comparing the resulting model behavior.
The removal rules are compared on TGB, BadCLIP, and BadCLIP++ to distinguish the effect of selecting particular examples from that of discarding the same amount of data.
These attacks cover composed retrieval and image-text classification, with trained image-text representations available for the gradient analysis below.
Random removal provides an equal-count comparison without feature-based selection.
Non-pathway removal instead selects the same number of examples using the mean percentile of the four non-pathway features from Section~\ref{sec:empirical-observations}, testing selection based on similarity and density at that count.
The unfiltered corpus establishes the attack behavior before removal.
We also remove every poisoned example while retaining all clean examples, a control termed the all-poison oracle that shows model behavior when no poisons remain.
For each attack, we train the victim model on the retained data under the same training configuration and matched random seeds, reporting ASR and clean-task performance over three seeds in Table~\ref{tab:fu-influence_removal}.

TraceGuard reduces mean ASR to \(1.61\%\), \(0.06\%\), and \(4.20\%\) on TGB, BadCLIP, and BadCLIP++, respectively, close to the corresponding all-poison oracle results of \(1.27\%\), \(0.06\%\), and \(5.60\%\).
Neither random removal nor non-pathway selection achieves comparable suppression despite removing the same number of examples.
For BadCLIP++, their ASRs remain above \(87\%\), compared with \(88.67\%\) without removal.
The difference therefore comes from which examples are excluded, rather than a general reduction in the amount of training data.
The clean-task results also depend on the selected examples.
On TGB, TraceGuard raises clean-task performance from \(39.46\%\) to \(50.32\%\), whereas random removal leaves it at \(39.82\%\).
On BadCLIP++, performance decreases from \(98.13\%\) to \(94.53\%\), close to the oracle's \(94.07\%\).
Thus, the selected set approaches the attack suppression obtained by removing all poisons in these comparisons, while its effect on normal learning remains task-dependent.

\begin{figure}[!htbp]
\centering\ArtifactFont
\includegraphics[width=0.76\linewidth]{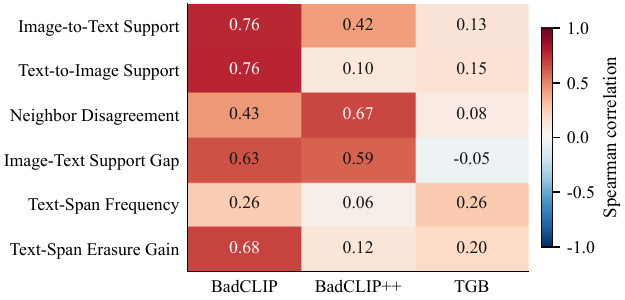}
\caption{Signed Spearman correlations between pathway features and gradient influence estimates.}
\label{fig:fu-influence_correlations}
\end{figure}

The removal comparison measures the effect of a selected set as a whole, whereas the gradient analysis estimates whether an individual example's training update reinforces the pairing shared by the poisons.
This local estimate gives a way to compare pathway features with the first-order influence described in Appendix~\ref{app:pathway-features}.
For each attack, we sample up to 512 clean and 512 poisoned examples to include both groups while limiting the cost of per-example gradient computation.
The reference objective measures the strength of the shared pairing through cosine similarity between the mean image and text representations of the sampled poisons.
We compute its gradient at the text projection layer, which maps text representations into the shared image-text space and instantiates the linear layer in Eq.~\eqref{eq:gradient-factorization}.
Following Eq.~\eqref{eq:gradient-bound}, the inner product between this objective gradient and an example's gradient descent direction for image-text alignment estimates how much that example's update would reinforce the pairing.
We compare the resulting influence estimates with the six feature percentiles using signed Spearman correlation, which measures agreement in how they rank examples without requiring a shared numerical scale (Figure~\ref{fig:fu-influence_correlations}).
Victim-model gradients and poison labels are used to construct this experimental reference, not to make TraceGuard's removal decisions.

The correlations vary with both the feature and the attack.
For BadCLIP, the two directional Neighbor Support features each reach \(0.76\), followed by Text-Span Erasure Gain at \(0.68\).
For BadCLIP++, Neighbor Disagreement has the largest correlation at \(0.67\), while Text-Span Frequency and Text-Span Erasure Gain fall to \(0.06\) and \(0.12\).
These differences connect the complementary relationships in Section~\ref{sec:pathway-features} to different aspects of the estimated training influence, rather than identifying one feature that tracks every attack equally well.
TGB provides a useful contrast with Table~\ref{tab:fu-influence_removal}.
Its correlations range from \(-0.05\) to \(0.26\), yet TraceGuard's selected set nearly matches the oracle's attack suppression.
The local gradient reference therefore does not fully explain the effectiveness of the removal set.
Read together, the two experiments support using pathway features to identify examples whose joint removal disrupts the attack, without requiring each feature to reproduce an individual influence ranking.

\subsection{Failure Analysis}
Residual ASR and reduced clean-task performance describe different failure outcomes, and either can depend on which poisoned and clean examples remain in the corpus.
ToxicTextCLIP and TokenSwap retain substantial ASR in the main comparison, while Shadowcast and MP-Nav show decreases in clean-task performance.
To distinguish these effects, we vary which poisoned and clean examples are removed and compare the resulting model behavior (Table~\ref{tab:fu-failure_controls}).
One control removes only the poisons detected by TraceGuard while retaining every clean example, so its comparison with TraceGuard measures the additional effect of removing clean examples.
An all-poison oracle also removes the missed poisons, showing how the model behaves when the remaining corpus is entirely clean.
We additionally remove randomly selected clean examples in the same number as TraceGuard to distinguish the effect of reducing the clean corpus from that of eliminating poisoned examples.
Each retained corpus is used to train the same victim model under a matched training configuration and seed, with the unfiltered corpus and TraceGuard as references.
Poison labels determine the diagnostic controls but do not enter TraceGuard's decisions.
\begin{table}[!htbp]
\centering\ArtifactFont\small
\setlength{\tabcolsep}{3pt}
\renewcommand{\arraystretch}{1.13}
\caption{Removal controls for TraceGuard's failure cases. ASR and clean zero-shot accuracy (CA) are percentages from one training seed. Random clean removal matches TraceGuard's removal count. The two poison-only controls use known poison labels.}
\label{tab:fu-failure_controls}
\begin{tabular*}{\linewidth}{@{\extracolsep{\fill}}>{\raggedright\arraybackslash}p{0.28\linewidth}rrrrrrrr@{}}
\toprule
 & \multicolumn{2}{c}{Shadowcast} & \multicolumn{2}{c}{ToxicTextCLIP} & \multicolumn{2}{c}{TokenSwap} & \multicolumn{2}{c}{MP-Nav} \\
\cmidrule(lr){2-9}
Removed examples & ASR & CA & ASR & CA & ASR & CA & ASR & CA \\
\midrule
No removal & 97.17 & 90.17 & 81.16 & 83.00 & 76.80 & 55.00 & 94.33 & 88.33 \\
TraceGuard & 0.17 & 72.00 & 91.72 & 81.55 & 6.27 & 93.93 & 0.83 & 74.50 \\
All poisons & 0.17 & 69.50 & 1.55 & 84.67 & 1.13 & 98.93 & 0.17 & 70.17 \\
Random clean examples & 98.17 & 93.00 & 94.42 & 84.73 & 73.80 & 28.20 & 76.00 & 82.33 \\
Detected poisons only & 0.17 & 69.67 & 78.98 & 84.95 & 6.53 & 93.87 & 0.17 & 69.67 \\
\bottomrule
\end{tabular*}
\end{table}

The poison-only controls separate two reasons why low ASR and high clean-task performance need not occur together.
For Shadowcast and MP-Nav, removing only the detected poisons reduces ASR to \(0.17\%\), but clean-task performance is \(69.67\%\) on both attacks even though every clean example is retained.
Removing all poisons produces similar clean-task results of \(69.50\%\) and \(70.17\%\).
The decreases observed with TraceGuard therefore cannot be attributed solely to false-positive removal.
In these settings, removing poisoned examples itself changes normal learning, so Clean FPR alone does not explain the clean-task outcome.

ToxicTextCLIP instead exposes the effect of the poisons left behind.
Removing only detected poisons leaves ASR at \(78.98\%\), whereas removing all poisons lowers it to \(1.55\%\).
The remaining poisons can thus sustain the attack even after most poisons have been detected.
Removing clean examples also matters here, as TraceGuard's ASR is \(91.72\%\), higher than that of the detected-poison-only control.
This case shows why Poison Recall must be considered together with residual ASR.
Counting the poisons removed does not establish that the missed poisons can no longer sustain the attack.

In the controlled TokenSwap comparison, TraceGuard reaches \(6.27\%\) ASR and \(93.93\%\) clean-task performance, close to removing only detected poisons at \(6.53\%\) and \(93.87\%\).
Random clean removal leaves ASR at \(73.80\%\) and reduces clean-task performance to \(28.20\%\).
Together with the equal-count comparisons in Table~\ref{tab:fu-influence_removal}, this contrast reinforces the importance of identifying the examples responsible for the attack.
The oracle result on ToxicTextCLIP also shows that the attack can be suppressed by removing the missed poisons as well.
\subsection{Ablation Study}\label{sec:ablation}
TraceGuard relies on pathway features to identify suspicious relationships and on adaptive rank-based filtering to decide which examples to remove and how much data to retain.
Our ablations examine these contributions separately by omitting groups of features or simplifying the selection procedure.
For the feature comparisons, each group represents one of the three complementary relationships introduced in Section~\ref{sec:pathway-features}.
We exclude both directional Neighbor Support features, Neighbor Disagreement together with Image-Text Support Gap, or Text-Span Frequency together with Text-Span Erasure Gain, and repeat the initial suspicious-set selection for each variant.
The initial set obtained with all six features provides the reference at the same selection stage, allowing us to examine what is lost when a relationship is omitted.

\begin{figure}[!htbp]
\centering\ArtifactFont
\includegraphics[width=\linewidth]{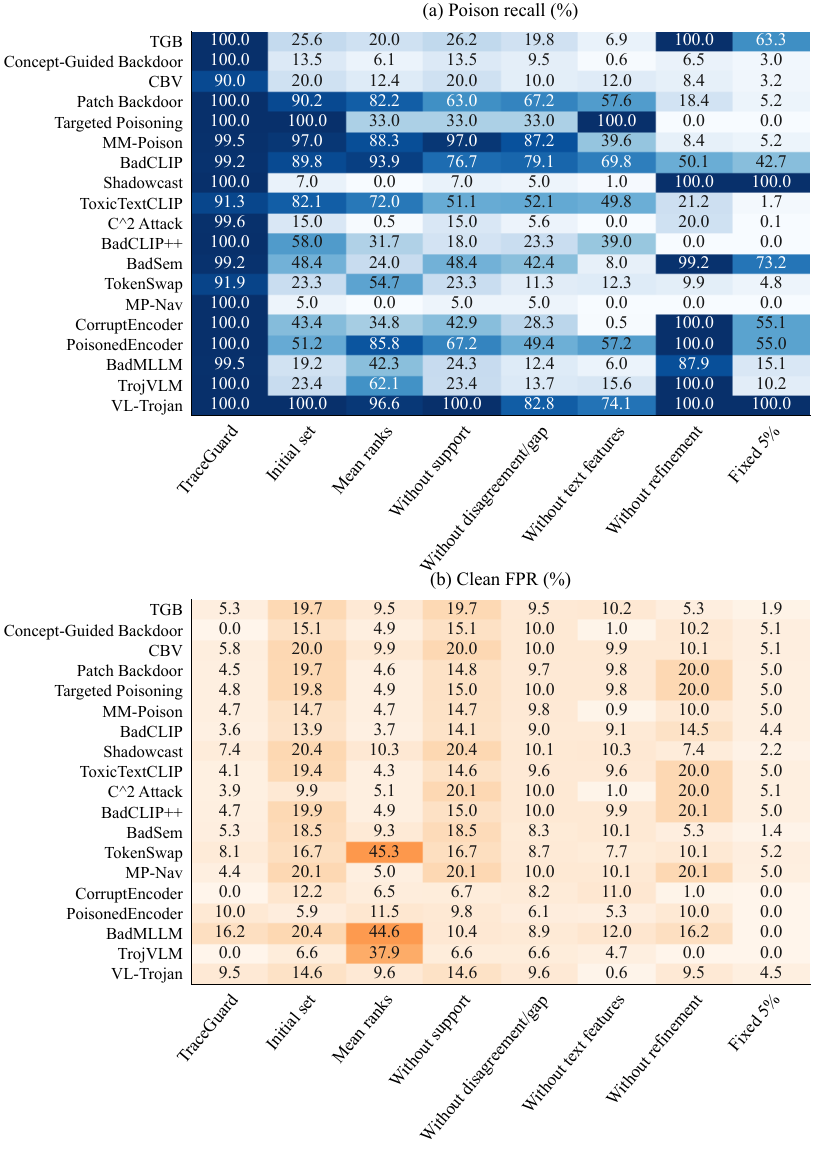}
\caption{Detection ablations on all 19 attack corpora. The two panels report poison recall and clean FPR for TraceGuard and changes to its feature families, ranking aggregation, initial-set selection, and feature-drop refinement. ``Without disagreement/gap'' removes Neighbor Disagreement and Image-Text Support Gap.}
\label{fig:fu-filtering_ablation}
\end{figure}

The initial suspicious set contains the examples selected from the base ranking before refinement or final acceptance.
Using this set directly for removal therefore tests the contribution of the subsequent selection stages.
Another variant omits refinement based on a drop in feature values, testing whether this step helps identify a smaller suspicious set.
Replacing the multiple rankings with the mean of the six feature percentiles tests ranking aggregation at TraceGuard's removal count, while a fixed 5\% removal rule tests a preset fraction in place of corpus-dependent selection.
Figure~\ref{fig:fu-filtering_ablation} reports Poison Recall and Clean FPR on all 19 attack corpora so that changes in detection can be considered alongside changes in clean-data removal.

The feature families contribute differently across attacks even at the initial selection stage.
On TGB, omitting the text-span features reduces Poison Recall from \(25.6\%\) to \(6.9\%\), whereas omitting Neighbor Support leaves it at \(26.2\%\).
For BadCLIP++, the initial set reaches \(58.0\%\) recall, which falls to \(18.0\%\) without Neighbor Support and \(23.3\%\) without Neighbor Disagreement and Image-Text Support Gap.
These contrasts agree with the attack-dependent correlations in Figure~\ref{fig:fu-influence_correlations}.
Different relationships expose different attacks, which motivates retaining the complementary features rather than selecting one relationship for every corpus.

Refinement and final selection improve more than the amount of poisoning detected.
For MP-Nav, moving from the initial set to the final removal set increases Poison Recall from \(5.0\%\) to \(100.0\%\) while reducing Clean FPR from \(20.1\%\) to \(4.4\%\).
Concept-Guided Backdoor shows the same change, from \(13.5\%\) recall and \(15.1\%\) Clean FPR to \(100.0\%\) recall with no clean examples removed.
This simultaneous gain in recall and clean-data retention cannot be obtained simply by increasing the removal fraction along an unchanged ranking, since that would retain the previously selected clean examples.
This improvement supports revising which examples are removed as well as how many, the purpose of shared-pattern refinement and adaptive removal selection in TraceGuard.
The tradeoff is not uniform, as PoisonedEncoder gains recall from \(51.2\%\) to \(100.0\%\) with Clean FPR increasing from \(5.9\%\) to \(10.0\%\).
Corpus-dependent selection improves the separation in many cases without eliminating the cost of recovering additional poisons in every corpus.

The training comparisons examine whether the differences in removal sets also change the target behavior learned by the model.
Table~\ref{tab:fu-ablation_outcomes} covers five attacks spanning image-text classification, composed retrieval, VLM fine-tuning, and encoder transfer.
For each attack, the victim training configuration and seed are fixed across variants.
Alongside the unfiltered reference, we include a matched-size oracle that maximizes the number of poisons removed while retaining the same amount of data as TraceGuard.
It removes poisoned examples first and fills any remaining removal count with clean examples.

Using all six features is not sufficient if their rankings are averaged into one.
At TraceGuard's removal count, the mean-rank variant leaves ASR between \(23.89\%\) and \(100.00\%\) across the five attacks, compared with \(0.00\%\)--\(3.20\%\) for TraceGuard.
Because the removal count is held fixed, this difference supports preserving the multiple rankings when selecting examples, rather than attributing the gain to more aggressive filtering.
A fixed \(5\%\) removal rule also leaves high ASR on every attack, from \(58.86\%\) to \(100.00\%\).
In combination with Figure~\ref{fig:fu-filtering_ablation}, these results show that both the selected examples and the removal fraction matter when the defender does not know the poison fraction.

The contribution of refinement also depends on the corpus.
Without refinement based on a drop in feature values, ASR rises from \(3.20\%\) to \(88.40\%\) on BadCLIP++ and from \(0.67\%\) to \(96.33\%\) on MP-Nav, but remains low on TGB and PoisonedEncoder.
This is consistent with using alternative refinements to capture different forms of shared attack patterns, rather than expecting each refinement to improve every attack.
The complete method yields ASR within \(1.00\) percentage point of the matched-size oracle on all five attacks.
Together with the equal-count removal experiment, this result ties TraceGuard's advantage to selecting examples that sustain the attack.
The gradient analysis and feature ablations explain why multiple corpus relationships are useful for finding those examples, while the set-selection ablations show why these relationships must also guide how much data is removed.

\FloatBarrier

\section{Stress Tests}
\label{app:stress-tests}

\label{sec:stress-tests}
The stress tests examine conditions that differ from evaluating each original attack separately with the default encoders.
They vary the attack's adaptation to the defense, the amount and composition of poisoning, and the encoders used to compute pathway features.

\subsection{Adaptive Attacks}
\label{app:adaptive}
An adaptive attacker knows how TraceGuard detects poisons and modifies the poisoned examples to evade it while retaining the target behavior.
Since TraceGuard uses relationships between image and text neighborhoods together with recurring text spans, we examine adaptations directed at each source of information and at both together.
We construct these variants from BadCLIP++, MP-Nav, and TGB, keeping the clean examples and poison fraction unchanged.
Image adaptation modifies poisoned images to increase cross-modal neighborhood agreement while retaining their alignment with the target text.
The optimization uses the same CLIP ViT-B/32 encoders as the defense and constrains pixel changes to an $\ell_\infty$ radius of $8/255$, with 40 optimization steps and two restarts.
Text adaptation distributes alternative target-side wordings across the poisons to reduce exact repetition, using two, four, or eight alternatives.
Joint adaptation optimizes the images after changing the paired text.

An adaptation may reduce detection by weakening the attack itself, so we measure ASR on models trained with and without filtering the adapted corpus.
Table~\ref{tab:fu-adaptive_attacks} compares the original and adapted attacks over three training seeds, alongside Poison Recall, Clean FPR, and the removal fraction.
\begin{FollowupTable*}
\centering\ArtifactFont\small
\setlength{\tabcolsep}{2.5pt}
\renewcommand{\arraystretch}{1.04}
\caption{Adaptive attacks against TraceGuard. Image adaptation targets neighborhood agreement, text adaptation varies target-side wording, and joint adaptation combines both. $k$ counts text alternatives. ASR and clean zero-shot accuracy (CA) are three-seed means $\pm$ standard deviations. All rates are percentages.}
\label{tab:fu-adaptive_attacks}
\begin{tabular*}{\linewidth}{@{\extracolsep{\fill}}>{\raggedright\arraybackslash}p{0.2\linewidth}rrrrrrr@{}}
\toprule
Attack variant & $k$ & ASR before & ASR after & CA after & Recall & FPR & Removed \\
\midrule
\multicolumn{8}{l}{\textbf{BadCLIP++}} \\
Original & 1 & 87.40\,$\pm$\,2.11 & 4.20\,$\pm$\,1.91 & 94.53\,$\pm$\,2.19 & 100.00 & 4.72 & 5.00 \\
Image adaptation & 1 & 8.80\,$\pm$\,6.41 & 9.40\,$\pm$\,1.59 & 88.87\,$\pm$\,0.64 & 99.67 & 4.72 & 5.00 \\
Text adaptation & 2 & 86.47\,$\pm$\,1.45 & 4.20\,$\pm$\,1.91 & 94.53\,$\pm$\,2.19 & 100.00 & 4.72 & 5.00 \\
Text adaptation & 4 & 87.33\,$\pm$\,2.16 & 9.00\,$\pm$\,0.72 & 89.87\,$\pm$\,0.61 & 100.00 & 4.72 & 5.00 \\
Text adaptation & 8 & 87.20\,$\pm$\,1.06 & 9.00\,$\pm$\,0.72 & 89.87\,$\pm$\,0.61 & 100.00 & 4.72 & 5.00 \\
Joint adaptation & 2 & 3.27\,$\pm$\,1.22 & 10.73\,$\pm$\,1.90 & 88.53\,$\pm$\,2.73 & 99.67 & 4.72 & 5.00 \\
Joint adaptation & 4 & 8.07\,$\pm$\,3.37 & 10.20\,$\pm$\,4.68 & 93.27\,$\pm$\,0.83 & 1.00 & 5.01 & 5.00 \\
Joint adaptation & 8 & 9.27\,$\pm$\,4.13 & 9.60\,$\pm$\,4.41 & 95.13\,$\pm$\,1.94 & 1.00 & 5.01 & 5.00 \\
\addlinespace[3pt]
\multicolumn{8}{l}{\textbf{MP-Nav}} \\
Original & 1 & 94.33\,$\pm$\,0.00 & 0.67\,$\pm$\,0.00 & 73.83\,$\pm$\,0.17 & 100.00 & 4.45 & 5.00 \\
Image adaptation & 1 & 3.89\,$\pm$\,0.25 & 0.83\,$\pm$\,0.00 & 74.89\,$\pm$\,0.38 & 75.00 & 4.59 & 5.00 \\
Text adaptation & 2 & 95.39\,$\pm$\,0.10 & 0.67\,$\pm$\,0.00 & 73.78\,$\pm$\,0.35 & 100.00 & 4.45 & 5.00 \\
Text adaptation & 4 & 95.44\,$\pm$\,0.38 & 0.67\,$\pm$\,0.00 & 73.89\,$\pm$\,0.19 & 100.00 & 4.45 & 5.00 \\
Text adaptation & 8 & 95.06\,$\pm$\,0.25 & 0.67\,$\pm$\,0.00 & 74.11\,$\pm$\,0.25 & 100.00 & 4.45 & 5.00 \\
Joint adaptation & 2 & 3.50\,$\pm$\,0.00 & 0.72\,$\pm$\,0.10 & 75.00\,$\pm$\,0.73 & 80.00 & 4.57 & 5.00 \\
Joint adaptation & 4 & 2.56\,$\pm$\,0.10 & 1.00\,$\pm$\,0.00 & 77.33\,$\pm$\,0.44 & 80.00 & 4.57 & 5.00 \\
Joint adaptation & 8 & 2.39\,$\pm$\,0.10 & 0.94\,$\pm$\,0.10 & 73.56\,$\pm$\,0.10 & 75.00 & 4.59 & 5.00 \\
\addlinespace[3pt]
\multicolumn{8}{l}{\textbf{TGB}} \\
Original & 1 & 26.46\,$\pm$\,3.35 & 1.55\,$\pm$\,0.31 & 51.44\,$\pm$\,0.83 & 100.00 & 5.26 & 10.00 \\
Image adaptation & 1 & 14.84\,$\pm$\,8.37 & 21.09\,$\pm$\,13.05 & 36.13\,$\pm$\,7.28 & 16.89 & 14.90 & 15.00 \\
Text adaptation & 2 & 33.88\,$\pm$\,18.91 & 16.49\,$\pm$\,8.15 & 33.44\,$\pm$\,5.02 & 12.67 & 9.86 & 10.00 \\
Text adaptation & 4 & 17.44\,$\pm$\,4.58 & 14.33\,$\pm$\,6.82 & 35.48\,$\pm$\,7.97 & 23.78 & 4.01 & 5.00 \\
Text adaptation & 8 & 11.54\,$\pm$\,8.69 & 28.80\,$\pm$\,20.30 & 43.80\,$\pm$\,3.67 & 29.33 & 3.72 & 5.00 \\
Joint adaptation & 2 & 19.08\,$\pm$\,12.80 & 11.88\,$\pm$\,7.21 & 33.07\,$\pm$\,5.63 & 12.67 & 9.86 & 10.00 \\
Joint adaptation & 4 & 18.80\,$\pm$\,14.14 & 20.11\,$\pm$\,10.75 & 34.47\,$\pm$\,8.33 & 23.78 & 4.01 & 5.00 \\
Joint adaptation & 8 & 17.28\,$\pm$\,14.13 & 30.26\,$\pm$\,10.82 & 41.17\,$\pm$\,5.96 & 29.33 & 3.72 & 5.00 \\
\addlinespace[3pt]
\bottomrule
\end{tabular*}
\end{FollowupTable*}

Changing target-side wording does not remove the detectable relationships in BadCLIP++ and MP-Nav.
Across two to eight text alternatives, their undefended ASR remains above \(86\%\) and \(95\%\), respectively, yet TraceGuard retains \(100\%\) Poison Recall.
Filtering reduces ASR to \(4.20\%\)--\(9.00\%\) for BadCLIP++ and \(0.67\%\) for MP-Nav without increasing their removal fractions or Clean FPR relative to the original attacks.
These results support using complementary pathway features rather than relying on exact repetition of one target phrase.
Image and joint adaptation produce a different outcome, reducing undefended ASR below \(10\%\) for both attacks.
For example, joint adaptation with four text alternatives lowers BadCLIP++ recall to \(1\%\), but also lowers its undefended ASR from \(87.40\%\) to \(8.07\%\).
This loss of attack effectiveness is consistent with a cost to concealing the measured relationships, although it does not establish that every successful adaptation must incur that cost.

For TGB, adaptation instead reduces detection without sacrificing attack effectiveness.
With two text alternatives, Poison Recall falls from \(100\%\) to \(12.67\%\), while mean undefended ASR is \(33.88\%\), compared with \(26.46\%\) for the original attack.
Residual ASR rises from \(1.55\%\) to \(16.49\%\), and Clean FPR increases from \(5.26\%\) to \(9.86\%\).
Several TGB adaptations also have large variation across training seeds, and filtering sometimes increases rather than decreases mean ASR.
Thus, the resilience to text variation observed for BadCLIP++ and MP-Nav does not extend to every attack, even with the same feature collection and selection rules.
\subsection{Varying Poisoning Rates}
\label{app:rates}
\begin{figure}[!htbp]
\centering\ArtifactFont
\includegraphics[width=0.92\linewidth]{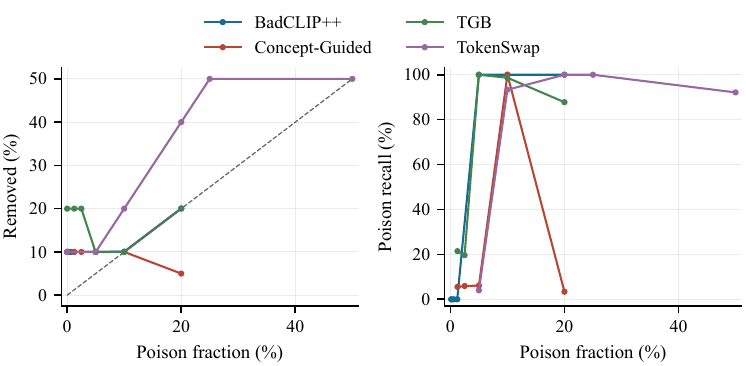}
\caption{TraceGuard's removal fraction and poison recall across all 26 measured poisoning rates. The dashed line denotes equal removal and poison fractions. Recall is undefined at zero poisoning.}
\label{fig:fu-removal_fraction}
\end{figure}

The poisoning rate is the fraction of training examples controlled by the attacker, which TraceGuard does not know when deciding how much data to remove.
We vary this fraction to test whether its adaptive removal selection follows changes in the amount of poisoning rather than favoring a fixed removal rate.
For BadCLIP++, TGB, Concept-Guided Backdoor, and TokenSwap, we resample clean and poisoned examples at a fixed corpus size for each attack and recompute the pathway features on each resulting corpus.
The evaluated fractions include poison-free corpora and extend to \(20\%\) for the first three attacks and \(50\%\) for TokenSwap.
Figure~\ref{fig:fu-removal_fraction} reports the removal fraction together with Poison Recall, since removing more data is useful only if it recovers poisoned examples.

\begin{figure}[!htbp]
\centering\ArtifactFont
\includegraphics[width=\linewidth]{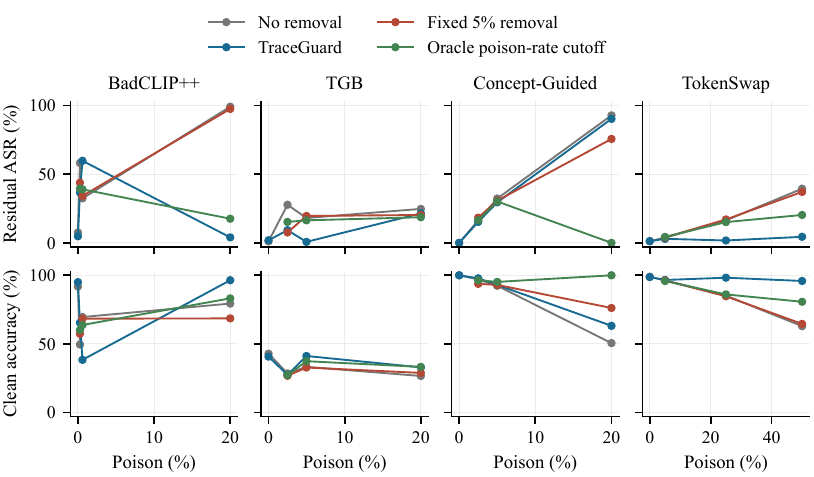}
\caption{ASR and clean zero-shot accuracy as the poisoning rate changes. All 56 trained conditions are shown. Poison-free endpoints compare no removal with TraceGuard. Each condition uses one training seed, and lines connect measured rates.}
\label{fig:fu-poison_rate_outcomes}
\end{figure}

To examine the effect on model behavior, Figure~\ref{fig:fu-poison_rate_outcomes} compares TraceGuard with no removal, fixed \(5\%\) removal, and an oracle poison-rate cutoff, using one training seed for each condition.
The two cutoff rules use the same ranking but differ in how many examples they remove.
The oracle removes a fraction equal to the true poison fraction, allowing us to distinguish knowledge of the poison rate from the ability to select the poisoned examples.

Adaptive removal is useful when the detected poison set exceeds a fixed removal budget.
At \(20\%\) BadCLIP++ poisoning, TraceGuard removes \(20.07\%\) of the corpus and recovers all poisons, reducing ASR from \(99.20\%\) to \(4.00\%\).
Fixed \(5\%\) removal instead leaves ASR at \(97.60\%\) (Figure~\ref{fig:fu-poison_rate_outcomes}).
The benefit also depends on which examples are selected, not just their number.
At \(50\%\) TokenSwap poisoning, TraceGuard and the oracle cutoff both remove half the corpus, but achieve Poison Recall of \(92.13\%\) and \(50.80\%\), respectively.
Their residual ASR is \(4.40\%\) versus \(20.33\%\), showing that knowing the poison fraction cannot replace selecting the relevant examples.

The removal fraction is not uniformly accurate across rates.
Figure~\ref{fig:fu-removal_fraction} shows zero recall at all four measured nonzero BadCLIP++ rates below \(5\%\), despite removing \(10\%\) of each corpus.
Concept-Guided Backdoor recall reaches \(100\%\) at \(10\%\) poisoning but falls to \(3.38\%\) at \(20\%\), where residual ASR remains \(90.35\%\) and the oracle cutoff reduces it to \(0.01\%\).
Conversely, TraceGuard recovers all TokenSwap poisons at a \(25\%\) poison fraction by removing \(50\%\) of the corpus, including one third of the clean examples.
Together, these results support adapting both the number and identity of removed examples, while showing that the current rules can miss small poison sets, underestimate larger ones, or recover them at substantial clean-data cost.

\subsection{Clean-Only Data}
Clean-only corpora contain no injected poisons, so every removed example counts as a false positive.
They test whether benign relationships among examples can trigger removal even when there is no attack to detect.
We apply TraceGuard to clean examples from CC3M and the COCO training and validation sets, as well as a mixed corpus containing equal numbers of examples from these sources and CIRR.
The mixed corpus tests whether combining data sources changes which clean examples are selected for removal.
Table~\ref{tab:fu-clean_only} reports overall Clean FPR and the rate within each source of the mixed corpus.
We also group examples by how many of the six pathway features place them in the highest-ranked \(10\%\), to examine whether false positives are associated with high ranks under several features rather than an isolated feature.

Table~\ref{tab:fu-clean_only} shows that TraceGuard removes \(5.00\%\)--\(10.00\%\) of the single-source clean corpora and \(20.00\%\) of the mixed corpus.
Within the mixture, Clean FPR ranges from \(5.73\%\) for COCO validation to \(41.42\%\) for CC3M, despite equal source sizes.
The errors are therefore concentrated in particular sources rather than distributed evenly across the mixture.
Nor does agreement among several features eliminate false positives.
On clean CC3M, \(43.46\%\) of examples in the upper tails of at least three features are removed, compared with \(5.00\%\) overall, although this trend does not hold on COCO validation.
These results distinguish identifying unusual relationships from establishing that those relationships serve an attack.
Pathway features provide evidence for the former, but the current selection rules do not reliably abstain from removal when the corpus contains no poisoning.
\begin{table}[!htbp]
\centering\ArtifactFont\fontsize{9}{11}\selectfont
\setlength{\tabcolsep}{2pt}
\renewcommand{\arraystretch}{1.08}
\caption{Clean-only false-positive rates (\%). The left panel reports overall FPR and FPR conditioned on the number of pathway features placing an example in their upper 10\% (0--3+). The right panel reports overall FPR by source within the mixed corpus.}
\label{tab:fu-clean_only}
\begin{minipage}[t]{0.67\linewidth}
\centering
\begin{tabular*}{\linewidth}{@{\extracolsep{\fill}}lrrrrr@{}}
\toprule
Corpus & Overall & 0 & 1 & 2 & 3+ \\
\midrule
CC3M & 5.00 & 2.81 & 1.57 & 8.71 & 43.46 \\
COCO train & 10.00 & 9.90 & 9.72 & 10.52 & 12.09 \\
COCO validation & 5.01 & 5.45 & 4.78 & 3.87 & 2.82 \\
Mixed sources & 20.00 & 15.39 & 24.07 & 32.59 & 34.81 \\
\bottomrule
\end{tabular*}
\end{minipage}\hfill
\begin{minipage}[t]{0.29\linewidth}
\centering
\begin{tabular*}{\linewidth}{@{\extracolsep{\fill}}lr@{}}
\toprule
Mixed-corpus source & FPR \\
\midrule
CIRR & 25.77 \\
COCO train & 7.07 \\
COCO validation & 5.73 \\
CC3M & 41.42 \\
\bottomrule
\end{tabular*}
\end{minipage}
\end{table}

\subsection{Encoder Robustness}
\label{app:encoders}
\begin{figure}[!htbp]
\centering\ArtifactFont
\includegraphics[width=\linewidth]{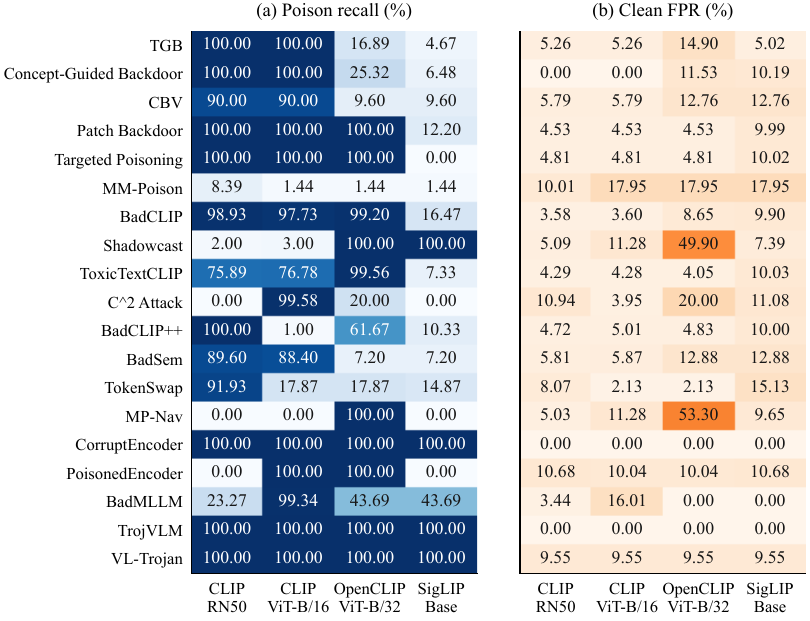}
\caption{TraceGuard filtering with four pretrained encoders on all 19 attack corpora. Poison recall and clean FPR are reported with the same filtering settings across encoders.}
\label{fig:fu-encoder_robustness}
\end{figure}

TraceGuard measures relationships among examples using pretrained image and text encoders, so the default CLIP ViT-B/32 checkpoint may affect which poisons it detects.
We test this dependence by substituting CLIP RN50, CLIP ViT-B/16, OpenCLIP ViT-B/32, and SigLIP Base while keeping the same 19 attack corpora and filtering settings.
For each encoder pair, we recompute all six pathway features and the resulting removal set rather than reusing decisions made with the default encoders.
Figure~\ref{fig:fu-encoder_robustness} reports Poison Recall and Clean FPR to measure how the choice of encoders affects both poison detection and clean-data retention.

Detection remains unchanged across all four encoders for CorruptEncoder and TrojVLM, which retain \(100\%\) Poison Recall with zero Clean FPR, and for VL-Trojan, which retains \(100\%\) recall with \(9.55\%\) Clean FPR.
The same filtering rules can therefore recover these poisons from representations produced by different pretrained models.
Other attacks depend strongly on the encoder, with BadCLIP++ recall changing from \(100\%\) with CLIP RN50 to \(1\%\) with CLIP ViT-B/16, whereas \(C^2\) Attack changes in the opposite direction, from \(0\%\) to \(99.58\%\).
For Shadowcast, OpenCLIP and SigLIP both achieve \(100\%\) recall, but their Clean FPR differs substantially, at \(49.90\%\) and \(7.39\%\).
Thus, the pathway-feature definitions are reusable across encoders, but the separation they provide is not encoder-independent.
The contrast with the stable attacks shows why the existence of an attack pathway alone does not determine how clearly a particular pretrained representation exposes it.

\subsection{Simultaneous Attacks}
\label{app:simultaneous}
\begin{figure}[!htbp]
\centering\ArtifactFont
\includegraphics[width=0.96\linewidth]{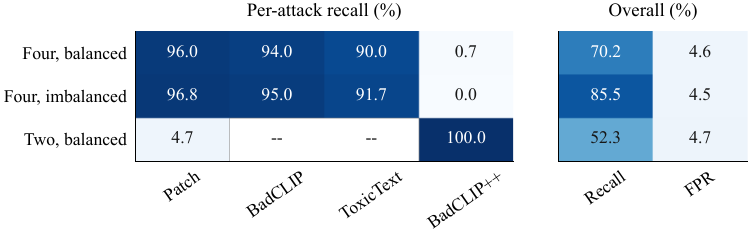}
\caption{TraceGuard detection when attacks coexist. Each mixture has 600 poisons among 100,000 examples and removes 5\% of the corpus. A dash denotes an absent attack. Per-attack recall and aggregate recall and FPR are percentages.}
\label{fig:fu-simultaneous_attacks}
\end{figure}

Simultaneous attacks introduce poisons from several attacks into the same corpus, so detecting one shared attack pattern may leave poisons associated with another pattern behind.
We test whether TraceGuard detects these patterns together by combining attacks that use the same CC3M corpus.
A two-attack mixture contains Patch Backdoor and BadCLIP++, while two four-attack mixtures also include BadCLIP and ToxicTextCLIP.
Each corpus contains 100,000 examples, of which 600 are poisoned, keeping the overall poison fraction fixed while changing the number of attacks and their relative contributions.
The two-attack mixture and one four-attack mixture allocate poisons equally across attacks.
The other four-attack mixture uses unequal counts to test whether attacks contributing fewer poisons are overlooked.
Because aggregate recall can conceal this difference, Figure~\ref{fig:fu-simultaneous_attacks} reports Poison Recall separately for each attack alongside the aggregate Poison Recall and Clean FPR.

At the same \(5\%\) removal fraction, TraceGuard detects Patch Backdoor, BadCLIP, and ToxicTextCLIP together in both four-attack mixtures, with per-attack recall of at least \(90\%\) and aggregate Clean FPR below \(4.61\%\) (Figure~\ref{fig:fu-simultaneous_attacks}).
This demonstrates that selection need not isolate only one attack type.
It nevertheless leaves BadCLIP++ almost entirely undetected, with recall of \(0.67\%\) in the balanced mixture and \(0\%\) in the imbalanced mixture.
Aggregate recall increases from \(70.17\%\) to \(85.50\%\) between these mixtures largely because the poorly detected attack contributes fewer poisons, rather than because its detection improves.

The two-attack mixture further shows that the number of attacks alone does not determine detection difficulty.
BadCLIP++ recall reaches \(100\%\), but Patch Backdoor recall falls to \(4.67\%\), reversing which attack is missed relative to the four-attack mixtures.
Since both poisons and their relative counts change between mixtures, this contrast points to a dependence on corpus composition rather than a uniformly harder problem as more attacks are added.
TraceGuard can recover several coexisting attack patterns, but selecting a single final suspicious set does not ensure coverage of every pattern present.

\FloatBarrier

\section{Computational Cost}
\label{app:cost}

TraceGuard's computational cost comes from computing the pathway features and selecting the suspicious set.
For a corpus of \(N\) examples, representation extraction takes \(O(N)\) time with a fixed encoder and bounded input lengths and text-span erasures per example.
Neighbor retrieval can be more costly because exhaustive search requires \(O(N^2)\) similarity computations.
We therefore use exact search only for small corpora and HNSW approximate search~\citep{malkov2018hnsw} for larger ones, as specified in the experimental setup.

Once the representations and neighbors are available, the feature computations depend on the local neighborhood size \(k\), rather than all pairs in the corpus.
Neighbor Support and Image-Text Support Gap take \(O(Nk)\) time at fixed representation dimension, while sparse evaluation of Neighbor Disagreement takes \(O(Nk^2)\) time because each example has at most \(k^2\) two-step paths in Eq.~\eqref{eq:neighbor-disagreement}.
Text-span counting and erasure comparisons take linear time for bounded spans and fixed \(k\).
Sorting the fixed collection of rankings costs \(O(N\log N)\), and subsequent selection and refinement require linear scans when the numbers of removal fractions and alternative sets are fixed.
These operations together cost \(O(N\log N+Nk^2)\), or \(O(N\log N)\) for the fixed \(k=10\) used here.
Encoding and neighbor retrieval must be added to this bound, with neighbor-index construction and query costs depending on the corpus and search settings.

\begin{figure}[!htbp]
\centering\ArtifactFont
\includegraphics[width=0.92\linewidth]{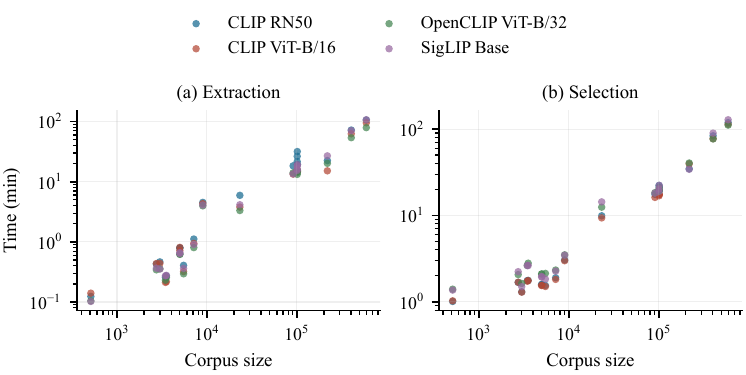}
\caption{Measured TraceGuard component times across encoders and attack corpora. Extraction includes representation extraction and initial scoring. Selection reports the sum of separately timed subsequent stages. All 76 extraction timings and 75 available selection timings are shown; the components are not an end-to-end timing measurement.}
\label{fig:fu-computational_cost}
\end{figure}

We measure runtime across four encoders and 19 attack corpora containing 512 to 591,753 examples in Figure~\ref{fig:fu-computational_cost}.
The figure separates representation extraction and initial scoring from the subsequent selection stages, which are timed separately and summed.
These component timings exclude victim-model training and are not end-to-end measurements.
On corpora of roughly 100,000 examples, these components take 13.2--31.8 and 16.2--22.3 minutes, respectively.
On the largest corpus, extraction and initial scoring take 79.2--107.1 minutes and the subsequent stages take 111.1--128.2 minutes across encoders.
Thus, the cost of examining corpus relationships and selecting samples remains substantial even after the initial representations are available.

Across corpus sizes, fitting log runtime against log corpus size separately for each encoder gives slopes of 0.995--1.056 for extraction and initial scoring and 0.682--0.754 for the subsequent stages, using all available measurements.
Extraction and initial scoring therefore grow approximately linearly over the evaluated range, while the subsequent stages grow more slowly in these measurements.
These trends describe differences across attack corpora rather than an asymptotic bound, but show no quadratic runtime growth on the evaluated corpora of up to about 600,000 examples.

\end{document}